\documentclass[12pt]{article}

\usepackage{epsf,amsfonts,hyperref}
\usepackage[dvips]{color}
\usepackage{amsmath}
\renewcommand{\appendix}[1]{
    \setcounter{equation}{0}
    \renewcommand{\thesection}{\Alph{section}}
    \section{Appendix: \protect\indent #1}
}

\newcommand\encadremath[1]{\vbox{\hrule\hbox{\vrule\kern8pt
\vbox{\kern8pt \hbox{$\displaystyle #1$}\kern8pt}
\kern8pt\vrule}\hrule}}
\def\enca#1{\vbox{\hrule\hbox{
\vrule\kern8pt\vbox{\kern8pt \hbox{$\displaystyle #1$}
\kern8pt} \kern8pt\vrule}\hrule}}

\newcommand\figureframex[3]{
\begin{figure}[bth]
\hrule\hbox{\vrule\kern8pt
\vbox{\kern8pt \vbox{
\begin{center}
{\mbox{\epsfxsize=#1.truecm\epsfbox{#2}}}
\end{center}
\caption{#3}
}\kern8pt}
\kern8pt\vrule}\hrule
\end{figure}
}
\newcommand\figureframey[3]{
\begin{figure}[bth]
\hrule\hbox{\vrule\kern8pt
\vbox{\kern8pt \vbox{
\begin{center}
{\mbox{\epsfysize=#1.truecm\epsfbox{#2}}}
\end{center}
\caption{#3}
}\kern8pt}
\kern8pt\vrule}\hrule
\end{figure}
}

\makeatletter
\@addtoreset{equation}{section}
\makeatother
\newtheorem{theorem}{Theorem}[section]

\newtheorem{remark}{Remark}[section]
\newtheorem{proposition}{Proposition}[section]
\newtheorem{lemma}{Lemma}[section]
\newtheorem{corollary}{Corollary}[section]
\newtheorem{definition}{Definition}[section]
\def\br{\begin{remark}\rm\small}
\def\er{\end{remark}}
\def\bt{\begin{theorem}}
\def\et{\end{theorem}}
\def\bd{\begin{definition}}
\def\ed{\end{definition}}
\def\bp{\begin{proposition}}
\def\ep{\end{proposition}}
\def\bl{\begin{lemma}}
\def\el{\end{lemma}}
\def\bc{\begin{corollary}}
\def\ec{\end{corollary}}
\def\beaq{\begin{eqnarray}}
\def\eeaq{\end{eqnarray}}
\newcommand{\proof}[1]{{\noindent \bf proof:}\par
{#1} $\square$}

\newcommand{\eq}[1]{eq.~(\ref{#1})}

\newcommand{\beq}{\begin{equation}}
\newcommand{\eeq}{\end{equation}}
\newcommand{\bea}{\begin{eqnarray}}
\newcommand{\eea}{\end{eqnarray}}

\renewcommand{\and}{{\qquad {\rm and} \qquad}}

\newcommand{\virg}{{\qquad , \qquad}}

\newcommand{\tr}{{\,\rm tr}\:}

\newcommand{\td}[1]{{\tilde{#1}}}
\newcommand{\ovl}[1]{{\overline{#1}}}

\renewcommand{\l}{\lambda}
\newcommand{\om}{\omega}

\newcommand{\ee}[1]{{{\rm e}^{#1}}}

\newcommand{\Pint}{{\int\kern -1.em -\kern-.25em}}

\renewcommand{\Re}{{\mathrm{Re}}}

\newcommand{\bcycle}{{\cal B}}
\newcommand{\acycle}{{\cal A}}

\newcommand{\genus}{{\mathfrak{g}}}

\renewcommand\l{\lambda}

\newcommand\Res{\mathop{{\rm Res}}}

\begin{document}

\sloppy


\pagestyle{empty}
\hfill SPhT-T09/175
\addtolength{\baselineskip}{0.20\baselineskip}
\begin{center}
\vspace{26pt}
{\large \bf {Topological expansion of the Bethe ansatz,\\ and quantum algebraic geometry}}
\newline
\vspace{26pt}

{\sl L.\ Chekhov}\hspace*{0.05cm}${}^1$
,
{\sl B.\ Eynard}\hspace*{0.05cm}${}^2$
,
{\sl O.\ Marchal}\hspace*{0.05cm}${}^2$
\\
\vspace{6pt}
${}^1\,\,$ Steklov Mathematical Institute, ITEP and Laboratoire Poncelet,\\ 
Moscow, Russia\\ 
\vspace{1pt}
${}^2\,\,$ CEA, IPhT, F-91191 Gif-sur-Yvette, France, \\
CNRS, URA 2306, F-91191 Gif-sur-Yvette, France.\\
\end{center}

\vspace{20pt}
\begin{center}
{\bf Abstract}:
In this article, we solve the loop equations of the $\beta$-random matrix model, in a way similar to what was found for the case of hermitian matrices $\beta=1$.
For $\beta=1$, the solution was expressed in terms of algebraic geometry properties of an algebraic spectral curve of equation $y^2=U(x)$.
For arbitrary $\beta$, the spectral curve is no longer algebraic, it is a Schr\"odinger equation $((\hbar\partial)^2-U(x)).\psi(x)=0$ where $\hbar\propto (\sqrt\beta-1/\sqrt\beta)$.
In this article, we find a solution of loop equations, which takes the same form as the topological recursion found for $\beta=1$.
This allows to define natural generalizations of all algebraic geometry properties, like the notions of genus, cycles, forms of 1st, 2nd and 3rd kind, Riemann bilinear identities, and spectral invariants $F_g$, for a quantum spectral curve, i.e. a D-module of the form $y^2-U(x)$, where $[y,x]=\hbar$.
Also, our method allows to enumerate non-oriented discrete surfaces.
\end{center}

\tableofcontents

\vspace{26pt}
\pagestyle{plain}
\setcounter{page}{1}


\section{Introduction}

\subsubsection*{Spectral invariants and algebraic geometry}

In \cite{Eyn1loop,EOFg}, was presented the definition of spectral invariants $F_g$ for any algebraic plane curve, i.e. given by a polynomial equation 
$$
{\cal E}(x,y)=\sum_{i,j} {\cal E}_{i,j}\, x^i y^j=0.
$$ 

Those invariants $F_g({\cal E})$ are defined in terms of algebraic geometry quantities defined on the Riemann surface of equation ${\cal E}(x,y)=0$. Their definition involves residues at branchpoints of some meromorphic forms.
Their definition provides a natural basis of meromorphic forms of 1st, 2nd and 3rd kind, and a natural framework for all algebraic geometry notions.

Moreover, the invariants $F_g$ of \cite{EOFg} have many nice properties, for instance their deformations under changes of the complex structure of ${\cal E}$ is given by some "special geometry" relations, and provide a natural form-cycle duality.
Also, they are invariants under changes of ${\cal E}$ which conserve the symplectic form $dx\wedge dy$ in $\mathbb C\times \mathbb C$, they have nice modular properties, and finally, they define the tau-function of some dispersionfull integrable system associated to ${\cal E}$.

Also, those invariants $F_g$ have deep relationships with enumerative geometry, for instance they have been related to the Kodaira-Spencer theory \cite{DVKS}, to combinatorics of discrete surfaces (maps), to intersection theory \cite{EOFg, EynVolmum}, and they are conjectured to be equal to the Gromov-Witten invariants of some toric Calabi-Yau target 3-folds \cite{BKMP}.

\subsubsection*{Algebraic geometry on "quantum" curves}

Here, our goal is to define those notions for a {\bf "quantum curve"}, where ${\cal E}(x,y)$ is a non-commutative polynomial of $x$ and $y$:
\beq\nonumber
{\cal E}(x,y)=\sum_{i,j} {\cal E}_{i,j}\, x^i\, y^j
\qquad , \quad
[y,x]=\hbar.
\eeq
The notion of quantum curve has arised in many ways in the litterature \cite{DDmodules}, and is also called D-modules, i.e. a space of functions quotiented by ${\rm Ker}\, {\cal E}(x,y)$, where $y=\hbar \partial/\partial x$.

In other words, one has to study functions $\psi(x)$ such that:
\beq\nonumber
{\cal E}(x,\hbar \partial_x).\psi(x)=0.
\eeq

In our attempt to define the spectral invariants analogous to those of \cite{EOFg} for such D-modules, we are naturally led to define all analogous properties of algebraic geometry.
For instance we define the notions of {\bf branch points}, {\bf sheets}, {\bf genus}, {\bf cycles}, {\bf forms}, {\bf Bergman kernel}, and so on...

Because of non-commutativity, some notions like branch-points, cuts and sheets, become "blurred" or "non-localized", i.e. the branchpoint is  no longer a point, but a "region" of the complex plane, and cuts are asymptotic accumulation lines of points. 

But, otherwise, it is surprising to find that almost all relationships of classical algebraic geometry, remained unchanged when $\hbar\neq 0$, for instance the Riemann bilinear identity, the Rauch variational formula, and the topological recursion defining the spectral invariants.

\medskip

Moreover, we shall find, that in order for our quantities to make sense, we must have a "vanishing monodromy" condition, which can be interpreted as a {\bf Bethe ansatz}, and this gives a geometrical interpretation of the Bethe ansatz.

\medskip

Let us also mention that in a previous article \cite{EynOM}, we treated a special case, where the Schr\"odinger potential $U(x)$ was quantized, and we shall see, under the light of this new work, that it was the case of a degenerate quantum surface, with no branchpoints.

\subsubsection*{Hyperelliptical case}

Here, for simplicity, we shall restrict ourselves  to polynomials of degree $2$ in $y$ (called hyperelliptical in algebraic geometry), of the form:
\beq\nonumber
{\cal E}(x,y) = y^2 - U(x)
\virg
[y,x]=\hbar
\eeq
i.e. to the Schr\"odinger equation:
\beq\nonumber
\hbar^2 \psi'' = U \psi.
\eeq
We leave the higher degree case for a further work.

\subsubsection*{Link with $\beta$ matrix models}

The spectral invariants $F_g$ were first introduced for the solution of loop equations arising in the 1-hermitian random matrix model \cite{Eyn1loop, ChekEynFg}. They were later generalized to other hermitian  multi--matrix models \cite{CEO, EPrats}.

There exist other matrix models, which are defined with non hermitian matrices. In fact it is well known since Wigner \cite{Mehtabook} that depending on the symmetry of the problem, it is sometimes interesting to have matrices that are not hermitian. (For example, real-symmetric, unitary, orthogonal or quaternionic, ...). Therefore, it seems reasonable to extend the definition of the spectral invariants for those other models. 
Those other matrix models are often called $\beta$-ensembles, and they are classified by an exponent $\beta$.
The 3 Wigner ensembles (see \cite{Mehtabook}, and we changed $\beta\to \beta/2$) correspond to $\beta=1$ (hermitian case), $\beta=1/2$ (real symmetric case), $\beta=2$ (real self-dual quaternion case), but it is easy to define a $\beta$ one-matrix model for any other value of $\beta$ (see section \ref{secMM} for more details).

\medskip

In \cite{ChekEynbeta}, a first attempt to generalize the solution of \cite{Eyn1loop} to other matrix models was proposed, but it was not as nice as the topological recursion of \cite{Eyn1loop}.
In \cite{ChekEynbeta}, it was assumed that $\beta=O(1)$ when the size $N$ of the random matrix becomes large, and it was found that all spectral invariants were related to a double series expansion of the form:
\beq\nonumber
\sum_{g,k}\, N^{2-2g-k}\,\,(\sqrt\beta-1/\sqrt\beta)^k F_{g,k}
\eeq
The coefficients $F_{g,k}$ were computed in \cite{ChekEynbeta}.
Here, in this article, we shall work at fixed $\hbar=(\sqrt\beta-1/\sqrt\beta)/N$, instead of fixed $\beta$, i.e. we shall define the resummed $F_g$'s as:
\beq\nonumber
F_g(\hbar) = \sum_k \, \hbar^k\, F_{g,k}.
\eeq
The $F_{g,k}$'s of  \cite{ChekEynbeta} can be recovered by computing the semi-classical small $\hbar$ expansion of $F_g(\hbar)$.
In this article we shall argue that $F_g(\hbar)$ is the natural generalization of the symplectic invariants of \cite{EOFg} for a "quantum spectral curve" ${\cal E}(x,y)$ with $[y,x]=\hbar$.

\medskip

The tool which we use for studying the $\beta$-matrix model, is the loop equation method. Loop equations are related to the invariance of an integral under change of variable. They can be obtained by integrating by parts. Loop equations for the $\beta$-matrix model have been written many times \cite{Dum, eynbeta}, and here we show how to solve them order by order in $1/N$, at fixed $\hbar$.

\medskip

The $\beta$-matrix model and its loop equations are explained in section \ref{secMM}.

\section{Schr\"odinger equation and Bethe ansatz}\label{secdef}

\subsection{Schr\"odinger equation, generalities and notation}
Let:
\beq\label{eqSchroedinger}
\hbar^2 \psi''(x) = U(x)\, \psi(x)
\eeq
be a Schr\"odinger equation with $U(x)$ a polynomial.
Let $U(x)$ be a polynomial of degree $2d$, and define the polynomial "potential" $V(x)$ of degree $d+1$ by its derivative:
\beq\label{defV'}
V'(x) = 2\,(\sqrt{U})_+ = \sum_{k=0}^d t_{k+1}\,x^k
\eeq
where $()_+$ means the polynomial part of the Laurent series at $x\to\infty$.
We also define:
\beq\label{defP}
P(x) = \frac{V'^2(x)}{4} - U(x) -\hbar \frac{V''(x)}{2}
\eeq
so that $P$ is a polynomial of degree $d-1$.

Eventually, we define:
\beq\label{deft0}
t_0 = \mathop{{\rm lim}}_{x\to\infty}\, \frac{xP(x)}{V'(x)}
\eeq

\br
Just in order to give names to those parameters, let us say that
in the language of integrable systems, the coefficients $t_0,t_1,t_2,\dots,t_{d+1}$ are called the ``Casimirs'', and the remaining coefficients of $P$ are the ``conserved charges''. They will play a special role later on in this article.
In matrix model language (see section \ref{secMM}), $t_1,\dots,t_{d+1}$ are called the times associated to the potential $V(x)$, $t_0$ is often called the temperature, and the remaining coefficients of $P$ are called "filling fractions".
In the language of algebraic geometry, parameters $t_k$ with $k\geq 1$ are coupled to 2nd kind meromorphic differential forms, $t_0$ is coupled to 3rd kind, and the remaining coefficients of $P$ are coupled to 1st kind holomorphic differentials, see section \ref{variations} about form-cycle duality.

\er

\subsubsection{Stokes Sectors}

From the study of the Schr\"odinger equation we know that the function $\psi(x)$ is subject to the Stokes phenomenon, i.e. although $\psi(x)$ is an entire function, its asymptotics look discontinuous near $\infty$. We therefore need to introduce properly the Stokes sectors by defining the following quantities:
Let
\beq\nonumber
\theta_0= {\rm Arg}(t_{d+1})
\eeq
be the argument of the leading coefficient of $V(x)$.

We define the Stokes lines going to $\infty$ as:
\beq
L_k = \left\{x\,\, / \,\, {\rm Arg}(x) =-\frac{\theta_0}{d+1} + \pi \,\frac{k+\frac{1}{2}}{d+1}\, \right\}
\eeq
Those are the lines where $\Re V(x)$ vanishes asymptotically.

We define the sectors:
\beq
S_k =  \left\{{\rm Arg}(x) \in ]-\frac{\theta_0}{d+1}+ \pi \,\frac{k-\frac{1}{2}}{d+1},-\frac{\theta_0}{d+1}+ \pi\,\frac{k+\frac{1}{2}}{d+1}[\, \right\}
\eeq
i.e. $S_k$ is the sector between $L_{k-1}$ and $L_k$.

Notice that in even sectors we have asymptotically $\Re V(x)>0$ and in odd sectors we have $\Re V(x)<0$.

\figureframex{9}{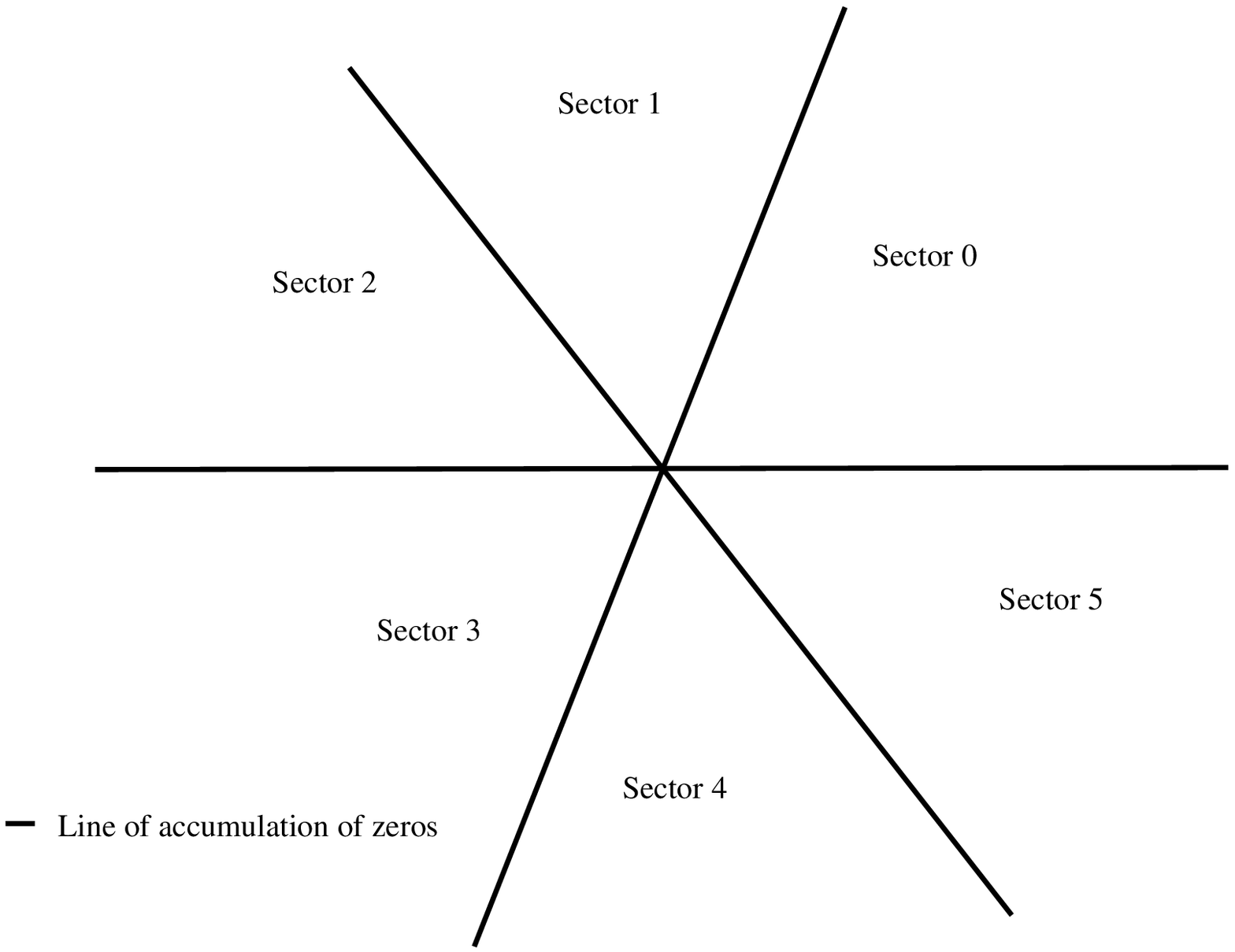}{\label{figstokessectors} Example of sectors for a potential of degree $\deg V=3$, i.e. $d=2$. If $\deg V=d+1$ there are $2d+2$ sectors.}

\subsubsection{Stokes phenomenon}

Any solution of a linear equation, is analytical where the coefficients of the equation are analytical, and it may possibly have essential singularities where the coefficients are singular. Here, $U(x)$ is an entire function with a singularity (a pole), only at $\infty$, thus, any solution $\psi$ is an entire function with a possible essential singularity at $\infty$. The asymptotics of $\psi$ near $\infty$ are subject to the Stokes phenomenon. This means that, although $\psi$ is analytical in the whole complex plane, its asymptotics at infinity may change from sectors to sectors.

\bigskip

From the study of the Schr\"odinger equation it is known that in each sector $S_k$, $\psi(x)$ has a large $x$ expansion:
\beq\nonumber
\psi(x)  \mathop{{\sim}}_{S_k} \ee{\pm\,{1\over 2\hbar}V(x)}\,x^{C_{k}}\,\, (A_k+\frac{B_k}{x}+\dots)
\eeq
and the sign $\pm$, may jump discontinuously from one sector to another as well as the numbers $A_k, B_k, C_k,\dots$  (and in general, all the coefficients of the series in $\frac{1}{x^j}$ at infinity).

\subsection{Decreasing solution}

Let us consider a specific solution $\psi(x)$ of the Schr\"odinger equation which is exponentially decreasing in some even sector at infinity. For writing convenience, \textbf{we will choose $\psi(z)=\psi_0(z)$ a decreasing solution in sector $S_0$. Without further indication, $\psi(z)$ is now understood to be $\psi_0(z)$ in the rest of the article.} Note that this choice is quite arbitrary at the moment, and one should wonder if the quantities we are about to compute depend on this choice, but we are presently not able to answer this question properly, and leave it for further study.

\bigskip

An important and useful result is the Stokes theorem which claims that if the asymptotics of $\psi(x)$ is exponentially small in some sector, then the same asymptotics holds in the two adjacent sectors (and therefore $\psi(x)$ is exponentially large in those two sectors). 

\bigskip

In the general case, (i.e. a generic potential $U(x)$) our solution $\psi(x)$ is decreasing only in sector $0$, and is exponentially large in all other sectors. But if the Schr\"odinger potential $U(x)$ is non-generic (quantized), then there may exist several sectors in which $\psi(x)$ is exponentially small. Due to Stokes theorem, if $\psi$ is exponentially small in some sectors then it must be exponentially large in the adjacent sectors, this implies that there are at most $d+1$ sectors in which $\psi$ is exponentially small.

\medskip
The case studied in \cite{EynOM} was the most degenerate case, such that $\psi$ is exponentially small in $d+1$ sectors.

\subsubsection{Zeroes of $\psi$}

The main difference with our previous article \cite{EynOM} is that we will not restrict ourselves to the case where $\psi(x)$ is a quasi-polynomial which can only be obtained with very non-generic potential $U(x)$. Here $\psi(x)$ is an entire function with an essential singularity at $\infty$, and with isolated zeroes labelled $s_i$:
\beq
\psi(s_{i})=0
\eeq
\textbf{In particular, the number of zeroes of $\psi$ may be finite or infinite.}

If $\psi(x)$ has an infinite number of zeroes, it is known that the zeroes may only accumulate near $\infty$, and only along the Stokes half-lines $L_j$'s bordering the sectors (see fig.\ref{stokeslines}). In fact, there is an accumulation of zeroes along the half--line $L_j$ if and only if $\psi$ is exponentially large on both sides of the half-line. 

For example in the case where $\psi(x)=\psi_k(x)$ is a solution that exponentially decreases in sector $k$ then it implies that there is no accumulation of zeroes along the half-lines $L_k$ and $L_{k-1}$.

\medskip

If $U(x)$ is generic, then $\psi$ has an infinite number of zeroes, and the zeroes accumulate at $\infty$ along all half-lines $L_j$ with $j\neq 1,2d+1$ (because remember that $\psi$ is implicitely assumed to be $\psi_0$ which decreases in sector $0$), i.e. there are generically $2d$ half-lines of zeroes. The situation is illustrated in fig \ref{stokeslines}.

\figureframex{9}{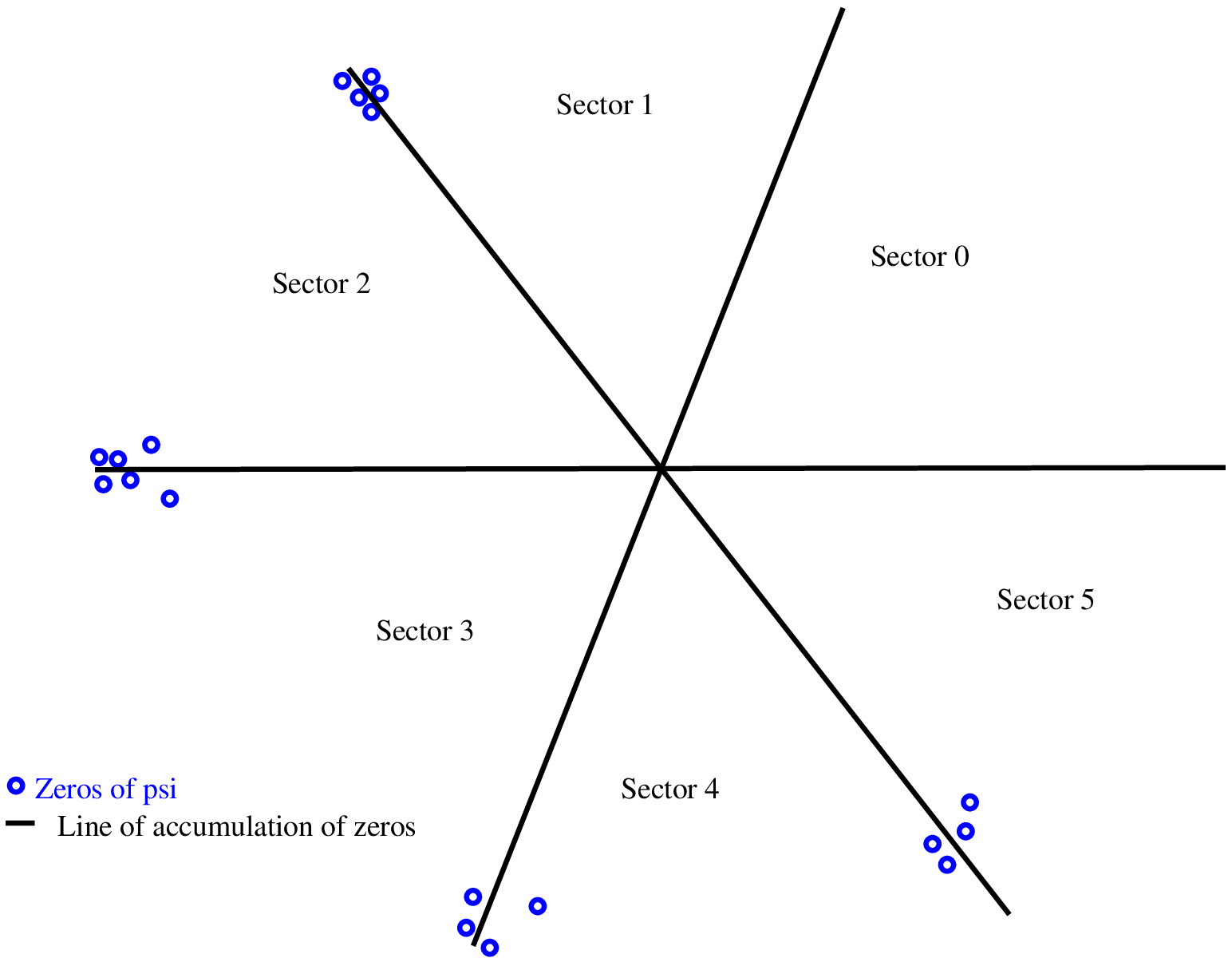}{\label{stokeslines} The zeroes of $\psi$ accumulate near $\infty$ along the half-lines bordering sectors where $\psi_0$ is exponentially large on both sides. In particular, there is no accumulation of zeroes along $L_0$ and $L_{2d+1}$.}

If $U(x)$ is non-generic (quantized), then there are additional sectors in which $\psi$ is exponentially small, and thus there can be no zeroes accumulating along the two half-lines bordering these sectors. Remember that from Stokes theorem, each time we have a new sector in which $\psi$ is decreasing, we have two half-lines less of zeroes.
Therefore, the number of half-lines of zeroes is always even, and we call it:

\bd The genus $\genus$ of the Schr\"odinger equation is defined by:
\beq
2 {\mathfrak g}+2=\#\,{\hbox{ half-lines\,\, of\,zeroes}}
\eeq
And if $\psi$ has a finite number of zeroes (i.e. there is no half-line of zeroes), we define ${\mathfrak g}=-1$. 
We have
\beq -1\leq \genus\leq d-1\eeq

\ed

\medskip
Note also that the definition of $\genus$ a priori depends on the choice of the solution $\psi=\psi_0$ since two different solutions of the same Schr\"odinger equation may have different numbers of semi-lines of zero accumulation.

An exception, is in the special cases $\genus=-1$ 
where it is easy to see that every choice of $\psi=\psi_{2k}$ would give the same value of $\genus$.

Indeed, consider $\genus_{{2k}}$ and $\genus_{{2k'}}$ be the genus defined from the solutions $\psi_{2k}$ exponentially small in sector $S_{2k}$ and $\psi_{2k'}$ exponentially small in sector $S_{2k'}$: 

\smallskip
if $\genus_{{2k}}=-1$, this means that $\psi_{2k}$ is exponentially small in all even sectors, in particular in sector $S_{2k'}$, and therefore $\psi_{2k}\propto \psi_{2k'}$, and therefore $\genus_{{2k'}}=-1$.


\subsubsection{Case $\genus=-1$}

The case $\genus=-1$ was studied in \cite{EynOM}.
This is the case where $\psi$ has only a finite number of zeroes, it is a quasipolynomial:
\beq
\psi(x)\,\ee{{\hbar\over 2}\,V(x)} = {\rm polynomial}.
\eeq

Notice that in order to diminish $\genus$ by $1$, we need to quantize one parameter of $U$, and therefore to reach $\genus$, we need to quantize $d-1-\genus$ parameters.
In particular, to reach $\genus=-1$, we need to quantize $d$ parameters, i.e. $P$ is completely fixed in terms of $V'$, and in particular, $t_0$ is quantized.

In the applications to random matrices, $t_0$ is usually a free parameter (called the temperature) and is never considered quantized, and therefore the case $\genus=-1$ is never obtained in random matrices.

Another way to say that, is that the case $\genus=-1$ has no $\hbar\to 0$ classical limit, and therefore in classical geometry we always have $\genus\geq 0$.

\subsection{Resolvent}

The first ingredient of our strategy is to define a resolvent similar to the one in matrix models.

\bd
We define the resolvent for a generic solution $\psi$ by:
\beq
\om(x) = \hbar \frac{\psi'(x)}{\psi(x)} +\frac{V'(x)}{2}
\eeq
\ed

It is clear that this function is analytical except at the zeros of $\psi(x)$ where it has \textbf{simple} poles with residue $\hbar$:
\beq
\om(x) \mathop{{\sim}}_{x\to s_{i}} {\hbar \over x-s_{i}} + {\rm reg}.
\eeq
It also has a possible essential singularity at infinity with the same location of discontinuities as $\psi(x)$. Eventually, note again that the definition of $\om(x)$ depends on the choice of $\psi(x)$.

\subsection{Sheets}

In sector $S_k$ we have the asymptotic:
\beq
\psi(x)  \mathop{{\sim}}_{S_k} \ee{{\eta_k\over 2\hbar}V(x)}\,x^{-{\eta_k t_0\over \hbar} - d\,{1+\eta_k\over 2}}\,\, (A_k+\frac{B_k}{x}+\dots)
\eeq
where $\eta_k=\pm 1$. That translates for the resolvent to:
\beq
\om(x) \mathop{{\sim}}_{x\to \infty_k}  {1+\eta_k\over 2}\,(V'(x)-\hbar {d\over x}) - \frac{\eta_k\,t_0}{x}  +O(1/x^2),
\eeq
Therefore it depends if the solution $\psi$ is exponentially big or small in sector $k$ (and of course on the parity of $k$). For a generic $\genus=d-1$ solution which is exponentially big in every sector except $S_0$ (and thus has an alternating sign in the exponential) then $\eta_k= (-1)^k$ (except $\eta_0=-1$). 

\bd
We call {\bf "physical sheet"}, the union of sectors where $\eta_k=-1$, in those sectors we have:
\beq
\om(x) \sim {t_0\over x}+O(1/x^2)
\eeq
Notice that the sectors $S_0,S_1$ and $S_{2d+1}$ are always in the physical sheet.

And we call {\bf "second sheet"}, the union of sectors where $\eta_k=+1$, in those sectors we have:
\beq
\om(x) \sim V'(x)+O(1/x)
\eeq

\ed
This definition comes from the analogy with the resolvent in matrix model (see section \ref{secMM} for details).

For a generic potential $U(x)$, all odd sectors are in the physical sheet, and all even sectors except $S_0$ are in the second sheet.

Notice that if $\genus=-1$, there is only the physical sheet, i.e. there is no second sheet.

\subsection{The Bethe ansatz}

In the polynomial case studied before \cite{EynOM}, a key ingredient for establishing results was the Bethe ansatz. This ansatz basically deals with the behaviour of $\om(x)$ around zeroes of $\psi$. 
The zeroes of $\psi$ are called "Bethe roots".

The Bethe ansatz can be formulated in many ways. One way to formulate it, is to say that $1/\psi^2$ has no residue at the $s_i$'s:
\beq\label{Betheansatz1}
\Res_{s_i} {1\over \psi^2(x)} = 0
\eeq
 in this way, it will play a key role in defining contour integrals, because all integrals of the type $\int dx/\psi^2(x)$ are insensitive to the exact location integration path with respect to the $s_i$'s, i.e. such integrals will depend only on the homotopy classes of paths.

\smallskip
Equation (\ref{Betheansatz1}) can also be formulated, in a form very similar to the Bethe ansatz in the Gaudin model \cite{Gaudin, BBTbook} as follows:
\bt
The roots $s_i$ of $\psi$ satisfy the Bethe ansatz:
\beq \label{bethe ansatz}
\forall\, i \, , \qquad \quad V'(s_{i}) = 2\, \mathop{{\rm lim}}_{x\to s_i}\,\, \left(\om(x)-{\hbar\over x-s_i}\right).
\eeq

\et

\smallskip

It is a regularized version of the Bethe equation for Gaudin model:
$$
\forall i\, ,\, \qquad V'(s_{i}) \,\, "="\,\, 2\hbar\,\sum_{j\neq i} {1\over s_{i}-s_{j}}
$$
when the number of zeros is infinite and the sum is ill-defined.

\medskip

\proof{This theorem is a classical result and is easy, it just consists in rewriting the Schr\"odinger equation as a Ricatti equation.
We proceed the same way as in \cite{EynOM} and compute:
\bea\label{eqBethe1}
&& V'(x)\om(x) - \om^2(x) - \hbar \om'(x) \cr
&=& V'(x)(\hbar \frac{\psi'(x)}{\psi(x)} +\frac{V'(x)}{2}) - \left({V'(x)^2\over 4}+\hbar V'(x) \frac{\psi'(x)}{\psi(x)}+\hbar^2 \frac{\psi'(x)^2}{\psi^2(x)}\right)\cr
&& -\hbar\left(\hbar{\psi''(x)\over \psi(x)}-\hbar {\psi'^2(x)\over \psi^2(x)} + {V''(x)\over 2}\right) \cr
&=& {V'(x)^2\over 4} - \hbar^2{\psi''(x)\over \psi(x)} -\hbar {V''(x)\over 2} \cr
&=& {V'(x)^2\over 4} - U(x) -\hbar {V''(x)\over 2} \cr
&=& P(x)
\eea
which is a polynomial in $x$, of degree $d-1$.

From its definition, it is clear that $\om^2+\hbar\om'$ has no double pole at the $s_{i}$'s, but it could have simple poles. Consider now a zero $s_{i}$ of $\psi$, and define:
\beq\nonumber
\bar\om_{i}(x) = \om(x) - {\hbar \over x-s_{i}}
\eeq
Then, $\bar\om_{i}(x)$ is regular at $x=s_{i}$, and we may compute $\bar\om_{i}(s_{i})$.
Compute:
\bea
\Res_{x\to s_{i}} \om^2(x)+\hbar \om'(x)
&=& \Res_{x\to s_{i}}\bar\om^2_{i}(x)+2\hbar {\bar\om_{i}(x)\over x-s_{i}} + {\hbar^2\over (x-s_{i})^2}  + \hbar \bar\om'_{i}(x) - {\hbar^2\over (x-s_{i})^2} \cr
&=& \Res_{x\to s_{i}} 2\hbar {\bar\om_{i}(x)\over x-s_{i}} \cr
&=& 2\hbar\, \bar\om_{i}(s_{i})
\eea
On the other hand we have, from \eq{eqBethe1} we have:
\bea
\Res_{x\to s_{i}} \om^2(x)+\hbar \om'(x)
&=& \Res_{x\to s_{i}} V'(x) \om(x) - P(x) \cr
&=& \Res_{x\to s_{i}} V'(x) \om(x)  \cr
&=& \hbar\,V'(s_{i})
\eea

Therefore we find :
\beq\nonumber
\forall\, i \, , \qquad \quad V'(s_{i}) = 2\, \bar\om_{i}(s_{i}).
\eeq
This equation is the Bethe equation for the roots $s_{i}$'s. Note that the potential $V'(x)$ is completely determined by the data of the potential $U(x)$ and does not depend on $\psi$. In particular, in the case where there are only a finite number of $s_{i}$'s we recognize the Bethe equation for Gaudin model \cite{EynOM}:
\beq\nonumber
\forall i\, ,\, \qquad V'(s_{i}) = 2\hbar\,\sum_{j\neq i} {1\over s_{i}-s_{j}}
\eeq
which were completely defining the $s_i$'s.
}

\section{Towards a "Quantum Riemann Surface"}

From the definition of our non-commutative spectral curve (i.e the Schr\"odinger equation), it is tempting to generalize the classical notions kwown in algebraic geometry and Riemann surfaces to our "quantum" case ("quantum" is not to be understood as "quantized" but as "non-commutative" $[y,x]=\hbar$). For a Riemann surface, the central notions are those of cuts, sheets, genus, cycles and meromorphic differentials forms of 1st, 2nd and 3rd kind. In our context, the picture needs a proper adaptation in order to recover the terminology of Riemann surfaces and algebraic geometry. 

In this section we will define the notions of genus, $\acycle$-cycles, $\bcycle$-cycles and the first kind differentials dual to them.
Here, let us assume that $\genus\geq 0$.

\subsection{Cuts}

First, we like to think of the 2 sheets, as the sectors which correspond to the 2 possible behaviors of the resolvent at $\infty$: $\om(x) \sim {t_0/x}$ (physical sheet) or $\om(x)\sim V'(x)$ (second sheet).

Then, we consider the cuts as sets of roots $s_i$'s. In some sense, each pair of half lines of accumulation of zeroes can be thought of as a cut.

\bd
We define cuts as pairs of half-lines of zeroes.

There is some arbitrariness in grouping the half-lines of zeroes by pairs.
\ed
There is $\genus+1$ cuts, like in classical algebraic geometry, and notice that the case $\genus=-1$ which has no classical counterpart, has no cuts.

\smallskip

Notice that, contrarily to classical geometry, where the endpoints of the cuts are zeroes of $U(x)$, here the endpoints are somehow blurred, we may move a finite number of $s_i$'s  from one cut to another.





\subsection{Cycles}

In standard algebraic geometry, the non-contractible $\acycle$-cycles are often thought of as surrounding cuts in the physical sheet, and their dual $\bcycle$-cycles are going through the cuts, from one sheet to the other, see fig \ref{figclassicalalgebraicgeometry}.

\figureframex{14}{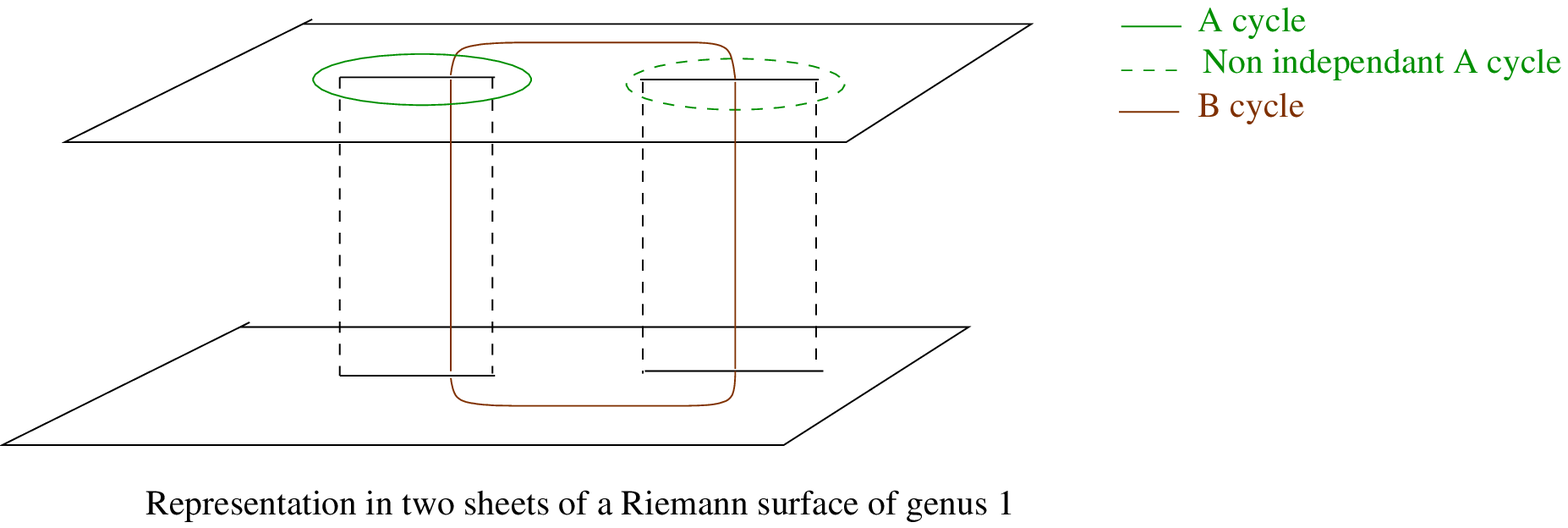}{Representation in two sheets of a Riemann surface of genus $1$.{
\label{figclassicalalgebraicgeometry}}}

\subsubsection{A-Cycles}

Consider the complex plane from which we remove the second sheet (sectors where $\om(x)\sim V'(x)$).
It is clear that it contains $\genus+1$ sectors near $\infty$, and there are $\genus$ homologically linearly independent contours which link them.

\bd
We define $\acycle$-cycles $\acycle_1,\dots,\acycle_{\genus}$ as $\genus$ linearly independent non-contractible contours going from $\infty$ to $\infty$ in the physical sheet.

A choice of $\acycle$-cycles is not unique.
\ed

Remark that this notion really makes sense only for $\genus\geq 1$.

\medskip

Notice that each time $\psi(x)\sim \ee{-V(x)/2\hbar}$ in an even sector, it means it is exponentially small and thus it also behaves like $\ee{-V(x)/2\hbar}$ in the neighboring odd sectors.
That means that we can always choose $\acycle$-cycles going from odd sector to odd sector.

\medskip

Since the first sheet and second sheet are separated by half-lines of accumulations of zeroes, every $\acycle$-cycle surrounds an even number of such half-lines of accumulations of zeroes, i.e. surrounds the cuts in the physical sheet.
Like in standard algebraic geometry, the cuts are identified as pairs of half-lines of zeroes accumulations and the $\cal{A}$ cycles are going enclosing these cuts. 

\subsubsection{Examples}

In the generic case $\genus=d-1$,  we can define $d$ $\cal{A}$-cycles but only $d-1$ are linearly independent. See picture where $d=7$:
\bigskip

{\mbox{\epsfxsize=14.truecm\epsfbox{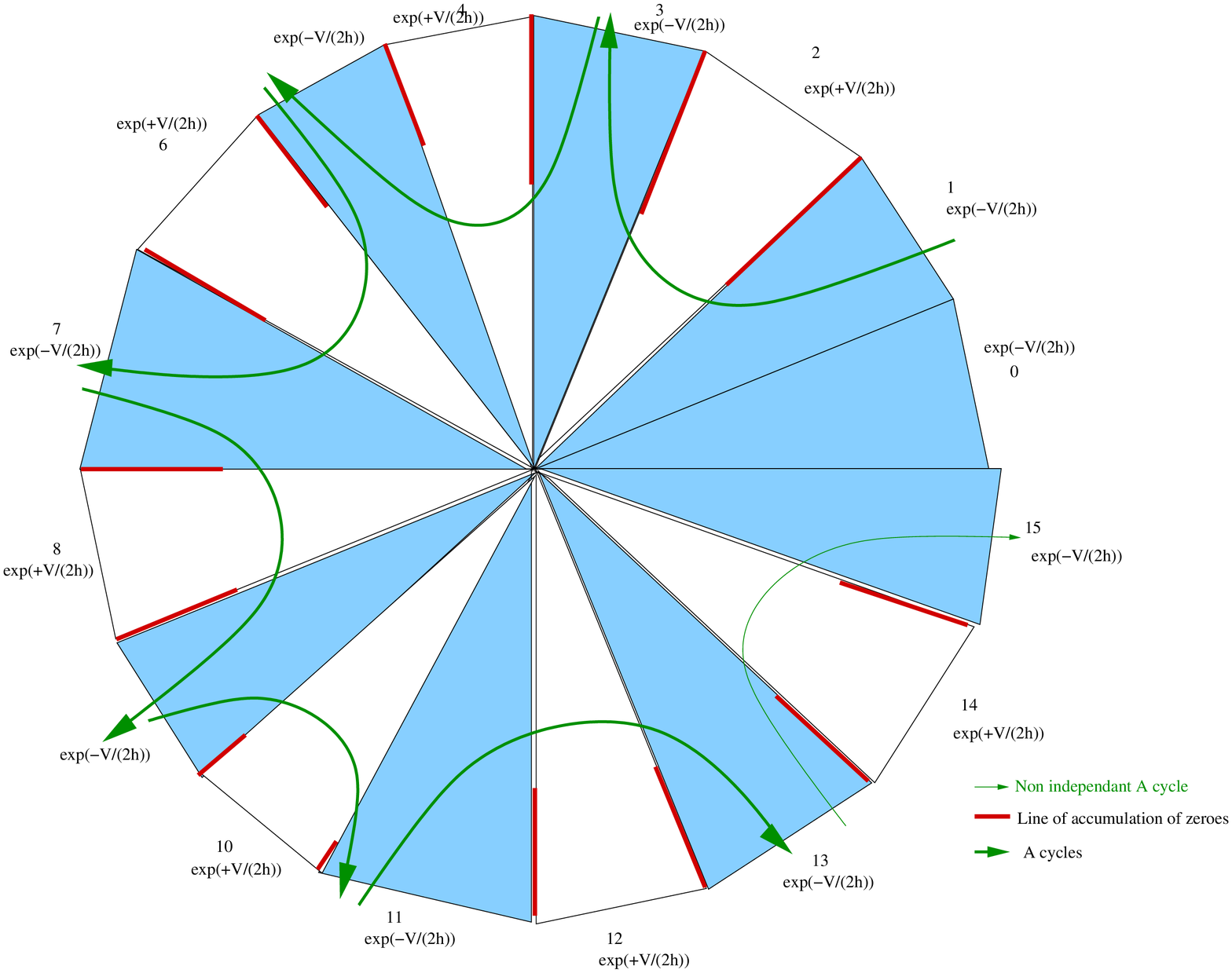}}}

We clearly see that the dashed contour is not linearly independent with the others since the global sum of the contours (dashed included) is contractible in the physical sheet.

For a non-generic case, there are sectors at infinity where $\psi$ is exponentially small. In these cases, the definition of the contours need some adaptations because these sectors correspond to "degenerate" cuts. Here are a few examples of how to deal with these cases. Basically, each time there are two sectors where $\psi$ is small we can replace one of the standard $\cal{A}$ cycle, by a $\hat{ \cal{A}}$ cycle (sometimes called also "degenerate" $\cal{A}$ cycles) that connect them. Here are some examples of the contours in more and more peculiar situations for $d=7$:
\bigskip

{\mbox{\epsfxsize=14.truecm\epsfbox{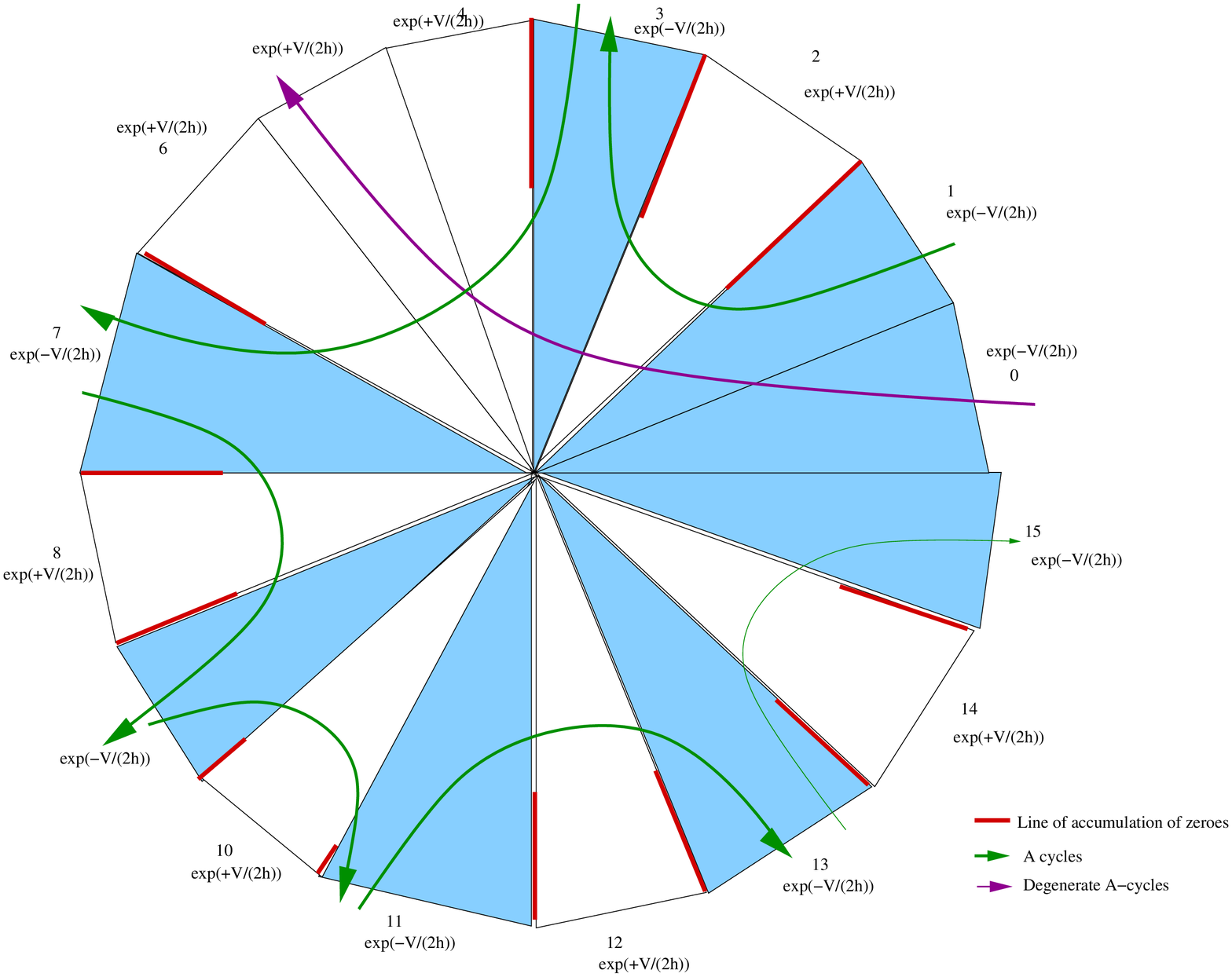}}}

From then it is easy to generalize into more complicated frames:
\bigskip

{\mbox{\epsfxsize=14.truecm\epsfbox{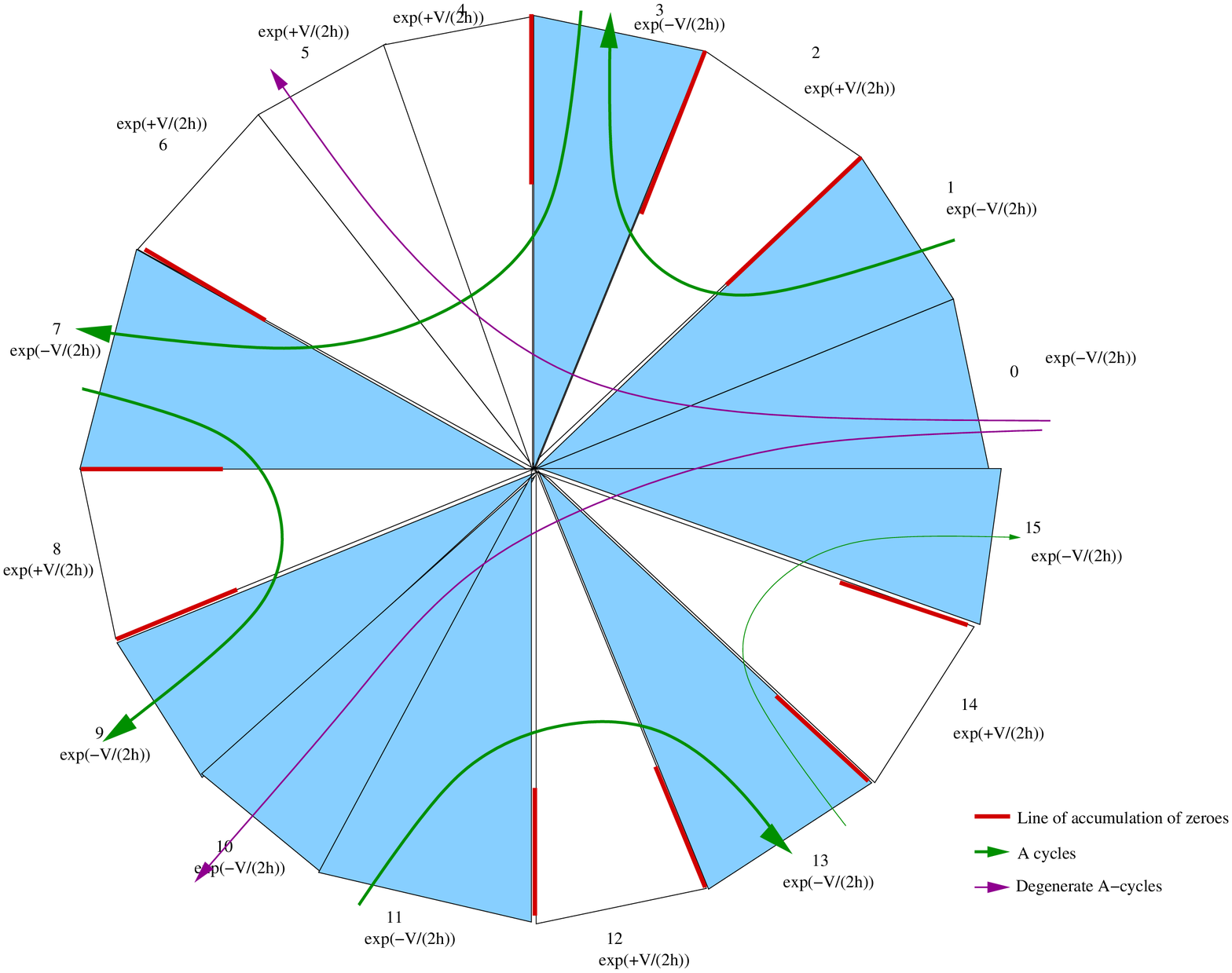}}}

It is then easy to generalize the method in more sophiticated situations. 

In the extreme  case where $\psi$ is exponentially small in all even sectors, there are only $d$ independant  "degenerate" $\hat{ \cal{A}}$ cycles and no $\cal{A}$ cycles, the genus is $\genus=-1$. 
This is the polynomial case studied in \cite{EynOM} where there are no $\cal{A}$ cycles. 

\medskip

From the definitions, it is easy to see that the genus $\genus$ defined above corresponds to the number of independant $\cal{A}$ cycles (we exclude the $\hat{\cal{A}}$ cycles). It is also obvious that the sum of independant $\cal{A}$ and $\hat{ \cal{A}}$ cycles always equals $d-1$.

\subsubsection{B-Cycles}

As in classical algebraic geometry, it is standard to define the $\cal{B}$ cycles with an origin lying in the non-independant cut. Moreover, although it would be possible to define $\hat{\cal{B}}$ cycles attached to the $\hat{\cal{A}}$ cycles, we prefer limiting ourselves to the definition of $\cal{B}$ cycles attached only to the $\cal{A}$ cycles. Basically, they start from the non-independant cut, goes through their corresponding $\cal{A}$ cycle and end at infinity in the same sector as their corresponding $\cal{A}$ cycle. As there are two sectors in which their corresponding $\cal{A}$ cycle ends, we double them so that one goes into one sector and the other one in the second sector. We also choose the whole so that they intersect only with their corresponding $\cal{A}$-cycles:
\beq
\acycle_\alpha\cap \bcycle_\beta=2 \delta_{\alpha,\beta}
\eeq

This definition is easier understandable with the following pictures: 

Generic case:

{\mbox{\epsfxsize=14.truecm\epsfbox{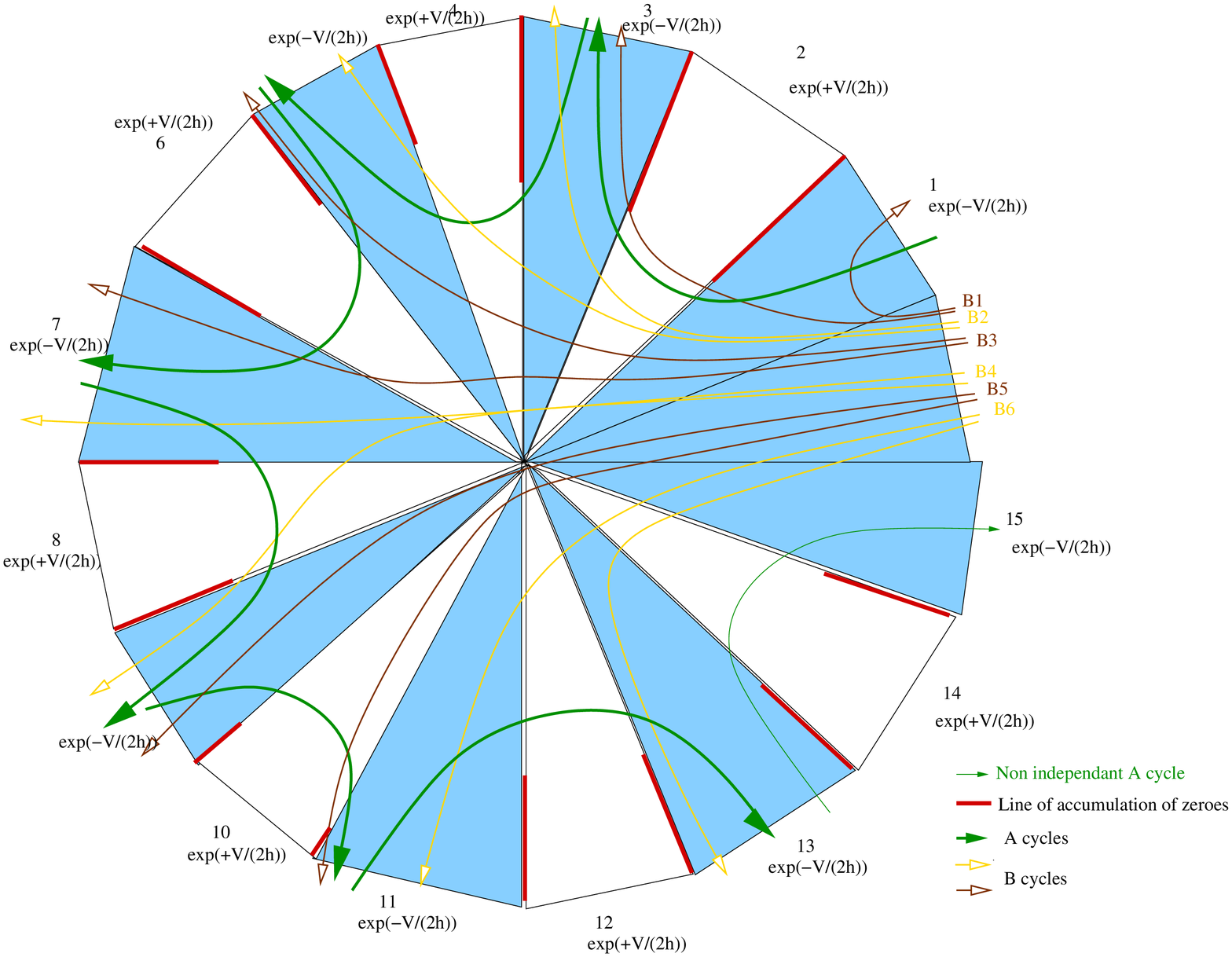}}}

And in a degenerate case:

{\mbox{\epsfxsize=14.truecm\epsfbox{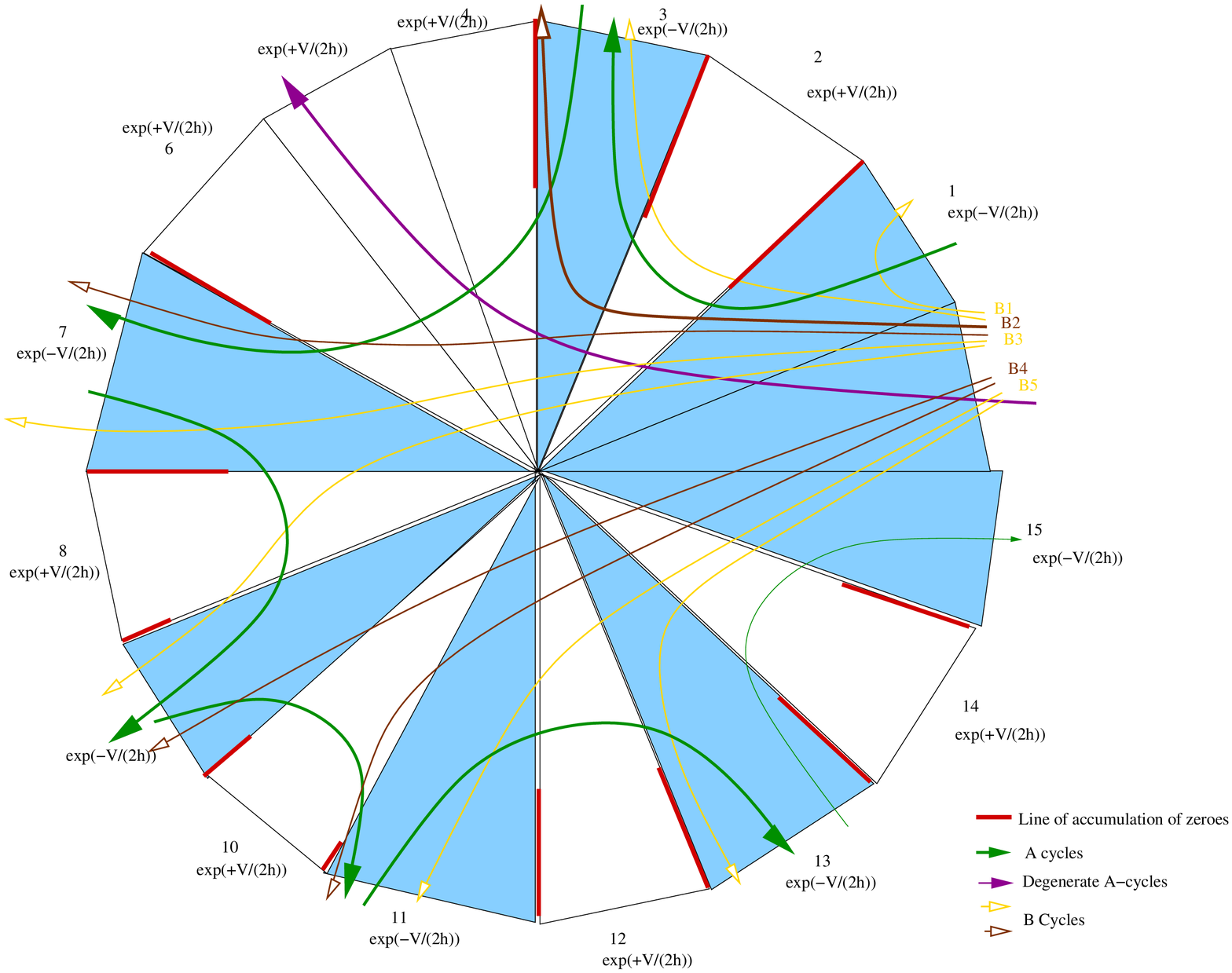}}}

\subsection{First kind functions}

After defining the cycles, another important step is to define the equivalent of the first, second and third kind differentials. In this section, we propose a definition of the first kind differentials. 

Let $h_k$, $k=1,\dots,d-1$, be a basis (arbitrary for the moment, but we will choose it orthonormal later on), of the complex vector space of polynomials of degree $\leq d-2$.
To have more convenient notation, we will label the $\hat{\cal{A}}$-cycles as $\cal{A}_\alpha$ , $\genus+1\leq \alpha \leq d-1$ and the standard $\cal{A}$ are labelled $ \cal{A}_\alpha$, $1\leq \alpha \leq \genus$.

Consider the following functions:
\beq
v_k(x) = {1\over \hbar\, \psi^2(x)}\,\int_{\infty_0}^x h_k(x')\,\psi^2(x')\,dx'
\virg
\deg h_k\leq d-2.
\eeq

Notice that, thanks to the Bethe ansatz, $v_{k}(x)$ has double poles with vanishing residues at the $s_{j}$'s (the zeroes of $\psi$), and behaves like $O(1/x^2)$ in sector $S_0$ and in sectors where $\psi$ is exponentially large. (because the polynomial is of degree less than $d-2$).
Therefore, the following integrals are well defined:
\beq
I_{k,\alpha} = \oint_{\acycle_\alpha} v_k(x)\,dx
\virg \alpha=1,\dots,\genus, \, k=1,\dots,d-1.
\eeq

For the degenerate contours $\hat{\cal{A}}_\alpha$, we cannot take the integral since it would not converge. We define instead:
\beq
I_{k,\alpha} = \int_{\hat{\acycle}_\alpha} h_k(x)\, \psi^2(x)\,dx
\virg \alpha=\genus+1,\dots,d-1, \, k=1,\dots,d-1.
\eeq

The matrix $I_{k,\alpha}$ with $k,\alpha=1,\dots,d-1$ is a square matrix, which gives a pairing between the set of paths \{ $\acycle_\alpha,\hat\acycle_\alpha$ \} and the space of polynomials of degree at most $d-2$. Let us choose a basis $h_k$, dual to the $\acycle$-cycles, i.e.:
\beq\label{eqdualvkacycle}
I_{k,\alpha}=\delta_{k,\alpha}.
\eeq

\bigskip

Choosing this set of polynomials gives then the following relations:
\beq \label{orthogonalbasis}
\forall\, i=1,\dots,\genus ,j=1,\dots,d-1\, , \qquad \quad
 \oint_{\acycle_i} v_j(x)\,\, dx = \delta_{i,j}
\eeq
\beq
\forall i=\genus+1,\dots,d-1, \, j=1,\dots,d-1 : \int_{\hat{\acycle}_i} h_j(x)\, \psi^2(x)\,dx =\delta_{j,i}
\eeq

Moreover, from the definitions, we get an asymptotic expression of $v_k(x)$ at infinity:

\bt\label{thvkxlarge}
The functions $v_k(x)$ with $k\leq \genus$ are such that:
\beq
k=1,\dots, \genus, \qquad v_k(x)=O(x^{-2})
\eeq
in all sectors at infinity.

And the functions $v_k(x)$ with $\genus+1 \leq k\leq d-1$ are such that:
\beq
k=\genus+1,\dots, d-1, \qquad v_k(x)=O(x^{-2})
\eeq
in all sectors except in the sector where $\hat\acycle_k$ ends, where we have:
\beq
v_k(x) = {1\over \hbar\psi(x)^2} +O(1/x^2).
\eeq

\et
\proof{
In sector $\infty_0$, we clearly have $v_k(x)\sim O(x^{\deg h_k-d}) = O(x^{-2})$.
And in a sector $S_i$ where $\psi$ is exponentially small we have:
\beq
v_k(x) = {1\over \hbar\psi^2(x)}\,\left[ \int_{\infty_i}^x h_k(x')\,\psi^2(x')\, dx' +  \int_{\infty_0}^{\infty_i} h_k(x')\,\psi^2(x')\, dx'  \right] ,
\eeq
and due to our choice of basis \eq{eqdualvkacycle}, we have
\beq
v_k(x) = {\delta_{i,i_k}\over \hbar\psi(x)^2} +{1\over \hbar\psi^2(x)}\, \int_{\infty_i}^x h_k(x')\,\psi^2(x')\, dx'  = {\delta_{i,i_k}\over \hbar\psi(x)^2} +O(1/x^2),
\eeq
in sector $S_i$.
}

We claim that the function $v_k(x) \, k=1,\dots,\genus$ are the generalization of holomorphic forms (1st kind differentials). 

\br Classical limit.

The small $\hbar$ BKW expansion $\psi\sim \ee{\pm {1\over 2\hbar}\,\int \sqrt U}$ gives:
\beq
v_k(x) \sim {\pm h_k(x)\over \sqrt{U(x)}} 
\eeq
and $v_k(x)dx$ are indeed the holomorphic forms on the algebraic curve $y^2=U(x)$.

\er

\subsection{Riemann matrix of periods}

An interesting quantity in standard algebraic geometry is the Riemann matrix of periods which is the integrations of the holomorphic differentials over $\cal{B}$-cycles. Now that we have defined properly the cycles, we can define a similar ``quantum'' Riemann period matrix $\tau_{i,j}$, $i,j=1,\ldots, \genus$ by:
\beq
\tau_{i,j} \stackrel{{\rm def}}{=} \oint_{\bcycle_i} v_j(x)\,\, dx.
\eeq

Note that this definition makes sense since $v_j(x)$ ($j=1,\dots, \genus$) behaves as $O(1/x^2)$ in the sectors where the $\cal{B}$-cycles go. 
Also, thanks to the Bethe ansatz, $v_j$ has no residue at the roots $s_i$'s, therefore those integrals depend only on the homology class of $\bcycle$-cycles, and not on a representent.

Like for the classical Riemann matrix of periods we have the following property: 

\begin{theorem}
The period matrix $\tau$ is symmetric: $\tau_{i,j}=\tau_{j,i}$. 
\end{theorem}

\proof{
We anticipate on results which shall be proved later, but which don't depend on this theorem.
The proof comes directly from theorem \ref{thBergmanABcycles} below, since:
$$
\oint_{\mathcal{B}_{\beta}} dx \oint_{\mathcal{B}_{\alpha}}B(x,z)dz=2i\pi \oint_{\mathcal{B}_{\beta}} dx v_{\alpha}(x)=2i\pi \tau_{\beta,\alpha}
$$
and from the symmetry theorem \ref{thBergmanSymmetry} for the Bergman kernel $B(x,z)=B(z,x)$:
$$
\oint_{\mathcal{B}_\beta} dx \oint_{\mathcal{B}_\alpha}B(x,z)dz=\oint_{\mathcal{B}_\alpha} dz \oint_{\mathcal{B}_\beta} dx B(x,z)=2i\pi \oint_{\mathcal{B}_\alpha} dz v_{\beta}(z)=2i\pi \tau_{\alpha,\beta}.
$$
}

\subsection{Filling fractions}

In random matrices, the notion of filling fractions, is just the $\acycle$-cycle integrals of the resolvent.
Here, we easily generalize it by the definition:

\bd
The filling fractions $\epsilon_1,\dots,\epsilon_d$ are defined as follows:
\beq
\alpha=1,\dots,\genus , \qquad
\epsilon_\alpha = {1\over 2i\pi}\,\oint_{\acycle_\alpha} \left(\om(x)-\frac{t_0}{x}\right) +\frac{t_0 n_\alpha}{(d+1)} 
\eeq
where the integer $n_\alpha$ is half the number of Stokes half-lines surrounded by the cycle $\cal{A}_\alpha$. In other words, $\frac{2n_\alpha}{2d+2}$ corresponds to the angular fraction of the complex plane defined by the cycle $\cal{A}_\alpha$.

For $\alpha=\genus+1,\dots,d-1$ we define
\beq
\alpha=\genus+1,\dots,d-1 , \qquad
\epsilon_\alpha =0
\eeq
And for $\alpha=d$, we choose a non-independent $\acycle$-cycle $\acycle_d$, which surrounds all the $s_i$'s which are not surrounded by $\acycle_1,\dots,\acycle_\genus$, and define:
\beq
\epsilon_d = {1\over 2i\pi}\,\oint_{\acycle_d} \left(\om(x)-\frac{t_0}{x}\right) +\frac{t_0 n_d}{(d+1)} 
\eeq

\ed

Note that this definition makes sense because all the cycles $\cal{A_\alpha}$ go from an infinity where $\om(x) -\frac{t_0}{x} \sim O\left(\frac{1}{x^2}\right)$. Note also that this definition depends on the exact locus of the contour $\acycle_\alpha$ and not only on its homotopy class, since $\om(x)$ has simple poles at the $s_i$'s with residue $\hbar$.
If we deform the contour $\acycle_\alpha$, the filling fractions can change by some integer times $\hbar$. 

In other words, the filling fractions are  "blurred" when $\hbar\neq 0$, they are defined modulo an integer times $\hbar$. In the classical limit $\hbar\to 0$, they become deterministic.

\medskip
We have:
\bt
\beq
\sum_{\alpha=1}^{d} \epsilon_\alpha = t_0
\eeq
\et

\proof{
When we perform the sum over the contours $\acycle_\alpha$, the contour ${\cal A}_d$ was defined as the "complementary" of the others, i.e. so that the sum is contractible. Since the function $x\to\om(x)- t_0/x$ is integrable at infinity, we find that its global integral is null. With the same argument, it is easy to see that $\sum_{\alpha=1}^{d} n_\alpha=(d+1)$ because we take all Stokes lines once and only once. Therefore we get:
$$
\sum_{\alpha=1}^{d} \epsilon_\alpha=0 + \frac{t_0}{d+1}\sum_{\alpha=1}^{d}n_\alpha=t_0.
$$
Note that it also tells us that only $d-1$ of the epsilon's are independant.
}

\br
In the case $\genus=-1$, the only filling fraction is $\epsilon_d=t_0$, and it is also the sum of residues of $\om$ at the $s_i$'s:
$$
\epsilon_d=t_0=\sum_i \Res_{s_i}\om = \hbar\,\#\{s_i\}
$$
This shows again, that $\genus=-1$ corresponds to a case where $t_0$ is quantized, namely $t_0$ is an integer times $\hbar$:
$$
t_0/\hbar \in \mathbb N.
$$

\er

\section{Kernels}

One of the key geometric objects in \cite{EynOM} and in \cite{EOFg}, is the "recursion kernel" $K(x,z)$.
It was used in the context of matrix models, to find a solution of loop equations.
Here, it will also allow us to define the 3rd and 2nd kind differentials.

\subsection{The recursion kernel $K$}

First we define:
\beq
\hat K(x,z) = {\frac{1}{\hbar}}{1\over \psi^2(x)}\, \int_{\infty_0}^x \psi^2(x')\,{dx'\over x'-z}
\eeq
and for each $\alpha=1,\dots,\genus$, we choose a point $P_\alpha\in \acycle_\alpha$ and  we define:
\beq
\hbar C_\alpha(z) 
=\oint_{\acycle_\alpha}\, {dx''\over \psi^2(x'')}\,\int_{\infty_0}^{P_\alpha} \psi^2(x')\,{dx'\over x'-z}
+\oint_{\acycle_\alpha}\, {dx''\over \psi^2(x'')}\,\int_{P_\alpha}^{x''} \psi^2(x')\,{dx'\over x'-z}
\eeq
where in the last integral, the integration contour between $P_\alpha$ and $x''$, is along $\acycle_\alpha$. This is described in fig.\ref{kernelkdefinition}.

\figureframex{9}{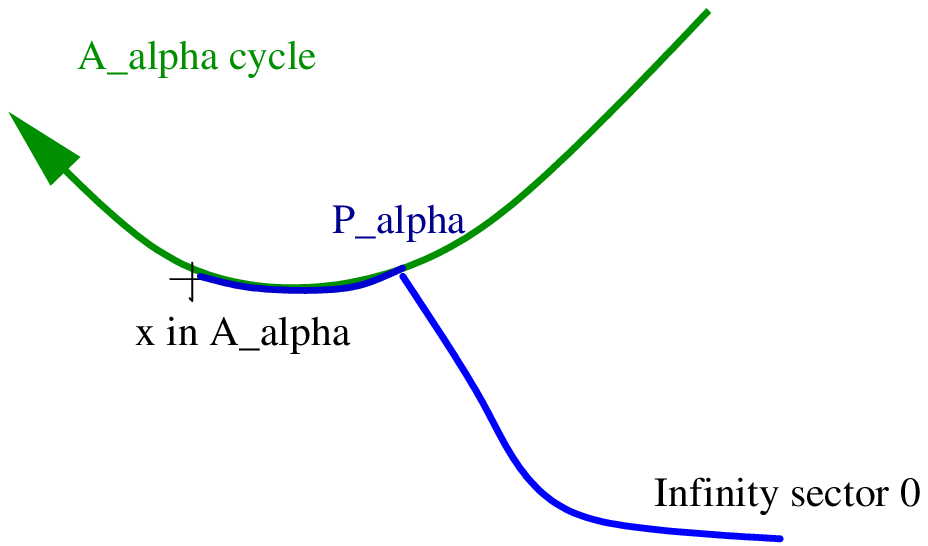}{\label{kernelkdefinition} Picture of the path of integration used for the definition of the kernel $K(x,z)$.}

For each $\alpha=\genus+1,\dots,d-1$, we define:
\beq
C_\alpha(z) 
=\int_{\hat{\cal{A}}_\alpha}\,  \psi^2(x')\,{dx'\over x'-z}.
\eeq

We now need to describe the domain of definition of these functions.

First, one can see that for a fixed $x$, these functions are defined for $z$ outside of some "cuts" (see figure \ref{kernelkdefinition})

$\bullet$ Choose a path between $\infty_0$ and $x$, then $\hat K(x,z)$ is defined for $z$ outside of this path. Across the path $]\infty_0,x]$, $\hat K(x,z)$ has a discontinuity:
\beq
\delta \hat K(x,z) = {2i\pi\over \hbar}\,\,{\psi^2(z)\over \psi^2(x)}
\eeq

$\bullet$ For each $\alpha=1,\dots \genus$, choose a path between $\infty_0$ and $P_\alpha$, then $C_\alpha(z)$ is defined for $z$ outside of this path, and outside $\acycle_\alpha$. 
Across the path $]\infty_0,P_\alpha]$, $C_\alpha(z)$ has a discontinuity:
\beq
\delta C_\alpha(z) = {2i\pi\,\,\psi^2(z)\over \hbar}\,\,\oint_{\acycle_\alpha} {dx''\over \psi^2(x'')}
\eeq
and across the path $\acycle_\alpha$, $C_\alpha(z)$ has a discontinuity:
\beq\label{discKxzAalpha}
\delta C_\alpha(z) = {2i\pi\,\,\psi^2(z)\over \hbar}\,\,\int_{P_\alpha}^z {dx''\over \psi^2(x'')}
\eeq

$\bullet$ For each $\alpha=\genus+1,\dots,d-1$, $C_\alpha(z)$ is defined for $z$ outside of the path $\hat{\acycle}_\alpha$. 
Across the path $\hat{\acycle}_\alpha$, $C_\alpha(z)$ has a discontinuity:
\beq\label{discKxzAalpha}
\delta C_\alpha(z) = {2i\pi\,\,\psi^2(z)}
\eeq

From these remarks, we now define the recursion kernel $K(x,z)$ by:
\bd Definition of the recursion kernel:
\beq
K(x,z) = \hat K(x,z) - \sum_{\alpha=1}^{d-1} v_\alpha(x)\,C_\alpha(z)
\eeq

it is defined for $z$ outside the cuts mentionned above.
\ed

For a fixed $z$, the analytical properties in $x$ of $K(x,z)$ are the same as those of $\hat{K}(x,z)$ since all $v_\alpha(x)$ are analytic. 
For a fixed $z$, the primitive of $\psi^2(x')\,{dx'\over x'-z}$ can be defined locally but not globaly on the complex plane. In fact there is a logarithmic cut to be arbitrarily chosen on $]\infty_0,z]$. Anywhere out of this cut the function $x \to K(x,z)$ is analytic.

\subsubsection{Properties of kernel $K$}

The definition of the kernel $K(x,z)$ might seem arbitrary at first glance. But in fact, the main reason for the introduction of such kernel is that it has many interesting properties:

It is clear from our definitions that:
\bt\label{thlargeKx}
For a given $z$, the kernel $K$ behaves like:
\beq
K(x,z) \mathop{\sim}\, O(x^{-2})
\eeq
when $x\to \infty$ in all sectors.
\et

\proof{ The result is obvious for sector $S_0$ and for sectors where $\psi$ is exponentially big. When it is not, the fact that we substract $C_\alpha, \alpha=\genus+1,\dots,d-1$ gives the result.}

\bt\label{thKlargez}
We have in all sectors at infinity :
\beq
K(x,z) \mathop{\sim}_{z\to\infty}\, O(z^{-d}).
\eeq
More precisely we have:
\beq
K(x,z) \sim -\,\sum_{k=d-1}^\infty {K_k(x)\over z^{k+1}}
\eeq
with
\beq
\hat K_k(x) = {1\over \hbar \psi^2(x)}\,\int^x_{\infty_0} x'^k\,\psi^2(x')\, dx',
\eeq
and
\beq
K_k(x) = \hat K_k(x) - \sum_{\alpha=1}^{\genus}\, v_\alpha(x)\, \oint_{\acycle_\alpha} \hat K_k(x')\,dx'
- \sum_{\alpha=\genus+1}^{d-1}\, v_\alpha(x)\, \oint_{\acycle_\alpha} \psi^2(x')\,x'^k\,dx'.
\eeq

\et

\proof{
It is clear that
\beq
\hat K(x,z) \sim -\sum_{k=0}^\infty {\hat K_k(x)\over z^{k+1}}
\eeq
where
\beq
\hat K_k(x) = {1\over \hbar \psi^2(x)}\,\int^x_{\infty_0} x'^k\,\psi^2(x')\, dx',
\eeq
and therefore
\beq
K_k(x) = \hat K_k(x) - \sum_{\alpha=1}^{\genus}\, v_\alpha(x)\, \oint_{\acycle_\alpha} \hat K_k(x')\,dx'
- \sum_{\alpha=\genus+1}^{d-1}\, v_\alpha(x)\, \oint_{\acycle_\alpha} \psi^2(x')\,x'^k\,dx'
\eeq
Now, if $k\leq d-2$, notice that $x'^k$ is a polynomial of degree $\leq d-2$, and it is thus a linear combinations of $h_\alpha(x)$'s:
\beq
x'^k = \sum_{\beta=1}^{d-1}\, b_{k,\beta}\, h_\beta(x')
\eeq
This implies:
\beq
\hat K_k(x) = \sum_{\beta=1}^{d-1}\, b_{k,\beta}\, v_\beta(x)
\eeq
Taking now the integral over an $\cal{A}$ cycle and using the normalization choice of $h_k(x)$ gives:
If $\alpha\leq \genus$
\beq
\oint_{\acycle_\alpha} \hat K_k(x')\,dx' = b_{k,\alpha}
\eeq
and if $\alpha> \genus$
\beq
\oint_{\acycle_\alpha} \psi^2(x')\,x'^k\,dx' = b_{k,\alpha}
\eeq
This implies that $K_k(x)=0$ if $k\leq d-2$, and therefore
\beq
K(x,z) = O(z^{-d}).
\eeq

}

\bt\label{thKointAcycles}
Let $\alpha=1,\dots,\genus$, and $z$ on the side of $\acycle_\alpha$ which does not contain $\infty_0$, then:
\beq
\oint_{\acycle_\alpha} K(x,z)\,dx = 0
\eeq
\et

\proof{
 Notice that if $z$ is on that side of $\acycle_\alpha$, we have $C_\alpha(z) = \oint_{\acycle_\alpha} \hat K(x,z)\,dx$, and therefore $\oint_{\acycle_\alpha} K(x,z)\,dx=0$. In fact one can see that the addition of the part with the $C_\alpha(z)$ was just put there to cancel out the $\cal{A}$-cycle integrals.
}

%



%

%


\subsection{Third kind differential: kernel $G(x,z)$}

The second important kernel to define is the equivalent of the third kind differential. 
In \cite{EynOM} this kernel was computed from $K$ by derivation, and we use the same definition. 

\bd We define the kernel $G(x,z)$ by:
\beq\label{eqdefG}
G(x,z) =- \hbar\,\psi^2(z)\, \partial_z\, {K(x,z)\over \psi^2(z)}=2\hbar\frac{\psi'(z)}{\psi(z)}K(x,z)-\hbar \partial_z\,K(x,z)
\eeq
\ed

From an easy integration by parts we find:
\bea
G(x,z)
&=& -  {1\over x-z}  + {2\over \psi^2(x)}\,\int^x_{\infty_0} {dx'\over x'-z}\, \psi^2(x') \left( {\psi'(x')\over \psi(x')} - {\psi'(z)\over \psi(z)}  \right) \cr
&& - \hbar \sum_\alpha v_{\alpha}(x)\,  \psi^2(z) \partial_z\, {C_{\alpha}(z)\over \psi^2(z)} \cr
\eea
which shows that near $x=z$ we have $G(x,z)\sim {1\over z-x}$, i.e. there is a simple pole of residue $1$ at $z=x$. Note in particular that ${1 \over x'-z}\left( {\psi'(x')\over \psi(x')} - {\psi'(z)\over \psi(z)}\right)$ has no singularity at $x'=z$ and therefore for a fixed $z$, there is no more any logarithmic cut $]\infty,z]$ as we had for $K(x,z)$.
\smallskip

Note again that a priori, this function of $z$ has the same lines of discontinuity as the kernel $K(x,z)$. 
But notice that the definition of $G$ ensures that all discontinuities of $K$ which are proportional to $\psi^2(z)$ cancel.

\bt\label{thdiscG}

$G(x,z)$ is an analytical function of $x$, with a simple pole at $x=z$ with residue $-1$, and double poles at the $s_{j}$'s (zeros of $\psi(x)$) with vanishing residue, and possibly an essential singularity around $\infty$.

$G(x,z)$ is an analytical function of $z$, with a simple pole at $z=x$ with residue $+1$, simple poles at $z=s_{j}$, and with a discontinuity across $\acycle_\alpha$-cycles with $\alpha=1,\dots,\genus$ (and thus no discontinuity on $\hat{\cal{A}}_\alpha$):
\beq
\delta G(x,z) = -2i\pi \, v_\alpha(x)
\eeq
\et
\proof{
$K(x,z)$ is discontinuous when $z$ crosses either $]\infty_0,x]$, $]\infty_0,P_\alpha]$ or $\acycle_\alpha$. However, the discontinuity of $K(x,z)$ across $]\infty_0,x]$, $]\infty_0,P_\alpha]$, and  $\hat{\cal{A}}_\alpha$ is proportional to $\psi^2(z)$, and this means by derivation that $G(x,z)$ is not discontinuous there.  
Across $\acycle_\alpha$ with $\alpha\leq \genus$, the discontinuity of $K(x,z)$ is given by \eq{discKxzAalpha}, and thus, the discontinuity of $G(x,z)$ is $\delta G(x,z) = -2i\pi \, v_\alpha(x)$.

Since $K(x,z)$ is regular when $z=s_{j}$, then it is clear that $G(x,z)$ has simple poles at $z=s_{j}$, with residue $-2\hbar K(x,s_{j})$.

In the variable $x$, it is clear from the definition and from the Bethe ansatz \ref{bethe ansatz}, that $K(x,z)$ has double poles at $x=s_{j}$ without residue, and this properties follows for $G(x,z)$.

}

\bt\label{thGlargexz}
\beq
G(x,z)=O(1/x^2)
\eeq
when $x\to \infty$ in all sectors. 

And at large $z$ in sector $S_k$:
\beq
\mathop{{\lim}}_{z\to \infty_k}\,\,  G(x,z) = G(x,\infty_k) =  \eta_k\,\,t_{d+1}\,K_{d-1}(x) 
\eeq
where $\eta_k=\pm 1$ is such that $\psi\sim \ee{\eta_k V/2\hbar}$ in sector $S_k$.
\et

\proof{The large $x$ behavior follows from theorem \ref{thlargeKx}. The large $z$ behavior is given by theorem \ref{thKlargez}, i.e. $G(x,z)\sim \eta_k V'(z) K(x,z) \sim \eta_k\, t_{d+1} K_{d-1}(x)$.
The sign depends on the behavior of the solution in this sector. }.

\bt\label{thGointAcycles}
Let $\alpha=1,\dots,\genus$, and $z$ on the side of $\acycle_\alpha$ which does not contain $\infty_0$, then:
\beq
\oint_{\acycle_\alpha} G(x,z)\,dx = 0
\eeq
\et

\proof{
Immediate from theorem \ref{thKointAcycles}}

\subsubsection{Semi-classical limit}

We claim that this kernel is the quantum version of the third kind differential. Indeed, in classical algebraic geometry a third kind differential is characterized by analyticity except a simple pole with non vanishing residue and a proper normalization on $\acycle$-cycles. Here, apart from the discontinuity along the $\acycle$-cycles which is expected since these contours represent the "quantum cuts", we have analyticity (apart from the $s_i$'s which also define the cuts), a simple pole with residue and a good normalization on $\cal{A}$-cycles. 

\medskip
In the BKW semiclassical expansion we have $\psi \sim \ee{{\pm 1\over \hbar}\int\sqrt{U}}$ and thus
\beq
\hat K(x,z) \sim {2\over x-z}\,{1\over \sqrt{U(x)}}
\eeq
and
\beq
K(x,z) \sim {1\over x-z}\,{1\over 2\sqrt{U(x)}} - \sum_\alpha\, v_\alpha(x)\,C_\alpha(z)
\eeq
and
\beq
G(x,z) \sim 2\sqrt{U(z)} K(x,z) \sim {1\over x-z}\,{\sqrt{U(z)}\over \sqrt{U(x)}} - 2\sum_\alpha\, v_\alpha(x)\,C_\alpha(z)\sqrt{U(z)}
\eeq
The form $G(x,z)dx$ has thus a simple pole at $x=z$, in the physical sheet with residue $+1$ and in the other sheet with residue $-1$, and it is normalized on $\acycle$-cycles $\oint_{\acycle_i} G(x,z)dx=0$.
This is indeed the usual 3rd kind differential in classical algebraic geometry.

\subsection{The Bergman kernel $B(x,z)$}

In classical algebraic geometry, the Bergman kernel is the fundamental second kind differential, it is the derivative of the 3rd kind differential, and it is another major tool in classical algebraic geometry. Following the same definition as in \cite{EynOM}, we define:
\beq
B(x,z) = -{1\over 2}\, \partial_z\, G(x,z).
\eeq
The kernel $B$ is going to be called the "quantum" Bergman kernel.

\bt
$B(x,z)$ is an analytical function of $x$, with a double pole at $x=z$ with no residue, and double poles at the $s_{j}$'s with vanishing residues, and possibly an essential singularity around $\infty$.

$B(x,z)$ is an analytical function of $z$, with a double pole at $z=x$ with no residue, and double poles at the $s_{j}$'s with vanishing residues, and possibly an essential singularity around $\infty$.
\textbf{In particular it has no discontinuity along the $\cal{A}$ cycles, it is defined analytically in the whole complex plane except at those double poles.}
\et

\proof{
Those properties follow easily from those of $G(x,z)$ of theorem \ref{thdiscG}. In particular, it is important to notice that the only discontinuity of $G(x,z)$ is along the $\acycle$-cycles, and is independent of $z$, therefore $B(x,z)$ has no discontinuity there.
}

\subsubsection{Properties of the Bergman kernel}

\bt\label{thBlargexz}
\beq
B(x,z) = O(1/x^2)
\eeq
when $x\to \infty$ in all sectors.

And
\beq
B(x,z) = O(1/z^2)
\eeq
when $z\to \infty$ in all sectors.
\et

\proof{Follows from the large $x$ and $z$ behaviors of $G(x,z)$.}

\bt\label{thloopeqB}
$B$ satisfies the loop equations:
\beq\label{loopeqBx}
(2{\psi'(x)\over \psi(x)}+\partial_x)\,\left(B(x,z)-{1\over 2(x-z)^2}\right) + \partial_z\,{{\psi'(x)\over \psi(x)}-{\psi'(z)\over \psi(z)}\over x-z} = P_2^{(0)}(x,z)
\eeq
where $P_2^{(0)}(x,z)$ is a polynomial in $x$ of degree at most $d-2$.
And
\beq\label{loopeqBz}
(2{\psi'(z)\over \psi(z)}+\partial_z)\,\left(B(x,z)-{1\over 2(x-z)^2}\right) + \partial_x\,{{\psi'(x)\over \psi(x)}-{\psi'(z)\over \psi(z)}\over x-z} =\td{P}_2^{(0)}(z,x)
\eeq
where  $\td{P}_2^{(0)}(z,x)$ is a polynomial in $z$ of degree at most $d-2$.
\et

\proof{This theorem is crucial for all what follows, and its proof is rather non-trivial. Since it is very long and technical, we present the proof in appendix \ref{BergmannLoopEquation}. Those equations are indeed the loop equations for the 2-point function in the $\beta$ matrix model, see section \ref{secMM}.}

\bt\label{thBergmanABcycles}
We have for every $\alpha=1,\dots,\genus$:
\beq
\oint_{\acycle_\alpha} B(x,z)\,dx = 0
\virg
\oint_{\acycle_\alpha} B(x,z)\,dz = 0
\eeq
and
\beq
\oint_{\bcycle_\alpha} B(x,z)\,dz = 2i\pi v_\alpha(x)
\eeq

\et

\proof{
The vanishing of $\acycle$-cycle integrals in the $x$ variable is by construction and can be seen as the consequence of the same result known for $G(x,z)$ on one side of $\cal{A}$ and the fact that $B(x,z)$ has no discontinuity along the $\cal{A}$-cycles. (Therefore, the nullity extend on both sides which no longer need to be treated separately).

For the $z$ variable, notice that if $\acycle_\alpha=]\infty_i,\infty_j[$ goes from $\infty_i$ to $\infty_j$, where both $\infty_i$ and $\infty_j$ are in the physical sheet, we have:
\beq
\oint_{\acycle_\alpha} B(x,z)dz = \int_{\infty_i}^{\infty_j} B(x,z)dz = -{1\over 2}\,(G(x,\infty_j)-G(x,\infty_i))
\eeq
and from theorem \ref{thGlargexz} $G(x,\infty_i) = \eta_i t_{d+1}K_{d-1}(x)$, we get:
\beq
\oint_{\acycle_\alpha} B(x,z)dz = \int_{\infty_i}^{\infty_j} B(x,z)dz = {\eta_i-\eta_j\over 2}\,t_{d+1}\,K_{d-1}(x)
\eeq
and since $\infty_i$ and $\infty_j$ are both in the physical sheet we have $\eta_i=\eta_j=-1$, and therefore
\beq
\oint_{\acycle_\alpha} B(x,z)dz = 0.
\eeq

And similarly, when performing the integral over $\bcycle_\alpha$, the contribution from infinities cancels out since the contour goes in the same sheet. But since $\bcycle_\alpha$ intersects its corresponding $\acycle_\alpha$ (and only this one) where the primitive $-\frac{1}{2}G(x,z)$ is discontinous, the result is the jump of $G(x,z)$ along this $\acycle_\alpha$, that is to say $i\pi v_\alpha(x)$. Eventually, since $\bcycle_\alpha$ and $\acycle_\alpha$ intersect twice, we find \eq{thBergmanABcycles}. 
}

\medskip

One of our key theorems is:
\bt \label{thBergmanSymmetry}
$B(x,z)$ is symmetric
\beq
B(x,z) = B(z,x)
\eeq

\et

\proof{
The proof relies essentially on the fact that $B(x,z)$ satisfies the loop equation in the two variables. We have:
\bea
&& (2{\psi'(z)\over \psi(z)}+\partial_z)\,(2{\psi'(x)\over \psi(x)}+\partial_x)\, (B(x,z)-{1\over 2(x-z)^2}) \cr
&=& (2{\psi'(z)\over \psi(z)}+\partial_z)\,\Big( P_2^{(0)}(x,z) - \partial_z\,{{\psi'(x)\over \psi(x)}-{\psi'(z)\over \psi(z)}\over x-z} \Big) \cr
&=& (2{\psi'(x)\over \psi(x)}+\partial_x)\,\Big( \td{P}_2^{(0)}(z,x) - \partial_x\,{{\psi'(x)\over \psi(x)}-{\psi'(z)\over \psi(z)}\over x-z} \Big) \cr
\eea
This implies:
\bea
&& (2{\psi'(z)\over \psi(z)}+\partial_z)\, P_2^{(0)}(x,z) -(2{\psi'(x)\over \psi(x)}+\partial_x)\, \td{P}_2^{(0)}(z,x) \cr
&=&  (2{\psi'(z)\over \psi(z)}+\partial_z)\partial_z\,{{\psi'(x)\over \psi(x)}-{\psi'(z)\over \psi(z)}\over x-z}  \cr
&&  - (2{\psi'(x)\over \psi(x)}+\partial_x) \partial_x\,{{\psi'(x)\over \psi(x)}-{\psi'(z)\over \psi(z)}\over x-z}  \cr
&=& 2 {U(x)-U(z)\over (x-z)^2} - {U'(x)+U'(z)\over x-z}
\eea
and therefore:
\bea \label{definitionofR(x,z)}
&& (x-z)^2\,(2{\psi'(z)\over \psi(z)}+\partial_z)\, P_2^{(0)}(x,z) +2U(z)+(x-z)U'(z) \cr
&=& (x-z)^2\,(2{\psi'(x)\over \psi(x)}+\partial_x)\, \td{P}_2^{(0)}(z,x) +2U(x)+(z-x)U'(x) \cr
&\stackrel{{\rm def}}{=}& R(x,z)
\eea
Here, the first line is a polynomial in $x$, whereas the second line is also a polynomial in $z$. Therefore, $R(x,z)$ is a polynomial in both variables, of degree at most $d$ in each variable.
Moreover, we must have:
\beq
R(x,x) = 2U(x)
\eeq
Therefore we must have:
\beq
R(x,z)
= {\frac{1}{\hbar^2}}\left({1\over {2}}\,V'(x)V'(z) -\hbar\,{V'(x) -V'(z)\over x-z} - P(x)-P(z) \right) + (x-z)^2 \td{R}(x,z)
\eeq
where $\td{R}(x,z)$ is a polynomial of both variables of degree at most $d-2$ in each variable.

Putting this back into \ref{definitionofR(x,z)} and using the symmetry $x \leftrightarrow z$ it implies that:
\beq \label{symm} 
 (2{\psi'(z)\over \psi(z)}+\partial_z)\, (P_2^{(0)}(x,z)-\td{P}_2^{(0)}(x,z))
=  \td{R}(x,z)-\td{R}(z,x)
\eeq
Then, we can decompose the r.h.s into the basis $h_\alpha(x)h_\beta(z)$ introduced in \ref{orthogonalbasis}:
\beq \td{R}(x,z)-\td{R}(z,x)=\sum_{\alpha,\beta=1}^{d-1} (\td R_{\alpha,\beta}-\td R_{\beta,\alpha}) h_\alpha(x)h_\beta(z)
\eeq
Integrating the differential equation \eq{symm} then gives:
\beq
P_2^{(0)}(x,z)-\td{P}_2^{(0)}(x,z) = \sum_{\alpha,\beta=1}^{d-1} (\td R_{\alpha,\beta}-\td R_{\beta,\alpha}) h_\alpha(x)\, v_\beta(z) + A_1(x)
\eeq
where $A_1(x)$ is some integration constant.

Then using the loop equations \ref{thloopeqB} we find by substraction that:
\beq \left(2\frac{\psi'(y)}{\psi(y)}+\partial_y\right)\left(B(y,z)-B(z,y)\right)=P_2^{(0)}(y,z)-\td{P}_2^{(0)}(y,z)\eeq
and again, integrating this differential equation we find:
\beq
B(x,z)-B(z,x) = \sum_{\alpha,\beta=1}^{d-1} (\td R_{\alpha,\beta}-\td R_{\beta,\alpha}) v_\alpha(x)\, v_\beta(z) + A(x)+\td A(z)
\eeq
where $(2\psi'/\psi+\partial)A=A_1$, and $\td A(z)$ is some other integration constant.

\smallskip

The large $x$ and large $z$ behavior of $B$ imply that $A(x)=\td A(z)=0$.
We thus get:
\beq \label{difference}
B(x,z)-B(z,x) = \sum_{\alpha,\beta} (\td{R}_{\alpha,\beta}-\td{R}_{\beta,\alpha})\, \td{v}_{\alpha}(x)\td{v}_{\beta}(z)
\eeq

Then, using theorem \ref{thBergmanABcycles}
\beq
\oint_{\acycle_\alpha} B(x,z)dx=0
=\oint_{\acycle_\beta} B(x,z)dz
\eeq
We find:
\beq
\forall \alpha,\beta , \qquad
\td{R}_{\alpha,\beta}=\td{R}_{\beta,\alpha}
\eeq
that is to say by \ref{difference} that the Bergman kernel is symmetric.
}

\bigskip

We claim that all these properties are essential to name this function a "quantum Bergman kernel". Indeed, the symmetry is absolutely necessary and is completely non-trivial. The fact that $B(x,z)$ has no  discontinuity is also essential since in standard algebraic geometry, it is defined everywhere on the Riemann surface. Using all these kernels and their properties, we can then generalize easily the recursion of \cite{Eyn1loop, EynOM} defining the correlation functions.

\subsection{Meromorphic forms and properties}

\subsubsection{Definition of meromorphic forms}

\bd
A meromorphic form ${\cal R}(x)$ is defined as:
\beq
{\cal R}(x) = {1\over \hbar\psi^2(x)}\,\int_{\infty_0}^x\,\, r(x')\,\psi^2(x')\,dx'
\eeq
where $r(x)$ is a rational function of $x$, which behaves at most like $O(x^{d-2})$ at large $x$, and whose poles $r_i$ are such that:
\beq
\Res_{x\to r_i}\, \psi^2(x)\,r(x)=0
\eeq
and for all degenerate $\hat\acycle_\alpha$ cycles
\beq
\int_{\hat\acycle_\alpha} \psi^2(x') r(x')\, dx' = 0.
\eeq

\ed

It is easy to see, that with this definition, the holomorphic forms $v_\alpha(x)$, the kernels $G(x,z)$ and $B(x,z)$ are meromorphic forms of $x$.

\medskip

\subsubsection{Analiticity properties}

A meromorphic forms ${\cal R}(x)$, has poles at $x=r_i$ the poles of $r(x)$, with degree 1 less than that of $r$, it behaves like $O(x^{-2})$ in all sectors of the physical sheet.
From the Bethe ansatz, it has double poles at the $s_i$'s, with vanishing residues.

In particular, it has an accumulation of poles along the half-lines $L_i$ of accumulations of zeroes of $\psi$.

Also, notice that the following integrals are well defined, and independent of homotopic deformations of $\acycle_\alpha$ (in particular independent of where are the $s_i$'s):
\beq
\oint_{\acycle_\alpha}\, {\cal R}(x)dx.
\eeq

\subsubsection{\label{integration contour}The integration contours around branch-points}

Let us choose some contour ${\cal C}_i$, such that each ${\cal C}_i$ surrounds (in the trigonometric direction) a half-line $L_i$ of accumulation of zeroes. In other words it surrounds a "branch point".
Let us also assume that $\sum_i {\cal C}_i$ surrounds all roots of $\psi$, i.e. each root of $\psi$ is enclosed in one ${\cal C}_i$.
We also assume that contours ${\cal C}_i$ and $\acycle_\alpha$ do not intersect (they have vanishing intersection numbers):
\beq
\forall i=1,\dots, 2\genus+2,\quad, \forall \alpha=1,\dots,d-1, \qquad \quad
{\cal C}_i\cap \acycle_\alpha = 0
\eeq

\medskip


{\mbox{\epsfxsize=14.truecm\epsfbox{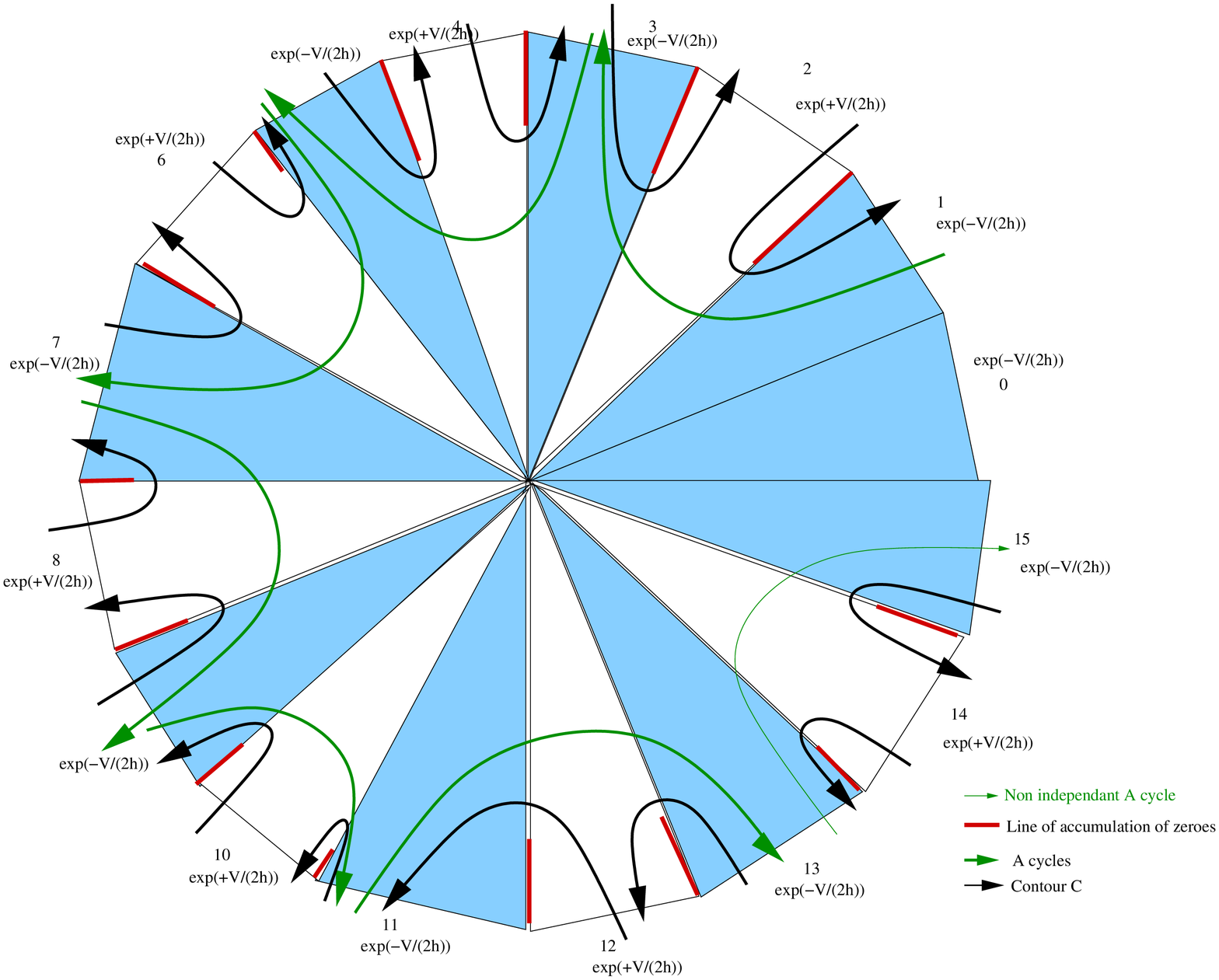}}}



\subsubsection{Riemann bilinear identity}

For the Riemann bilinear identity, we need the following useful lemma,  which we shall use very often in this article:

\bl \label{NullityOfIntegrals}
For every analytical function $f(x)$ which behaves at infinity at most like $f(x) =O\left(x^{d-2}\right)$ in all directions, and such that it has no singularities inside every contour ${\cal C}_i$ (and thus must be regular at the root $s_{j}$'s) we have, for $x_0$ outside of all $\acycle$-cycles (i.e. on the same side as $\infty_0$) :
$$\forall i\, , \qquad \quad  \,{1\over 2i\pi}\, \oint_{{\cal C}_i}\, dx\,\,  K(x_0,x)\, f(x)=0$$
\el

\proof{
Clearly, the contours ${\cal C}_i$ enclose no singularity of $K(x_0,x)f(x)$ and can be contracted to $0$.
}

\bigskip

Then we can write the bilinear Riemann identity:
\bt {\bf Riemann bilinear identity}

Consider a meromorphic form ${\cal R}(x)$, with poles $r_i$.

Then we have for $x$ outside of all $\acycle$-cycles (i.e. on the same side as $\infty_0$):
\beq
{\cal R}(x) 
=  -\sum_i \Res_{r_i} G(x,z){\cal R}(z)dz +  \sum_{\alpha=1}^{\genus}\, v_\alpha(x)\oint_{\acycle_\alpha} {\cal R}(z)\,dz.
\eeq

\et

\proof{
Since $G(x,z)=1/(z-x)+\dots$, we write Cauchy formula:
\beq
{\cal R}(x) = \Res_{z\to x} G(x,z)\,{\cal R}(z)\,dz
\eeq
and we deform the contour of integration from a small circle around $x$, to contours enclosing all other singularities, i.e. the $r_i$'s and the $s_i$'s.
By doing so, $G(x,z)$ has to cross the $\acycle$-cycles, and picks a discontinuity equal to $2i\pi\,v_\alpha(x)$ i.e. independent of $z$, so the contour integral of the product factorizes for each ${\acycle}_\alpha$.
We thus arrive to:
\bea
{\cal R}(x) 
&=&  -\sum_i \Res_{r_i} G(x,z){\cal R}(z)dz - \sum_i {1\over 2i\pi}\oint_{{\cal C}_i} G(x,z){\cal R}(z)dz \cr
&& +  \sum_{\alpha=1}^{\genus}\, v_\alpha(x)\oint_{\acycle_\alpha} {\cal R}(z)\,dz.
\eea
Then, we need to compute
$$ \oint_{{\cal C}_i} G(x,z){\cal R}(z)dz.$$
Write that $G(x,z)=\psi^2(z)\,\partial_z\, K(x,z)/\psi^2(z)$, and integrate by parts:
$$ \oint_{{\cal C}_i} G(x,z){\cal R}(z)dz = - \oint_{{\cal C}_i} K(x,z)r(z)dz$$
and using lemma \ref{NullityOfIntegrals}, we see that this vanishes.

}

\section{Definition of correlators and free energies}
\label{secdefWngFg}

In this section, we define the quantum deformations of the correlation functions introduced in \cite{Eyn1loop, EOFg}.
Although the following definitions are inspired from (non hermitian) matrix models (see section \ref{secMM}), they are valid in the present framework of an arbitrary Schr\"odinger equation, not necessarily linked to a matrix model. The special case of their application to matrix models will be discussed in section \ref{secMM}.

\subsection{Definition of correlators}

\bd\label{defWng}
We define the following functions $W_n^{(g)}(x_1,\dots,x_n)$ called {\bf $n$-point correlation function of "genus" $g$} by the recursion\footnote{here $g$ is any given integer, it has nothing to do with the genus $\genus$ of the spectral curve.}: 
\beq
W_1^{(0)}(x) = \om(x)
\virg
W_2^{(0)}(x_1,x_2)=B(x_1,x_2)
\eeq
\bea\label{mainrecformula}
 W^{(g)}_{n+1}(x_0,J)  
&=&   {1\over 2i\pi}\,\sum_{i=1}^{2\genus+2}\, \oint_{{\cal C}_i}\, dx\,\,  K(x_0,x)\, \Big( \ovl{W}_{n+2}^{(g-1)}(x,x,J) \cr
&& + \sum_{h=0}^g\sum'_{I\subset J} {W}_{|I|+1}^{(h)}(x,x_I) {W}_{n-|I|+1}^{(g-h)}(x,J/I) \Big)\cr
\eea
where $J$ is a collective notation for the variables $J=\{ x_{1},\dots,x_{n} \}$, and where $\sum\sum'$ means that we exclude the terms $(h=0,I=\emptyset)$ and $(h=g,I=J)$, and where:
\beq
\ovl{W}_{n}^{(g)}(x_1,...,x_n) = W_{n}^{(g)}(x_1,...,x_n) - {\delta_{n,2}\delta_{g,0}\over 2}\, {1\over (x_1-x_2)^2}
\eeq
Here $x_0$ and all the $x_i's$ are outside of the $\acycle$-cycles, i.e. on the same side as $\infty_0$.
The contour ${\cal C}_i$ (defined in section \ref{integration contour}) is a contour which surrounds the branchpoint $L_i$, i.e. a half-line of accumulation of zeroes, and chosen such that every $s_j$ is surrounded by exactly one ${\cal C}_i$, and such that ${\cal C}_i$ doesn't intersect any $\acycle$-cycle.
Very often we shall write
\beq
{\cal C}=\sum_{i=1}^{2\genus+2} {\cal C}_i.
\eeq
\ed

Appart from the precise definition of the kernel $K$, this definition is exactly the same topological recursion as in \cite{EOFg}, a sum of residues around all branchpoints of the same expression. In other words, the topological recursion is independent of $\hbar$.

\medskip

To shorten equation we will introduce the notation:
\bea
U_n^{(g)}(x,J)&=&\ovl{W}_{n+2}^{(g-1)}(x,x,J)
+ \sum_{I\subset J} \ovl{W}_{|I|+1}^{(h)}(x,x_I) \ovl{W}_{n-|I|+1}^{(g-h)}(x,J/I) \cr
&&+ \sum_{j}
\partial_{x_j} \left( {{\ovl{W}_n^{(g)}(x,J/\{j\})-{\ovl{W}_n^{(g)}(x_j,J/\{j\})}} \over {(x-x_j)}}\right)
\eea

To get:
\bt\label{thWngdefintU}
\beq W^{(g)}_{n+1}(x_0,J) ={1\over 2i\pi}  \oint_{{\cal C}}\, dx\,\,  K(x_0,x)\,U_n^{(g)}(x,J)
\eeq
\et

\proof{The only difference with the definition, is when we face a term like $B(x,x_j)W_n^{(g)}(x,J/\{j\})$. (note that there are twice this term). It can be split into two terms:
$\ovl{B}(x,x_j)W_n^{(g)}(x,J/\{j\})$ and ${1\over (x-x_j)^2}W_n^{(g)}(x,J/\{j\})$. The second term compensate exactly the $\partial_{x_j}  {{\ovl{W}_n^{(g)}(x,J/\{j\})} \over {(x-x_j)}}$. Thus, the only difference between the two definitions is the term: ${1\over 2i\pi} \oint_{{\cal C}}\, dx\,\,  K(x_0,x)\sum_{j}
\partial_{x_j} {{\ovl{W}_n^{(g)}(x_j,J/\{j\})} \over {(x-x_j)}}$. Therefore the definitions are only the same if these terms are null. This is the case because of Lemma \ref{NullityOfIntegrals}.
}


%

%

%
%

\subsection{Properties of correlators}

The main reason of definition. \ref{defWng}, is because the $W_n^{(g)}$'s have many beautiful properties, which generalize those of \cite{EOFg}, and in particular they provide a solution of loop equations. We shall prove the following properties:

\bt\label{thpolessiWng}
Each  $W_n^{(g)}(x_1,\dots,x_n)$ with $2-2g-n<0$, is an analytical functions of all its arguments, with poles only when $x_i\to s_{j}$.
Moreover, it vanishes at least as $O\left(1/{x_i^2}\right)$ when $x_i\to\infty$ in all sectors.
It has no discontinuity across $\acycle$-cycles.
\et
\proof{ in appendix \ref{approofthpolessiWng}}

\bt For all $(n,g)\neq (0,0)$ we have
\beq  
\forall \alpha \leq \genus: \label{nullityOfAcycle} \oint_{\cal{A}_\alpha} W_{n+1}^{(g)}(x_0,x_1,...,x_n) dx_1=0
\eeq
\beq  
\forall \alpha \leq \genus: \label{nullityOfAcycle} \oint_{\cal{A}_\alpha} W_{n+1}^{(g)}(x_0,x_1,...,x_n) dx_0=0
\eeq
\et

\proof{ We clearly have these properties for $W_2^{(0)}(x_0,x_1)$. By an easy recursion, the first property holds for $x_1,\dots,x_n$. The case of the variable $x_0$ is special and requires explanation. Indeed for fixed values of $x_1,...,x_n$, the dependance in $x_0$ comes from $K(x_0,x)$. The theorem then comes from a permutation of integrals. Indeed, since the contour $\cal{C}$ never crosses any $\cal{A}$-cycles by prescription then we can permute the integrals in $x$ and $x_0$. The nullity of the integral for $K(x_0,x)$ in \ref{thKointAcycles} then gives the result.
}  

\bt\label{thWngPng}
For $2-2g-n<0$, the $W_n^{(g)}$'s satisfy the loop equation, i.e. Virasoro-like constraints.
This means that the quantity:
\bea\label{loopeqPng}
 P_{n+1}^{(g)}(x;x_1...,x_n)
 &=&
2\hbar\frac{\psi'(x)}{\psi(x)}\overline{W}_{n+1}^{(g)}(x,x_1,...,x_n) + \hbar \partial_{x}{\overline{W}_{n+1}^{(g)}(x,x_1...,x_n)} \cr
&& + \sum_{I\subset J} \ovl{W}_{|I|+1}^{(h)}(x,x_I) \ovl{W}_{n-|I|+1}^{(g-h)}(x,J/I) +
\ovl{W}_{n+2}^{(g-1)}(x,x,J)  \cr
& &+ \sum_{j}
\partial_{x_j} \left( {{\ovl{W}_n^{(g)}(x,J/\{j\})-{\ovl{W}_n^{(g)}(x_j,J/\{j\})}} \over {(x-x_j)}}\right) \cr
\eea
is a polynomial in the variable $x$, of degree at most $d-2$.

\et
\proof{ in appendix \ref{approofthWngPng}}

\bt\label{thsym}
 Each $W_n^{(g)}$ is a symmetric function of all its arguments.
\et
\proof{ in appendix \ref{approofthsym}, with the special case of $W_3^{(0)}$ in appendix \ref{approofthW3Krich}.}

\bt\label{thW3Krich}
The 3 point function $W_3^{(0)}$ can also be written:
\beq
W_3^{(0)}(x_1,x_2,x_3) = {4\over 2i\pi}\,\sum_i \oint_{{\cal C}_i}\,\, {B(x,x_1)B(x,x_2)B(x,x_3)\over Y'(x)}\,dx
\eeq
(this can be seen as a quantum version of Rauch variational formula)
\et
\proof{ in appendix \ref{approofthW3Krich}}

\bt\label{thhomogeneity}
For $2-2g-n<0$, $W_n^{(g)}(x_1,\dots,x_n)$ is homogeneous of degree $2-2g-n$:
\beq
\left( \hbar\,{\partial\over \partial \hbar}+ \sum_{j=0}^{d+1} t_j\,{\partial\over \partial t_j} + \sum_{i=1}^\genus \epsilon_i\,{\partial\over \partial \epsilon_i} \right) W_n^{(g)}(x_1,\dots,x_n) = (2-2g-n)\,W_n^{(g)}(x_1,\dots,x_n)
\eeq

\et

\proof{Under a change $t_k\to \l t_k$, $\hbar\to \l \hbar$, $\epsilon_i\to \l\epsilon_i$, the Schr\"odinger equation remains unchanged, and thus $\psi$ is unchanged. The kernel $K$ is changed to $K/\l$ and nothing else is changed. By recursion, $W_n^{(g)}$ is changed by $\l^{2-2g-n}$.}





\section{Deformations \label{variations}}

In this section, we will consider the variations of correlators $W_n^{(g)}$ under infinitesimal variations of the Schr\"odinger potential $U(x)$ or $\hbar$. 
Infinitesimal variations of the resolvent $\om(x)$ can be decomposed on the basis of "meromorphic forms", and forms can be put in duality with cycles. The duality kernel pairing is the Bergman kernel.
We will find in this section, that the classical $\hbar=0$ formulae remain valid for $\hbar\neq 0$, and generalize the corresponding form-cycle duality in special geometry.

\subsection{Variation of the resolvent}

Let's consider an infinitesimal polynomial variation:
$$
U\to U+\delta U
\virg
\hbar\to \hbar+\delta \hbar
$$
where $\delta U$ is a polynomial of degree: $\deg \delta U\leq 2d$.
Since we have written $U=V'^2/4 - \hbar V''/2 -P$, we have:
\beq
\delta U = {V'\over 2}\delta V' - {\hbar\over 2}\delta V'' - {\delta \hbar\over 2}\delta V'' - \delta P
\eeq
with
\beq
\delta V'(x) = \sum_{k=1}^{d+1} \delta t_k\,x^{k-1},
\eeq
and $\delta P$ is of degree at most $d-1$:
\beq
\delta P = (t_{d+1} \delta t_0 +t_0 \delta t_{d+1})x^{d-1} + {\rm lower\, degree}.
\eeq

\bigskip

Let us compute $\delta\psi$, or more precisely $f=\delta\ln\psi=\delta\psi/\psi$, let us write it:
\beq
\delta \psi(x)= f(x)\psi(x)
\eeq

The Schr\"odinger equation $\hbar^2\psi''=U\psi$ implies:
\beq\label{deltapsifeq}
\hbar^2 (f\psi)'' - U f \psi = \delta U \psi - \delta\hbar^2\,\psi''
\eeq
i.e.
\beq
\hbar^2(f''\psi+2f'\psi') = (\delta U - 2\,{\delta \hbar\over \hbar}\,U)\,\psi
\eeq
Multiplying by $\psi$ we get:
\beq
\hbar^2(f'\psi^2)' = (\delta U - 2\,{\delta \hbar\over \hbar}\,U)\,\psi^2
\eeq
i.e.:
\beq
\delta (\psi'/\psi) = f'(x) = {1\over \hbar^2\,\psi^2(x)}\,\int_{\infty_0}^x\, \psi^2(x')\, (\delta U(x')- 2\,{\delta \hbar\over \hbar}\,U(x'))\,\, dx'.
\eeq
therefore, since $\om = V'/2+\hbar \psi'/\psi$:
\beq \label{variationw(x)}
\encadremath{
\delta \om(x) = {\delta V'(x)\over 2} + \delta \hbar \, {\psi'(x)\over \psi(x)}  + {1\over \hbar\,\psi^2(x)}\,\int_{\infty_0}^x\, \psi^2(x')\, (\delta U(x')- 2\,{\delta \hbar\over \hbar}\,U(x'))\,\, dx'.
}\eeq

If we write:
\beq
\delta U = {V'\over 2}\delta V'  - {\hbar\over 2}\delta V'' -  {\delta\hbar\over 2} V'' - \delta P
\eeq
where $\delta P$ is of degree at most $d-1$, and $V'/2 = \om-\hbar \psi'/\psi$, we have by integration by parts:
\bea\label{variation2w(x)}
\delta \om(x) 
&=&  {1\over \hbar\,\psi^2(x)}\,\int_{\infty_0}^x\, \psi^2(x')\, \Big(\om(x')\delta V'(x')-\delta P(x') \cr
&& \qquad -\,{\delta \hbar}\,(\om'(x')-{1\over 2}\,V''(x'))\Big)\,\, dx'.
\eea

\subsection{Decomposition of variations}

$U(x)$ is a polynomial of degree $2d$, it has $2d+1$ independent coefficients.
If we assume that we have a solution of genus $\genus<d-1$, this means that $U$ is non generic, and satisfies $d-1-\genus$ constraints.
In the space of all possible $U$'s, we shall consider the submanifold corresponding to $U$ of genus $\genus$, which is a submanifold of dimension
\beq
{\rm dim}= d+2+\genus
\eeq
and we shall consider variations of $U$ within that submanifold.
Variations transverse to the genus $\genus$ submanifold, are variations of higher genus and should be computed within a higher genus submanifold.

\medskip

Instead of the $d+2+\genus$ independent coefficients of the polynomial $U$, it is more convenient to choose a system of "flat" coordinates in our genus $\genus$ submanifold, given by:
\beq
t_0,t_1,\dots, t_{d+1},\,\,\epsilon_1,\dots,\epsilon_\genus.
\eeq
We have indeed $d+2+\genus$ coordinates.

\medskip
Let us write the variations as:
\beq
\delta U = \sum_{k=0}^{d+1} U_{t_k}\, \delta t_{k} + \sum_{i=1}^\genus U_{\epsilon_i}\delta \epsilon_i + U_{\hbar} \delta \hbar.
\eeq 

\subsubsection{Variations relatively to the filling fractions}

For the filling fraction $\delta \epsilon_\alpha$ we have $\delta V'=0$ and thus:
\beq
\delta U(x) = -\delta P(x)
\eeq
where $\deg \delta P\leq d-2$, so we decompose it on the basis of $h_\alpha$'s:
\beq
\delta P(x) = \sum_{\alpha'} c_{\alpha'}\,h_{\alpha'}.
\eeq
and therefore, from \eq{variationw(x)}:
\beq
\delta \om(x) = -\sum_{\alpha'} c_{\alpha'} \,v_{\alpha'}(x).
\eeq
Since $2i\pi \epsilon_{\alpha'} = \oint_{\acycle_{\alpha'}} \om$, we have:
\beq
2i\pi \,\delta_{\alpha,\alpha'} = \oint_{\acycle_{\alpha'}} \delta \om = -\sum_{\alpha''} \oint_{\acycle_{\alpha'}} c_{\alpha''}\,v_{\alpha''}
= -c_{\alpha'}
\eeq
This implies:
\beq
U_{\epsilon_\alpha}(x) = 2i\pi h_\alpha(x)
\eeq
and
\beq \encadremath{
\delta_{\epsilon_i}\om(x) = 2i\pi \, v_\alpha(x) = \oint_{\bcycle_\alpha}\, B(x,z)\,dz.}
\eeq

We shall say that the flat coordinate $\epsilon_\alpha$ is dual to the holomorphic form $v_\alpha$, which is itself dual to the cycle $\bcycle_\alpha$:
\beq
\epsilon_\alpha "=" {1\over 2i\pi}\,\oint_{\acycle_\alpha}\, \om
\qquad , \qquad
\delta_{\epsilon_\alpha} \om = 2i\pi\, v_\alpha = \oint_{\bcycle_\alpha}\, B.
\eeq

\subsubsection{Variations relatively to $t_0$}

We have:
\beq
\delta U(x) = -\delta P(x) = -t_{d+1}\,x^{d-1} + Q(x)
\eeq
where $\deg Q\leq d-2$.
Using \eq{variationw(x)} we get:
\beq
\delta \om(x) = {1\over \psi^2(x)}\,\int_{\infty_0}^x\, (-t_{d+1}x'^{d-1}+Q(x'))\,\psi^2(x')\,dx'
\eeq
and the polynomial $Q$ is chosen such that $\oint_{\acycle_i} \delta \om=0$ so that when decomposing $Q(x)$ on the basis $v_\alpha(x)$ and performing integrals over $\acycle$-cycles one finds the coefficients of the decomposition as integrals.
Therefore we have:
\bea
\delta \om(x) 
&=& -\,t_{d+1}\,K_{d-1}(x) \cr
&=& -t_{d+1}\,\Big( \hat K_{d-1}(x) - \sum_{\alpha=1}^{\genus}\, v_\alpha(x)\, \oint_{\acycle_\alpha} \hat K_{d-1}(x')\,dx'
- \sum_{\alpha=\genus+1}^{d-1}\, v_\alpha(x)\, \oint_{\acycle_\alpha} \psi^2(x')\,x'^{d-1}\,dx'\Big) \cr
\eea
where
\beq
\hat K_k(x) = {1\over \psi^2(x)}\,\int^x_{\infty_0} x'^k\,\psi^2(x')\, dx',
\eeq
and $K_k(x)$ is the $k^{\rm th}$ term in the large $z$ expansion of $K(x,z)= -\sum_{k=0}^\infty {K_k(x,z)\over z^{k+1}}$ computed in theorem \ref{thKlargez}.
From theorem \ref{thGlargexz} we have
$G(x,\infty_i) = \eta_i t_{d+1}\,K_{d-1}(x) $.
This shows that
\beq
\encadremath{
\delta_{t_0}\om(x)=G(x,\infty_0) = {1\over 2}\,(G(x,\infty_0)-G(x,\infty_-)) = \int_{\infty_0}^{\infty_-} B(x,z)\,dz
}\eeq
where $\infty_0$ is in the physical sheet, and $\infty_-$ is any infinity chosen in the second sheet.

We shall say that the flat coordinate $t_0$ is dual to the 3rd kind meromorphic form $-2G(x,\infty_0)$, which is itself dual to the chain $[\infty_0,\infty_-]$:
\beq \encadremath{
t_0=\Res_{\infty_0} \om
\qquad , \qquad
\delta_{t_{0}} \om = -2G(x,\infty_0) = \int_{\infty_0}^{\infty_-}\, B(x,z)\,dz}
\eeq
where $\Res$ means the coefficient of $1/z$ in the given sector.

\subsubsection{Variation relatively to $t_k, k=1\dots d$}

For $k=1,\dots, d$ we have:
\beq
U_{t_k}(x) = {V'(x)\over 2}\,x^{k-1} - Q(x)
\virg \deg Q\leq d-2
\eeq
and $Q$ is chosen such that $\oint_{\acycle_i} \delta \om =0$.
Using \eq{variationw(x)} we write:
\beq
\delta \om(x) = \delta \hat\om(x) - \sum_\alpha v_\alpha(x)\,\oint_{\acycle_\alpha} \delta\hat\om(x')\,dx'
\eeq
where
\beq
\delta \hat\om(x)= {1\over \psi^2(x)}\,\int_{\infty_0}^x\, {V'(x')\over 2}\,x'^{k-1}\,\,\psi^2(x')\,dx'
\eeq
Since $V'(x')=\sum_j t_{j+1}\,x'^j$,  we have:
\beq
2 \delta \om(x) =\sum_{j=0}^{d} t_{j+1}\,K_{k+j-1} 
\eeq
Let us compare it with the large $z$ behaviour of $G(x,z)$ in the physical sheet. We have:
\beq
G(x,z) = V'(z)\,K(x,z) +O(z^{-d-1})
\eeq
which means that the large $z$ expansion of $G(x,z)=\sum_k G_k(x)\,z^{-k}$ is given for $k=1,\dots,d$ by:
\beq
G_{k}(x) = -\sum_{j=0}^{d} t_{j+1} K_{k+j-1} 
\eeq
and therefore
\beq
\delta\om(x) = -{1\over 2} G_{k}(x)\,
\eeq
If we write the large $z$ expansion of $B(x,z)$ in the physical sheet, we have
\beq
B(x,z) =\sum_k B_k(x)\,z^{-k-1}\, = -\,{1\over 2}\sum_k k\,G_k(x,z) z^{-k-1}
\eeq
and thus
\beq \encadremath{
\delta_{t_{k}}\om(x) = {1\over k}\,B_{k}(x) = \Res_{\infty_0}\, {z^{k}\over k}\,B(x,z)\,dz}
\eeq

We shall say that the flat coordinate $t_k$ is dual to the 2nd kind meromorphic form ${1\over k}\,B_{k}(x)$, which is itself dual to a residue of $B$.

\subsubsection{Variations relatively to $t_{d+1}$}

When $k=d+1$, we have a few additional terms of degree $> d-2$:
\beq
U_{t_{d+1}}(x) = {V'(x)\over 2}\,x^{d} -{d\,\hbar\over 2}\,x^{d-1} -t_0 x^{d-1} - Q(x)
\virg \deg Q\leq d-2
\eeq
and $Q$ is chosen such that $\oint_{\acycle_i} \delta \om =0$.
Using \eq{variationw(x)} we write:
\beq
\delta \om(x) = \delta \hat\om(x) - \sum_\alpha v_\alpha(x)\,\oint_{\acycle_\alpha} \delta\hat\om(x')\,dx'
\eeq
where
\beq
\delta \hat\om(x)= {1\over \psi^2(x)}\,\int_{\infty_0}^x\, \left({V'(x')\over 2}\,x'^{d} -{d\,\hbar\over 2}\,x'^{d-1} -t_0 x'^{d-1}\right)\,\,\psi^2(x')\,dx'
\eeq
In other words we have:
\beq
2 \delta \om(x) =\sum_{j=0}^{d} t_{j+1}\,K_{d+j} - d\hbar\, K_{d-1} -2t_0 K_{d-1}
\eeq
Let us compare it with the large $z$ behaviour of $G(x,z)$. We have:
\beq
G(x,z) = (V'(z)-{2t_0\over z})\,K(x,z)-\hbar \partial_z K(x,z) +O(z^{-d-2})
\eeq
which means that the large $z$ expansion of $G(x,z)=\sum_k G_k(x)\,z^{-k}$ is given for $k=d+1$ by:
\beq
G_{d+1}(x) = -\sum_{j=0}^{d} t_{j+1} K_{d+j} + \hbar\,d\,K_{d-1} + 2t_0 K_{d-1} 
\eeq
and therefore
\beq
\delta\om(x) = -{1\over 2} G_{d+1}(x)\,
\eeq
If we write the large $z$ expansion of $B(x,z)$, we have
\beq
B(x,z) =\sum_k B_k(x)\,z^{-k-1}\, = -\,{k\over 2}\sum_k G_k(x,z) z^{-k-1}
\eeq
and thus
\beq
\encadremath{
\delta_{t_{d+1}}\om(x) = {1\over d+1}\,B_{d+1}(x) = \Res_{\infty_0}\, {z^{d}\over d}\,B(x,z)\,dz}
\eeq

We shall say that the flat coordinate $t_{d+1}$ is dual to the 2nd kind meromorphic form ${1\over d+1}\,B_{d+1}(x)$, which is itself dual to a residue of $B$.

\subsection{Variation relatively to $\hbar$}

We have:
\beq
\delta_{\hbar} \om(x) 
= -\, {1\over \hbar\,\psi^2(x)}\,\int_{\infty_0}^x\, \psi^2(x')\, \Big( \om'(x')-{1\over 2}\,V''(x')-\delta_\hbar P(x')\Big)\,\, dx'
\eeq
where $\delta_{\hbar} P$ is a polynomial of degree $\leq d-2$ chosen so that $\oint_{\acycle_i} \delta\om=0$.
For the moment, we have not found a good way of writing this expression as an integral with $B$, and we leave that question for a future work.

\subsection{Form-cycle duality}

Notice that in all cases, except $\delta_{\hbar}$, there exist a cycle $\delta\om^*$ and a function $\Lambda_{\delta\om}^*$ such that:
\beq
\delta\om(x) = \int_{\delta\om^*}\, B(x,z)\,\Lambda_{\delta\om}^*(z)\,dz.
\eeq
We will use this generic notation later on in order to avoid specifying the $3$ different cases.

Under a suitable reparametrization $z\to z'$ such that $dz'=\Lambda_{\delta\om}^*(z)\,dz$, we say that $\delta\om^*$ in the variable $z'$ is the cycle dual to the "meromorphic form" $\delta\om$.

\subsection{Variation of higher correlators}

The following theorem allows to compute the infinitesimal variation of any $W_n^{(g)}$ under a variation of the Schr\"odinger equation.
It tells about the "complex structure deformation" of our quantum Riemann surface.
It can be regarded as special geometry relations.

\bt
Under an infinitesimal deformation $U\to U+\delta U$, 
we have:
\beq
\delta W_n^{(g)}(x_1,\dots,x_n) = \int_{\delta\om^*}\, W_{n+1}^{(g)}(x_1,\dots,x_n,x')\,\Lambda^*(x')\,dx'
\eeq
where $(\delta\om^*,\Lambda_{\delta\om}^*)$ is the dual cycle to the deformation of the resolvent $\om\to \om+\delta \om$.
\et

\proof{
The loop equation  for $W_n^{(g)}(x,J)$ is:
\beq
(2\om(x)-V'(x)+\hbar \partial_x)\,W_n^{(g)}(x,J) + U_n^{(g)}(x,x;J) = P_n^{(g)}(x,J)
\eeq
taking a variation $\delta$ we have:
\bea
&&(2\om(x)-V'(x)+\hbar \partial_x)\,\delta W_n^{(g)}(x,J)
+ (2\delta \om(x)-\delta V'(x)) W_n^{(g)}(x,J)
 + \delta U_n^{(g)}(x,x;J) \cr
 &&=\delta P_n^{(g)}(x,J) 
\eea
notice that $\delta P_n^{(g)}(x,J)$ is a polynomial in $x$, of degree at most $d-2$.

\medskip
On the other hand, consider the loop equation for $W_{n+1}^{(g)}(x,J,x')$ and multiply it by $\Lambda^*(x')$ and integrate $x'$ along $\om^*$, one gets:
\bea
&&(2\om(x)-V'(x)+\hbar \partial_x)\,\int_{\om^*} W_{n+1}^{(g)}(x,J,x')\Lambda^*(x')dx'
 + \int_{\om^*}\delta U_{n+1}^{(g)}(x,x;J,x')\Lambda^*(x')dx' \cr
 &&=  \int_{\om^*}P_{n+1}^{(g)}(x,J,x')\Lambda^*(x')dx'  \cr
\eea

That gives by recursion hypothesis for the computation of $\int_{\om^*}\delta U_{n+1}^{(g)}(x,x;J,x')\Lambda^*(x')dx'$ and using \ref{variationw(x)}:
\bea
&&(2\om(x)-V'(x)+\hbar \partial_x)\left(\int_{\om^*} W_{n+1}^{(g)}(x,J,x')\Lambda^*(x')dx' - \delta W_n^{(g)}(x,J)\right) \cr
 &=& \delta P_n^{(g)}(x,J)-\int_{\om^*}P_{n+1}^{(g)}(x,J,x')\Lambda^*(x')dx' \cr
 &=& \sum_i \alpha_i(J)\, h_i(x)
\eea
where the right hand side is a polynomial of degee at most $d-2$ in $x$, which can be decomposed on the basis $h_i(x)$.

Solving the differential equation gives:
\beq
\int_{\om^*} W_{n+1}^{(g)}(x,J,x')\Lambda^*(x')dx' - \delta W_n^{(g)}(x,J)
= \sum_i \alpha_i(J)\, v_i(x)
\eeq
but since $W_n^{(g)}$ and $W_{n+1}^{(g)}$ are normalized on $\acycle$-cycles, this implies $\alpha_i=0$, i.e.:
\beq
\int_{\om^*} W_{n+1}^{(g)}(x,J,x')\Lambda^*(x')dx' = \delta W_n^{(g)}(x,J)
\eeq

}

\section{Free energies}

We use the variations and theorem \ref{thhomogeneity} to define the $F_g$'s.

Theorem \ref{thhomogeneity} gives:
\beq
(2-2g-n-\hbar\,\partial_{\hbar})\,W_n^{(g)} = \left(t_0\,\partial_{t_0}+\sum_{k=1}^{d+1}\,t_k\,\partial_{t_k}+\sum_{i=1}^\genus \epsilon_i \partial_{\epsilon_i}\right)\,W_n^{(g)}
\eeq
And in the previous section, we have seen how to write the derivatives of $W_n^{(g)}$ as integrals of $W_{n+1}^{(g)}$, that gives:
\beq
(2-2g-n-\hbar\,\partial_{\hbar})\,W_n^{(g)} = \hat H.\,W_{n+1}^{(g)}
\eeq
where $\hat H$ is the linear operator acting as follows:
\beq
\hat H.f(x) = t_0\,\int_{\infty_0}^{\infty_-} f +\sum_{j=1}^{d+1} \Res_{\infty_0}\, {t_{j}\,x^j\over j}\,f + \sum_{i=1}^\genus \epsilon_i\,\oint_{\bcycle_i} f.
\eeq

Those equations allow to define $W_0^{(g)}=F_g$ for $n=0$ and $g\geq 2$ as:

\bd
We define $F_g$ for $g\geq 2$ such that:
\beq
(2-2g-\hbar\,\partial_{\hbar})\,F_g = \hat H.\,W_{1}^{(g)}
\eeq

\ed

\medskip

It would remain to find the correct definitions of $F_0$ (called the prepotential) and $F_1$.
$F_0$ and $F_1$ should be such that under every deformation $\delta=\partial_{t_k},\partial_{t_0}, \partial_{\epsilon_i}$ we should have
\beq
\delta  F_g = H_{\delta}\, W_1^{(g)}.
\eeq
For example $\partial F_0/\partial t_k=\Res x^k \om(x)/k$ i.e. the coefficient of the term $1/x^{k-1}$ in the expansion of $\om(x)$ near $\infty_0$.

\smallskip
We leave the definitions of $F_0$ and $F_1$ for a future work.

\section{Classical and quantum geometry: summary}

Let us summarize the comparison between classical algebraic geometry, and its quantum counterpart introduced here.

\begin{figure}[bth]
\hrule\hbox{\vrule\kern8pt
\vbox{\kern8pt \vbox{
\begin{center}
{\mbox{\epsfxsize=7.truecm\epsfbox{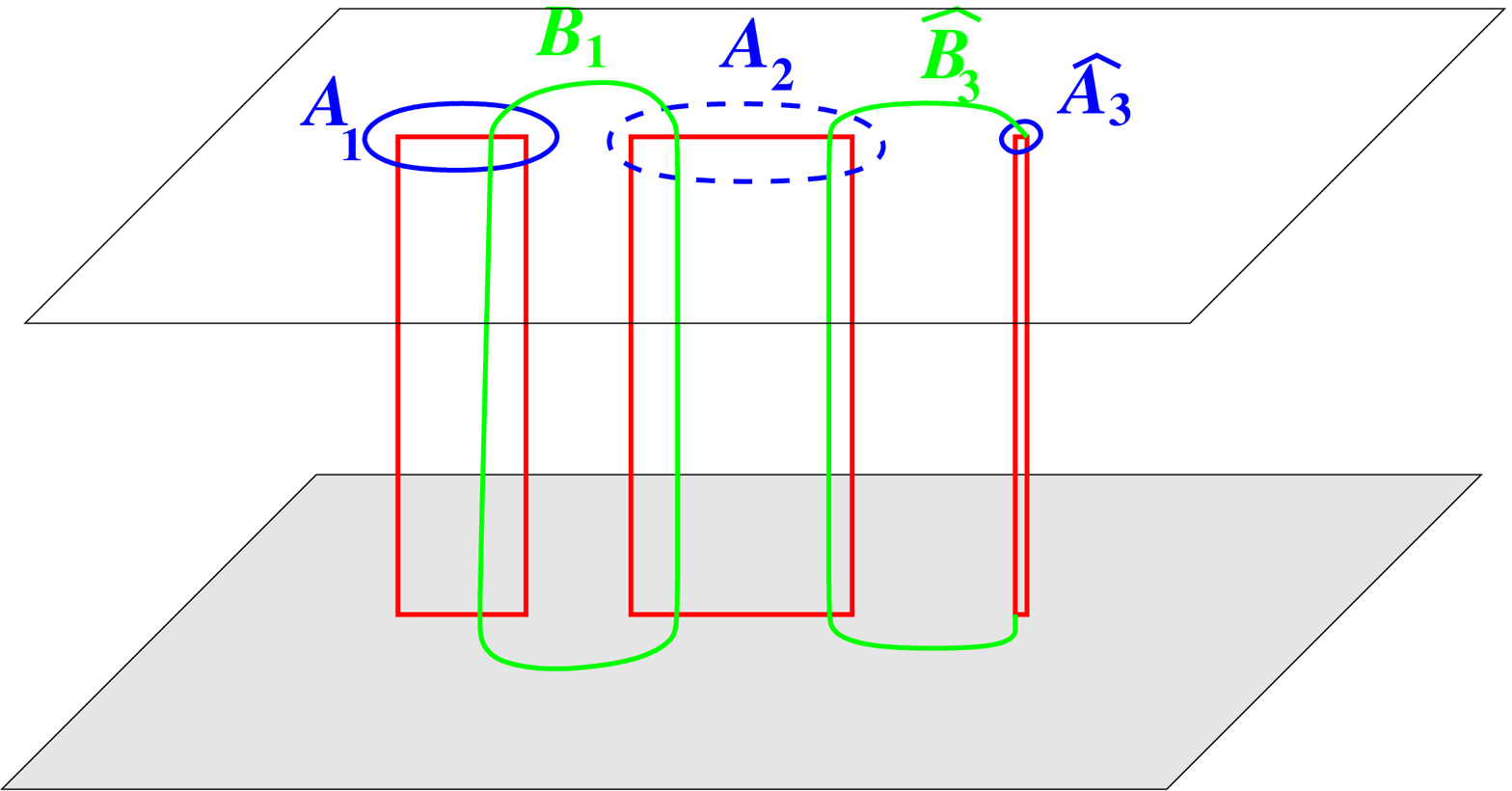}}}
{\mbox{\epsfxsize=7.truecm\epsfbox{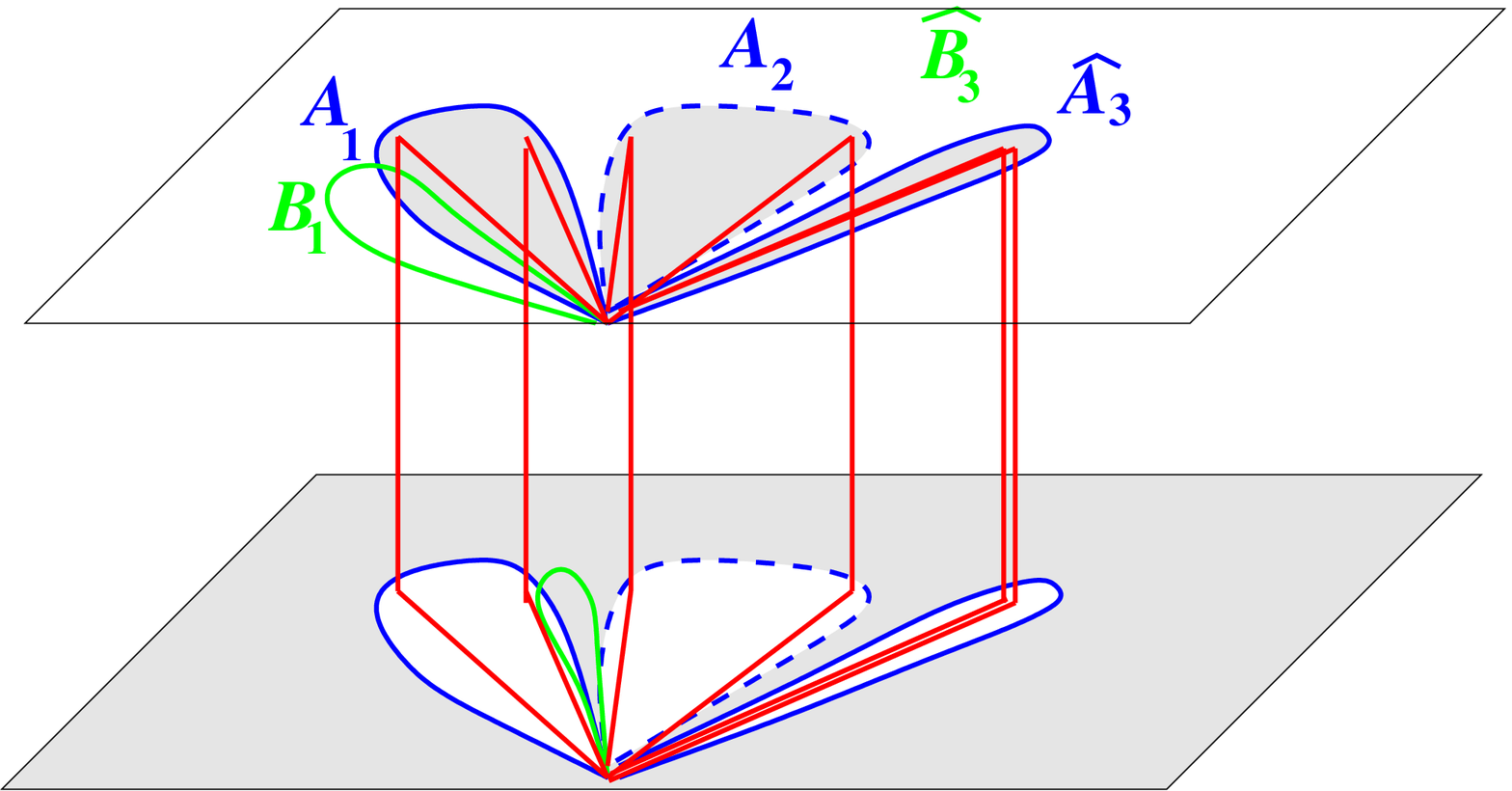}}}
\end{center}
\caption{Classical case $\hbar=0$ of a two sheeted Riemann surface. The branchpoints are paired (in an arbitrary way) to form cuts, and the two sheets are glued along the cuts. Another possibility, is to draw a cut from each branchpoint to $\infty$.
The $\acycle$-cycles surround pairs of branchpoints in the physical sheet.
There are also some degenerate branchpoints, which correspond to cuts of vanishing length.
}
}\kern8pt}
\kern8pt\vrule}\hrule
\end{figure}

\noindent
\begin{tabular}{|l
@{ $\,\, | \,\,$ } l
@{ $\,\, | \,\,$ } 
p{220pt}|}
\hline
\multicolumn{3}{|c|}{
\underline{\bf Summary}} \\
\hline
& {\bf classical $\quad \hbar=0$} & {\bf quantum} \\
\hline
 plane curve:  & 
$E(x,y)=\sum_{i,j} E_{i,j} x^i\,y^j $ &
$E(x,y)=\sum_{i,j} E_{i,j} x^i\,y^j \, , \quad [y,x]=\hbar$ \\
& $E(x,y)=0$ &  $E(x,\hbar \partial_x)\psi=0$  \\
\hline
hyperelliptical  &  $y^2=U(x) $ & $y^2-U(x) \, , \quad [y,x]=\hbar$, 
 \\
plane curve: &  $\deg U=2d$ &
$\hbar^2\psi'' = U\, \psi$  \\
\hline
Potential: &   \multicolumn{2}{c|} { $V'(x)=2(\sqrt{U(x)})_+$ } \\
\hline
2 sheets: & $y\sim_\infty \pm {1\over 2}V'(x) $
& 
$\hbar \psi'/\psi \sim_\infty \pm {1\over 2}V'(x) \sim_{\infty_k}  {\eta_k\over 2}V'(x) $\\
& & choice $\psi=\psi_0\searrow$ in sector $\infty_0$, $\eta_0=-1$\\
\hline
resolvent: & 
$\om(x) = {V'(x)/2} + y $.
& 
$\om(x) = {V'(x)/2} +\hbar {\psi'\over \psi} $.\\
\hline
physical sheet: &  $y\sim_\infty - {1\over 2}V'(x) $, $\om \sim t_0/x$
& 
$\hbar \psi'/\psi \sim_\infty - {1\over 2}V'(x) $, $\om \sim t_0/x$\\
&& sectors where $\psi\sim e^{-{V\over 2\hbar}}$\\
\hline
branchpoints: & simple zeroes of $U(x)$ & half-lines of accumulations \\
& $U(a_i)=0$, $U'(a_i)\neq 0$ & of zeroes of $\psi$ \\
& $i=1,\dots,2\genus+2$ &  $L_i$, $i=1,\dots,2\genus+2$ \\
\hline
genus $\genus$: &  \multicolumn{2}{c|}{$2\genus+2 = \#$ branch points} \\
& \multicolumn{2}{c|}{ $-1\leq \genus\leq d-1$ } \\
\hline
double points: & double zeroes of $U(x)$ & half-lines without accumulations \\
& $U(\hat a_i)=0$, $U'(\hat a_i)= 0$ & of zeroes of $\psi$ \\
\hline
genus $\genus=-1$ & degenerate surface & $\psi\,\ee{V/2\hbar} =$polynomial \\
\hline
$\acycle_\alpha$-cycles  & surround pairs of  & surround pairs of half-lines \\
$\alpha=1,\dots,\genus$  & branchpoints  & of accumulating zeroes \\
\hline
false $\hat\acycle_\alpha$-cycles & surround double  & links 2 sectors \\
$\genus<\alpha<d$ & points  & where $\psi \searrow$ \\
\hline
extra $\acycle_d$-cycle & surrounds last pair of  & surrounds last pair of half-lines  \\
$\alpha=d$ & branchpoints  &  of accumulating zeroes \\
\hline
$\bcycle$-cycles & \multicolumn{2}{c|} { $\acycle_i\cap \bcycle_j=\delta_{i,j}$  } \\
\hline
Holomorphic  & $v_i(x) = {-h_i(x) \over 2\sqrt{U(x)}}\,$ & $v_i(x) = {1\over \hbar \psi^2(x)}\int^x_{\infty_0} \psi^2(x')\,h_i(x')\,dx' $ \\ 
forms, & \multicolumn{2}{c|} {$h_i=$polynomials, $\deg h_i\leq d-2$} \\
1st kind & \multicolumn{2}{c|} {normalized: $\oint_{\acycle_\alpha} v_i(x)\,dx = \delta_{\alpha,i}, \, \alpha=1,\dots,\genus$ }\\
differentials & $h_i(\hat a_\alpha)=-{1\over 2}\delta_{\alpha,i}\,\sqrt{U''(\hat a_\alpha)} $ & $\oint_{\hat\acycle_\alpha} \psi^2 h_i = \delta_{\alpha,i} ,\, \alpha=\genus+1,\dots,d-1$  \\
\hline
Period matrix & \multicolumn{2}{c|} {$\tau_{i,j} = \oint_{\bcycle_j} v_i \quad $, $i,j=1,\dots,\genus$ , $\qquad \tau_{i,j}=\tau_{j,i}$} \\
\hline
Filling fractions & \multicolumn{2}{c|}{ $2i\pi\,\epsilon_\alpha = \oint_{\acycle_\alpha} \om(x)dx\quad $, $\alpha=1,\dots,\genus$, $\qquad \epsilon_{d}=t_0-\sum_{\alpha=1}^\genus \epsilon_\alpha$} \\
\hline
\end{tabular}
\bigskip

\noindent
\begin{tabular}{|l
@{ $\,\, | \,\,$ } l
@{ $\,\, | \,\,$ } 
p{220pt}|}
\hline
\multicolumn{3}{|c|}{
\underline{\bf Summary}} \\
\hline
& classical $\quad \hbar=0$& quantum \\
\hline
3rd kind form &  \multicolumn{2}{c|} {$G(x,z) \sim_{x\to z} 1/(z-x)$} \\
&  \multicolumn{2}{c|} {  $G(x,z) = (2\om(z)-V'(z)-\hbar\partial_z)K(x,z)$  } \\
\hline
Recursion kernel &  \multicolumn{2}{c|} {$K(x,z)$} \\
&  \multicolumn{2}{c|} { $K(x,z) = \hat K(x,z) - \sum_\alpha v_\alpha(x)\,C_\alpha(z)$ } \\
&  \multicolumn{2}{c|} { $C_\alpha(z) = \oint_{\acycle_\alpha}  \hat K(x',z) dx'$ } \\
& $\hat K(x,z) = {1\over z-x}\,{1\over 2\sqrt{U(x)}}$ & $\hat K(x,z) = {1\over \hbar \psi^2(x)}\int^x_{\infty_0} {\psi^2(x')dx'\over x'-z}$ \\
\hline
Bergman kernel & \multicolumn{2}{c|} {$B(x,z) = -{1\over 2}\,\partial_z\, G(x,z)$} \\
2nd kind & \multicolumn{2}{c|} { $B(x,z)\sim 1/2(x-z)^2$}\\
\hline
Symmetry: & \multicolumn{2}{c|} { $B(x,z)=B(z,x)$}\\
\hline  
& \multicolumn{2}{c|} {$\oint_{\acycle_\alpha} B(x,z)dx=0$ } \\
& \multicolumn{2}{c|} {$\oint_{\bcycle_\alpha} B(x,z)dx=2i\pi\, v_\alpha(z)$ } \\
\hline
Meromorphic  & ${\cal R}(x)dx = {r(x)dx\over 2\sqrt{U(x)}}$ & ${\cal R}(x) = {1\over \hbar\psi^2(x)}\,\int_{\infty_0}^x r(x')\,\psi^2(x')\,dx'$ \\
forms & \multicolumn{2}{c|} {$r(x)=$rational with poles $r_i$, $r(x)=O(x^{d-2})$}\\
&  & $\Res_{r_i} r(x')\psi^2(x')=0$\\
\hline
Higher & \multicolumn{2}{c|} {$W_{n+1}^{(g)}(x,J) = \sum_i {1\over 2i\pi}\oint_{{\cal C}_i} K(x,z)\, dz\,\Big( W_{n+2}^{(g-1)}(z,z,J)$ } \\
correlators & \multicolumn{2}{c|} {$\qquad \quad + \sum'_{h+h'=g,\, I\uplus I'=J} W_{1+|I|}^{(h)}(z,I) W_{1+|I'|}^{(h')}(z,I') \Big)$ } \\
& \multicolumn{2}{c|} { where ${\cal C}_i$ surrounds the branchpoint $L_i$} \\
\hline
Symmetry & \multicolumn{2}{c|} {$W_{n}^{(g)}(x_1,x_2,\dots,x_n) = W_{n}^{(g)}(x_{\sigma(1)},x_{\sigma(2)},\dots,x_{\sigma(n)})\, , \qquad\sigma\in S_n$} \\
\hline
Variations and & \multicolumn{2}{c|} {$U(x) \to U(x)+ \delta U(x)$}\\
dual cycle & \multicolumn{2}{c|} {  $\delta U^*$: $\delta \om(x) = \int_{\delta U^*} B(x,x') \,\, \Lambda_{\delta U}(x')\,\,dx'$}\\
\hline
$\delta V'=\sum \delta t_{k}\, x^{k-1}$ & \multicolumn{2}{c|} {$\delta_{t_k} \om(x) = \Res_\infty B(x,x')\,{x'^k\over k}\,dx' $} \\
\hline
variation $\delta t_0$ & \multicolumn{2}{c|} {$\delta_{t_0} \om(x) = \int_{\infty_0}^{\infty_-} B(x,x')\,dx' $} \\
\hline
variation $\delta \epsilon_i$ & \multicolumn{2}{c|} {$\delta_{\epsilon_i} \om(x) = \oint_{\bcycle_i} B(x,x')\,dx' $} \\
\hline
Variations of & \multicolumn{2}{c|} {$\delta W_{n}^{(g)}(x_1,\dots,x_n) = \int_{\delta U^*} W_{n+1}^{(g)}(x_1,\dots,x_n,x') \,\, \Lambda_{\delta U}(x')\,\, dx'$}\\
higher correlators & \multicolumn{2}{c|} {} \\
\hline
Rauch formula & \multicolumn{2}{c|} { $W_3(x_1,x_2,x_3) = \oint_{{\cal C}} {B(x_1,z)B(x_2,z)B(x_3,z)\over 4 Y'(z)} \,\, dz$ } \\
\hline
\end{tabular}

\section{Application: Matrix models \label{secMM}}

The reason why we introduced those $W_n^{(g)}$'s is because they satisfy the loop equations for $\beta$-random matrix ensembles.

Consider a (possibly formal) matrix integral:
\beq
Z = \int_{E_{N,\beta}}\, dM\,\, \ee{-{N\sqrt\beta\over t_0}\,\tr V(M)}
\eeq
where $V(x)$ is some polynomial, and 
where $E_{N,1}=H_N$ is the set of hermitian matrices of size $N$, $E_{N,1/2}$ is the set of real symmetric matrices of size $N$ and $E_{N,2}$ is the set of quaternion self dual matrices of size $N$ (see \cite{Mehtabook}).

Alternatively, we can integrate over the angular part and get an integral over eigenvalues only \cite{Mehtabook}:
\beq
Z= \int d\lambda_1\dots d\lambda_N\,\, \Delta(\lambda)^{2\beta}\,\, \prod_{i=1}^N\, \ee{-{N\sqrt\beta\over t_0}\,V(\lambda_i)}
\eeq
where $\Delta(\lambda)=\prod_{i<j}(\l_j-\l_i)$ is the Vandermonde determinant.

This allows to generalize the matrix model to arbitrary values of $\beta$.
In particular, we shall choose $\beta$ of the form:
\beq
\sqrt \beta = {\hbar N\over 2 t_0}\,\left( 1 \pm \sqrt{1+{4 t_0^2\over \hbar^2\,N^2}}\right)
\eeq
i.e.
\beq
\hbar = {t_0\over N}\,\left(\sqrt\beta-{1\over \sqrt\beta}\right).
\eeq
Notice that $\hbar=0$ correspond to the hermitian case $\beta=1$, and $\hbar\to -\hbar$ corresponds to $\beta\to 1/\beta$.

\subsection{Correlators and loop equations}

Then we define the correlators:
\beq
W_k(x_1,\dots,x_k) =  \beta^{k/2}\,\, \left< \sum_{i_1,\dots,i_k} {1\over x_1-\lambda_{i_1}} \dots {1\over x_k-\lambda_{i_k}} \right>_c
\eeq
and
\beq
W_0 = F = \ln{Z}.
\eeq

And we assume (this is automatically true if we are considering formal matrix integrals), that there is a large $N$ expansion of the type (where we assume $\hbar=O(1)$):
\beq\label{WkgdevtopMM}
W_k(x_1,\dots,x_k) = \sum_{g=0}^\infty (N/t_0)^{2-2g-k} W_k^{(g)}(x_1,\dots,x_k)
\eeq
\beq
W_0 = F = \sum_g (N/t_0)^{2-2g} W_0^{(g)} = \sum_g (N/t_0)^{2-2g} F_g.
\eeq
The loop equations are obtained by integration by parts, for example:
\beq
0 = \sum_i \int d\l_1\dots d\l_N {\partial\over \partial \l_i}\left( {1\over x-\l_i}\, \Delta(\l)^{2\beta}\, \prod_j \ee{-{N\sqrt\beta\over t_0}\,V(\l_j)}\right)
\eeq
gives:
\bea
0 &=& \sum_i \left<{1\over (x-\l_i)^2} + 2\beta\sum_{j\neq i}{1\over x-\l_i}{1\over \l_i-\l_j} - {N\sqrt\beta\over t_0}\,{V'(\l_i)\over x-\l_i}\right> \cr
&=& \sum_i \left<{1\over (x-\l_i)^2} + \beta\sum_{j\neq i}{1\over x-\l_i}{1\over x-\l_j} - {N\sqrt\beta\over t_0}\,{V'(\l_i)\over x-\l_i}\right> \cr
&=& \sum_i \left<{1-\beta\over (x-\l_i)^2} + \beta\sum_{j}{1\over x-\l_i}{1\over x-\l_j} - {N\sqrt\beta\over t_0}\,{V'(\l_i)\over x-\l_i}\right> \cr
&=& (\beta-1){1\over \sqrt\beta} W'_1(x) + \beta ({1\over \beta}W_1^2(x) + {1\over \beta}W_2(x,x)) \cr
&& \qquad - {N\sqrt\beta\over t_0}\, \left({1\over\sqrt\beta} V'(x)W_1(x) - \sum_i \left< {V'(x)-V'(\l_i)\over x-\l_i}\right>\right) \cr
\eea
We define the polynomial
\beq
P_1(x) = {\sqrt\beta}\, \sum_i \left< {V'(x)-V'(\l_i)\over x-\l_i}\right> = (V' \, W_1)_+.
\eeq
We thus have:
\beq
W_1^2(x) + \hbar{N\over t_0}\, W_1'(x) + W_2(x,x) = {N\over t_0}\, \left(V'(x)W_1(x)-P_1(x)\right)
\eeq

Using the expansion \eq{WkgdevtopMM}, that gives the Ricatti equation
\beq
{W^{(0)}_1}(x)^2 + \hbar\,{\partial_x} W_1^{(0)}(x)  = V'(x)W^{(0)}_1(x)-P^{(0)}_1(x)
\eeq
which is satisfied by $\om(x)$:
\beq
W_1^{(0)}(x)=\om(x).
\eeq
generalizing to the integration by parts of
\beq
0 = \sum_i \int d\l_1\dots d\l_N {\partial\over \partial \l_i}\Big( {1\over x-\l_i}\sum_{i_1,\dots,i_k}\prod_{j=1}^k {1\over x_j-\l_{i_j}} 
 \quad  \, \Delta(\l)^{2\beta}\, \prod_j \ee{-{N\sqrt\beta\over t_0}\,V(\l_j)}\Big)
\eeq
and using the expansion \eq{WkgdevtopMM} to higher orders in $t_0/N$, one gets the loop equations of theorem \ref{thWngPng},
where
\bea
P_{k+1}(x;x_1,\dots,x_k) 
&=& \sum_g (N/t_0)^{2-2g-k}\, P_{k+1}^{(g)}(x;x_1,\dots,x_k) \cr
&=& \beta^{k/2}\,\left<\sum_i {V'(x)-V'(\l_i)\over x-\l_i}\sum_{i_1,\dots,i_k}\prod_{j=1}^k {1\over x_j-\l_{i_j}} \right>_c.
\eea
In other words, {\bf the correlation functions of $\beta$ matrix models, obey the topological recursion of def. \ref{defWng}}.

\bigskip

{\bf Remark:}

In \cite{EynOM}, a solution of loop equations for the $\beta$-matrix ensemble was proposed, but that solution was such that $U(x)$ was non-generic, corresponding to $\genus=-1$, and that $\psi(x)$ had only a finite number of zeroes. This case implied that $t_0$ was quantized.
Generic matrix models cannot correspond to that situation.

That solution was thus not very useful for actual matrix models.
Here instead, we have the solution for every $U(x)$, i.e. every contour of integration for the $\l_i$'s, and therefore we have the solution of loop equations for the actual matrix model.

\subsection{Example: real eigenvalues}

Very often, we are interested in a matrix model with real potential  $V(x)$ of even degree (i.e. $d$ is odd) and such that the eigenvalues are integrated along the real axis.
The resolvent $\om(x)$ is the Stieljes transform of the density of eigenvalues:
\beq
\om(x) = {t_0\sqrt\beta\over N}\, \int_{\mathbb R} {\rho(x')\,dx'\over x-x'}
\virg
\rho(x) = \left< \sum_i \delta(x-\l_i)\right>
\eeq
Let us consider that it is defined by this integral in the upper half-plane for $x\in \mathbb H_+$, and it is extended to the lower half-plane by analytical continuation.

By definition, $\om(x)$ is regular in the upper half-plane, therefore we look for a $\psi(x)$ which has no zero in the upper half-plane, i.e. no zero on the half-lines $L_0, L_1, \dots, L_{d-1}$.
I.e. it has at most $d+1$ half-lines of zeroes , and thus:
$$
\genus\leq {d-1\over 2}.
$$

\section{Non-oriented Ribbon graphs}

\begin{figure}[bth]
\hrule\hbox{\vrule\kern8pt
\vbox{\kern8pt \vbox{
\begin{center}
{\mbox{\epsfxsize=10.truecm\epsfbox{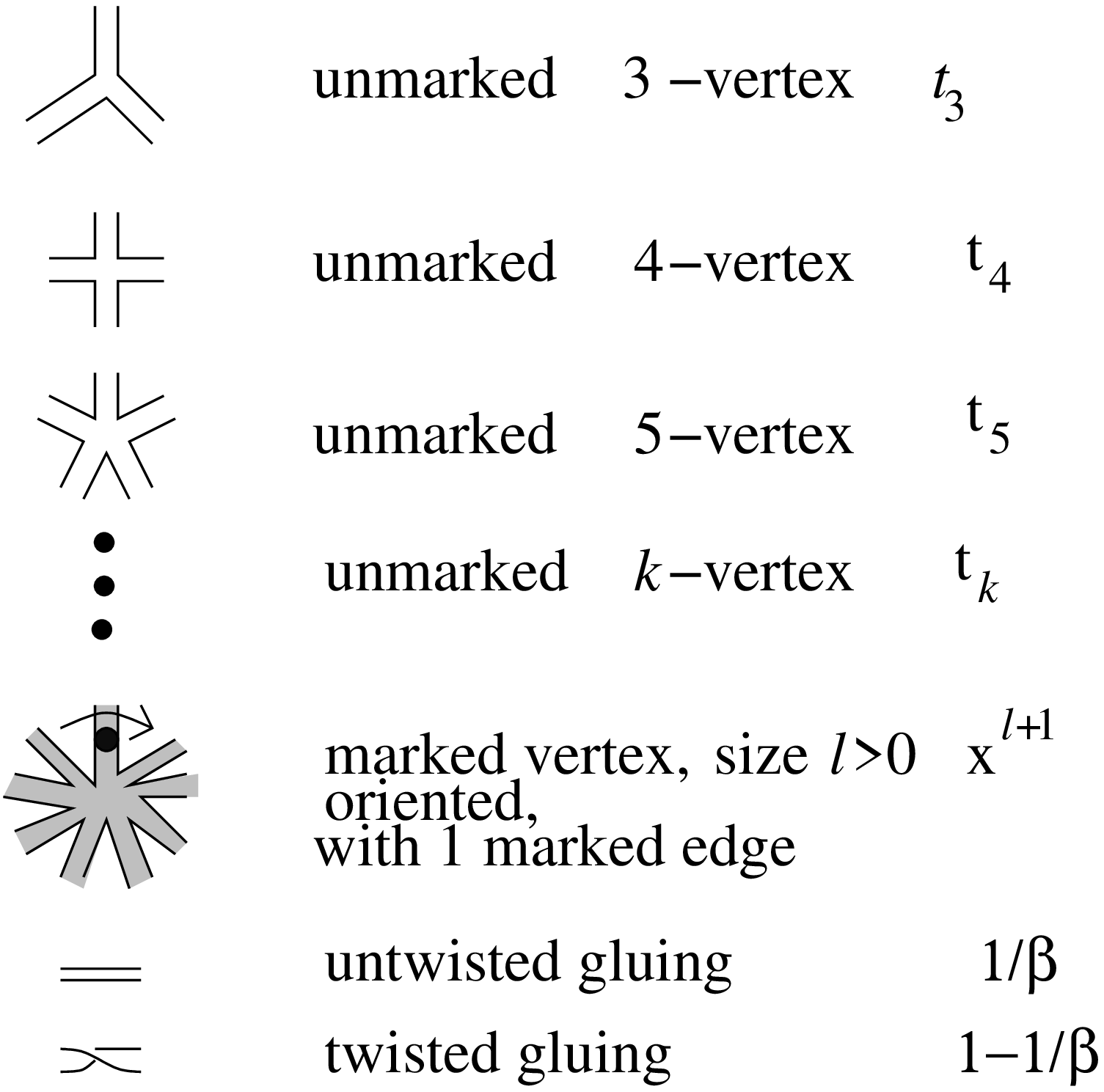}}}
\end{center}
\caption{Consider the set of ribbon graphs obtained by gluing those vertices.
Marked vertices are of degree $l\geq 1$, they are oriented and have one marked half-edge.
Unmarked vertices are unoriented, and are of degree $\geq 3$.
Vertices are glued together by their half-edges, either twisted (with weight $1/\beta$) or untwisted (with weight $1-1/\beta$).\label{figribbonpieces}}
}\kern8pt}
\kern8pt\vrule}\hrule
\end{figure}

\begin{figure}[bth]
\hrule\hbox{\vrule\kern8pt
\vbox{\kern8pt \vbox{
\begin{center}
{\mbox{\epsfxsize=10.truecm\epsfbox{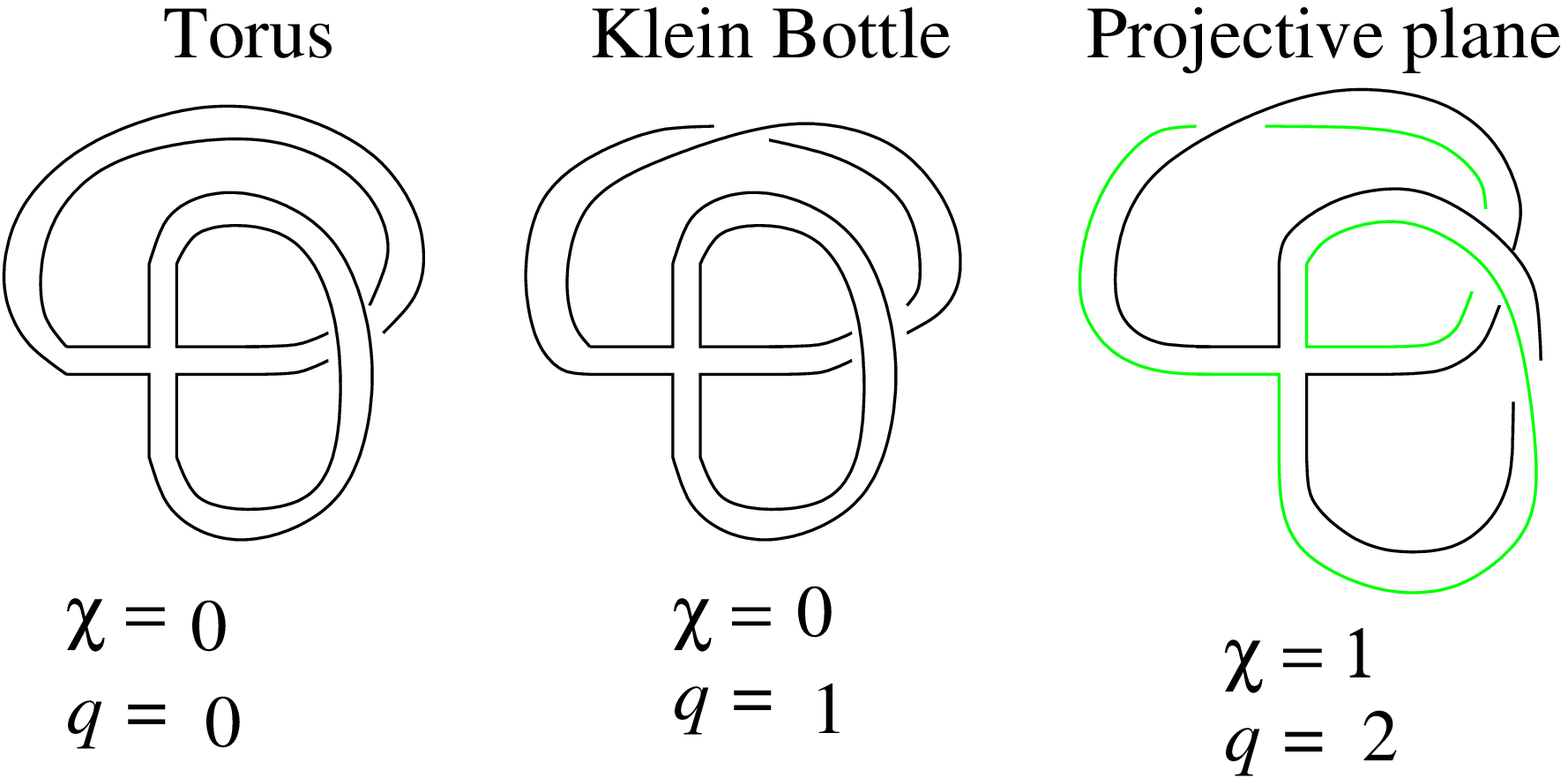}}}
\end{center}
\caption{Examples of ribbon graphs of genus $g=1$.\label{figexribbongraphs}}
}\kern8pt}
\kern8pt\vrule}\hrule
\end{figure}

Consider the set of all closed connected ribbon graphs obtained by gluing the pieces represented in fig. \ref{figribbonpieces}.
Closed means every half-edge is glued to another half-edge.
Connected means every vertex is connected to any other vertex.
See for example fig.~\ref{figexribbongraphs}.

\smallskip

We define the genus of such a ribbon graph ${\cal G}$ as follows.
We replace every twisted edge of ${\cal G}$ by a non-twisted one, we thus obtain another ribbon graph ${\cal G}'$, which is oriented. We define the genus of ${\cal G}$ equal to that of ${\cal G}'$:
$$g({\cal G})=g({\cal G}').$$
The genus of ${\cal G}'$ is computed as usual for oriented ribbon graphs, from the Euler characteristics of ${\cal G}'$:
$$\chi({\cal G}') =2-2g = \#\{{\rm vertices}({\cal G})\}-\#\{{\rm edges}({\cal G})\}+\#\{{\rm single\,lines}({\cal G}')\}$$
where single lines are the lines bordering each side of the ribbon edges. One should follow single lines and see how many connected single lines a graph contains.
Obviously ${\cal G}$ and ${\cal G}'$ have the same number of fat vertices and fat edges (each edge containing two single lines), but they may have different number of single lines.

This defines what we call the genus $g$ of a ribbon graph.

\bigskip

For a given Ribbon graph ${\cal G}$ we call:

$\bullet\,$ $n_i({\cal G})=\#$unmarked vertices of degree $i$, for $3\leq i\leq d+1$,

$\bullet\,$ $l_i({\cal G})=$size of the $i^{\rm th}$ marked vertex, we have $l_i({\cal G})\geq 1$.

$\bullet\,$ $e({\cal G})=\#$edges,

$\bullet\,$ $q({\cal G})=\#$twisted edges,

$\bullet\,$ $v({\cal G})=\#$connected single lines,

$\bullet\,$ $g({\cal G})=$genus,

$\bullet\,$ $\#Aut({\cal G})=$symmetry factor of ${\cal G}$.

\bd
Let $\mathbb M^{(g)}_k(v')$, be the set of ribbon graphs ${\cal G}$ with $k$ marked vertices, $q$ twisted edges, and of genus $g$, and  such that ${\cal G}'$ has $v({\cal G}')=v'$ connected single-lines, .
\ed

\bp
$\mathbb M^{(g)}_k(v')$ is a finite set.
\ep

\proof{
The number of vertices of ${\cal G}'$ is:
$$
\#\{{\rm vertices}\} = k+\sum_{i\geq 3} n_i
$$
The number of edges is twice the number of half-edges, i.e.
$$
2\,\#\{{\rm edges}\} = \sum_{i\geq 3} i\,n_i + \sum_{i=1}^k l_i
$$
That gives:
$$
2-2g = \#\{{\rm vertices}\}-\#\{{\rm edges}\}+v' = k-{1\over 2}\,\sum_{i\geq 3} (i-2) n_i-{1\over 2}\sum_{i=1}^k l_i + v
$$
i.e.
$$
k+v'+2g-2 = {1\over 2}\,\sum_{i\geq 3} (i-2) n_i+{1\over 2}\sum_{i=1}^k l_i 
$$
Since the left hand side is fixed, we see that the number and size of vertices are bounded, so that there is only a finite number of possible oriented ribbon graphs ${\cal G}'$.
Since ${\cal G}'$ has a bounded number of edges, there is only a finite number of possibilities of twisting them, i.e. there are also only a finite number of graphs ${\cal G}$.
}

\subsection{Generating functions}

In order to enumerate the sets $\mathbb M^{(g)}_k(v')$, we define the following generating functions:

\bd
We define:
\bea
&& W^{(g)}_k(x_1,\dots,x_k;t_3,\dots,t_{d+1},\beta;t_0) \cr
&=& \beta^{-k/2}\,\sum_{v'\geq 1}\, t_0^{v'}\,\, \sum_{{\cal G}\in \mathbb M^{(g)}_k(v')}
{1\over \#{\rm Aut}({\cal G})}\,\, {t_3^{n_3({\cal G})}\,t_4^{n_4({\cal G})}\,\dots\, t_{d+1}^{n_{d+1}({\cal G})}\over x_1^{l_1({\cal G})}\, x_2^{l_2({\cal G})}\, \dots \, x_k^{l_k({\cal G})}}\,\, \beta^{-e({\cal G})}\,\, (\beta-1)^{q({\cal G})}\,  \cr
&& + {\delta_{k,1}\delta_{g,0}\delta_{q,0}}\,\, {t_0\over x_1}
 + {\delta_{k,2}\delta_{g,0}\delta_{q,0}}\,\, {1\over 2\,(x_1-x_2)^2}. 
\eea
It is a formal series in powers of $t_0$.

Most often, for readability, we shall write only the dependence in the $x_i$'s:
$$ W_k^{(g)}(x_1,\dots,x_k;t_3,\dots,t_{d+1},\beta;t_0) \equiv W_k^{(g)}(x_1,\dots,x_k).$$
Also, for $k=0$ we write
$$
W_0^{(g)}=F_{g}.
$$
\ed

\subsection{Tutte's recursive equations}

Tutte's equation is a recursion on the number of edges to construct the ribbon graphs.
It consists in finding a bijection between ribbon graphs of various ensembles, by recursion on the number of edges.
Let
$\mathbb M^{(g)}_{l_1,\dots,l_k}$ be the set of ribbon graphs of genus $g$, and with $k$ marked vertices of size $l_1,\dots,l_k$.

\bigskip

Consider a ribbon graph ${\cal G}\in \mathbb M^{(g)}_{l_0+1,L}$ where $L=\{l_1,\dots,l_k\}$, with marked vertices of degrees $l_0+1,L$.

\begin{figure}[bth]
\hrule\hbox{\vrule\kern8pt
\vbox{\kern8pt \vbox{
\begin{center}
{\mbox{\epsfxsize=7truecm\epsfbox{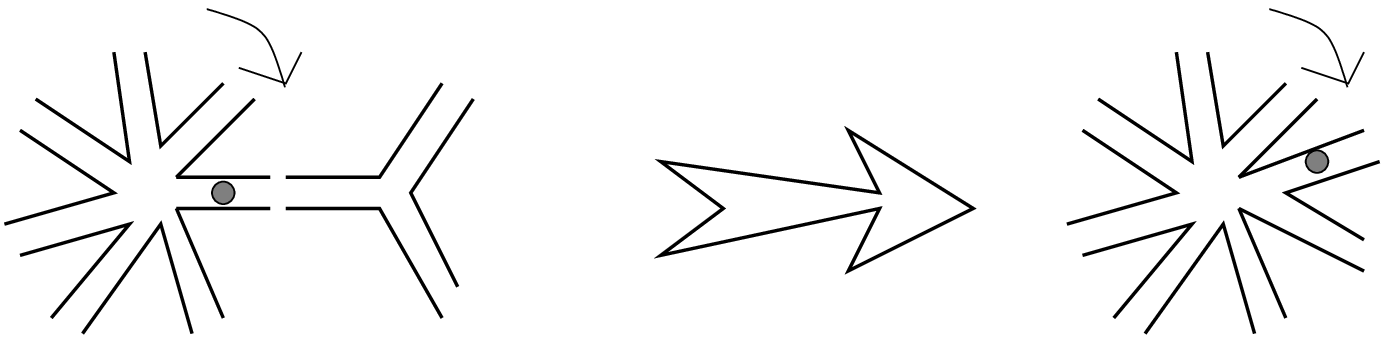}}}$\quad;\quad$
\vspace{2pt}
{\mbox{\epsfxsize=7truecm\epsfbox{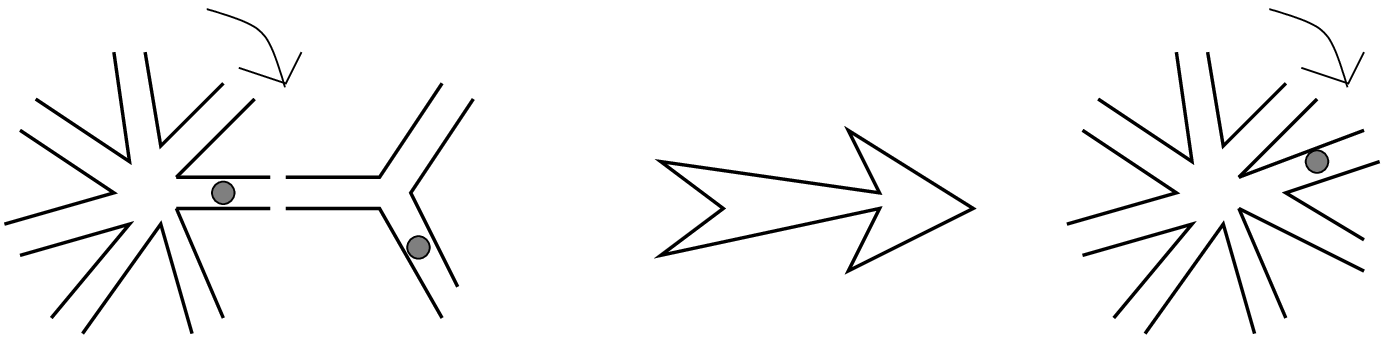}}}
\vspace{2pt}
{\mbox{\epsfxsize=9truecm\epsfbox{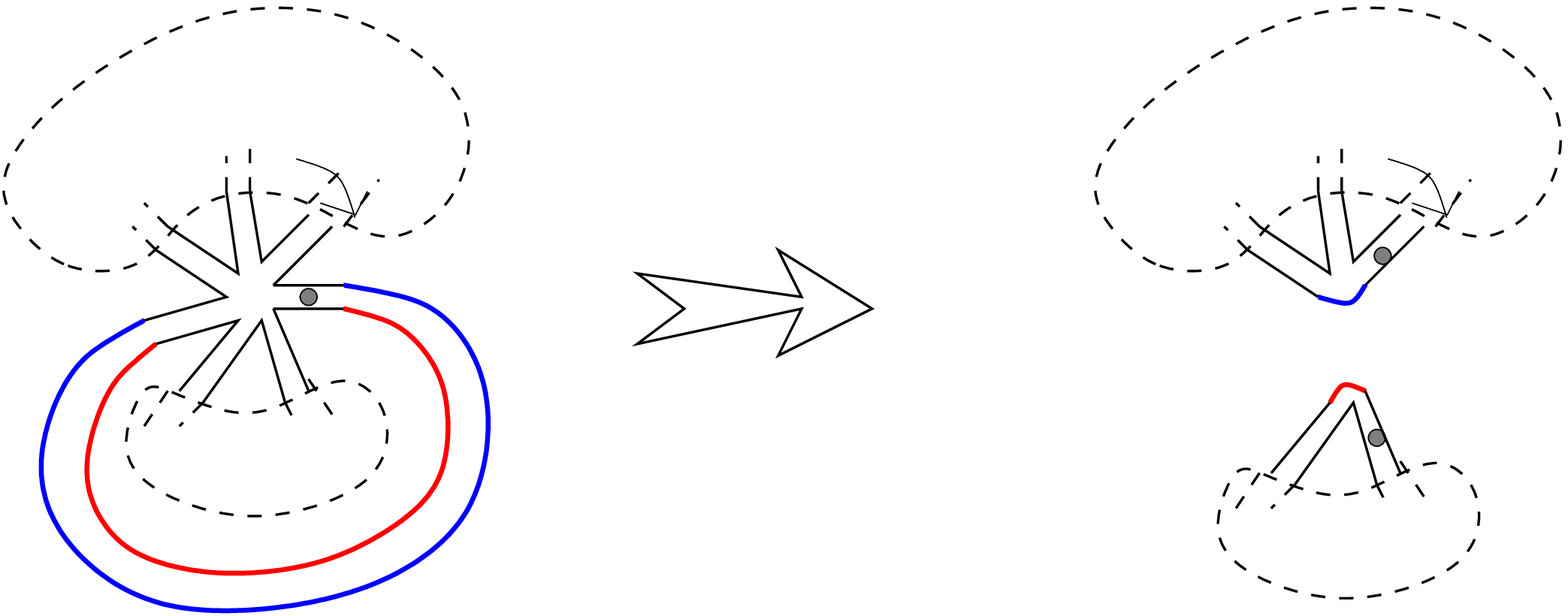}}}
\vspace{2pt}
{\mbox{\epsfxsize=9truecm\epsfbox{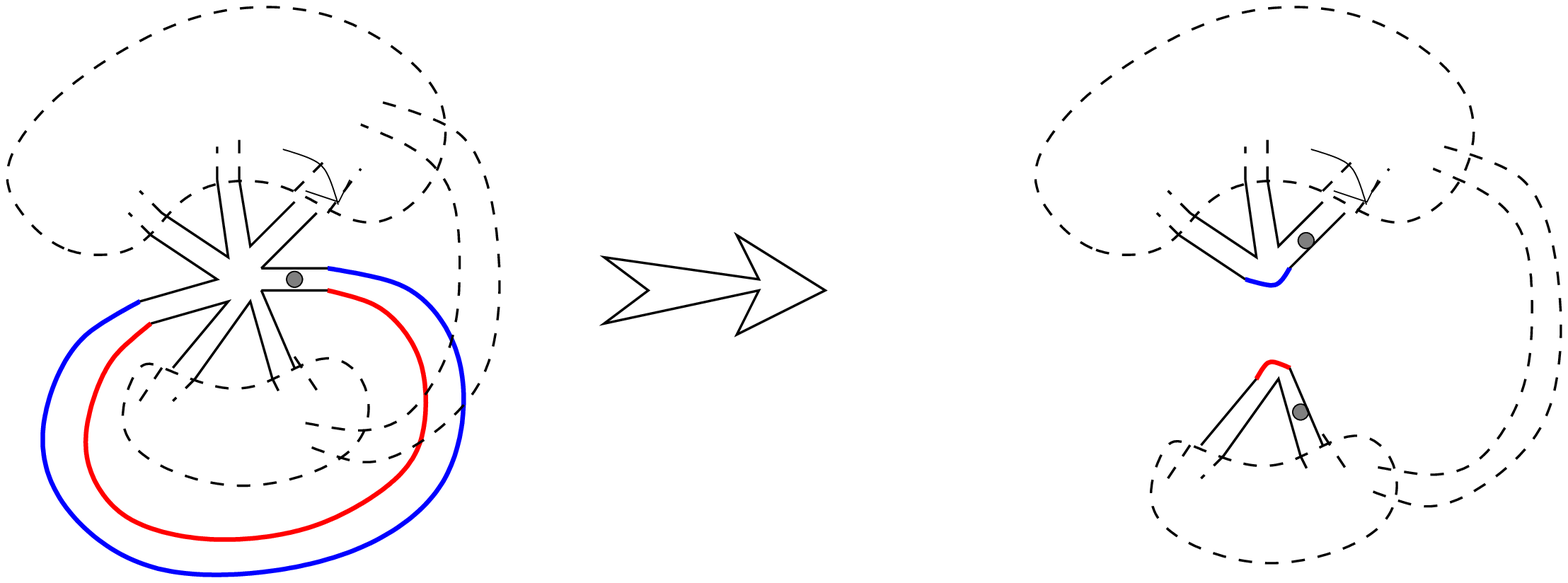}}}
\vspace{2pt}
{\mbox{\epsfxsize=9truecm\epsfbox{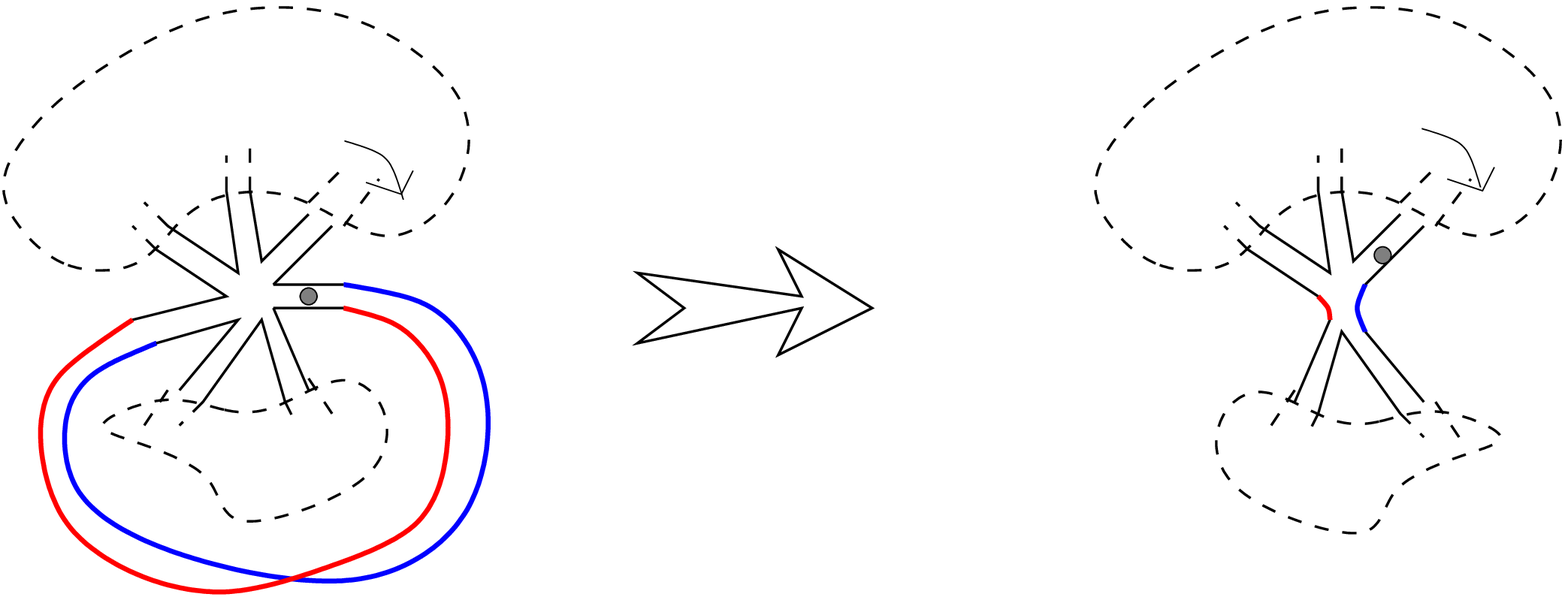}}}
\end{center}
\caption{When we shrink the single lines of a marked edge, several possibilities may occur:
1) the other side is an umarked vertex of order $j+1$, we get a new vertex of order $l_0+j$;
2) the other side is a marked vertex of order $l_i$, we get a new vertex of order $l_0+l_i-1$;
3)4) the other side is the same vertex and the edge is untwisted. Then shrinking the edge splits the vertex into two vertices, this may disconnect the graph or not;
5) the other side is the same vertex and the edge is twisted. Then shrinking the edge doesn't disconnect the vertex.
\label{figtutteseqs}}
}\kern8pt}
\kern8pt\vrule}\hrule
\end{figure}

Consider the marked edge of marked face $0$. It is either twisted or untwisted.
Several mutually exclusive situations may occur (see fig \ref{figtutteseqs}):

$\bullet$ on the other side of the marked edge, there is an unmarked vertex of size $j+1$ with $j\geq 2$.
We then shrink the marked edge to concatenate the two vertices into one marked vertex of degree $l_0+j$. The orientation is inherited from the initial marked vertex, and the marked edge is chosen as the first edge to the left of the shrinked edge. 
It is clear that we don't change the number of single lines in ${\cal G}$ or ${\cal G}'$. We decrease the number of vertices and edges by 1, so we don't change the genus.
We thus get a ribbon graph in $\mathbb \mathbb M^{(g)}_{l_0+j,L}$, and this is weighted with weight $t_{j+1}\,(1/\beta+(1-1/\beta))=t_{j+1}$.

\medskip

$\bullet$ on the other side of the marked edge, there is the marked vertex $i\neq 0$, of size $l_i\geq 1$. We then shrink the marked edge to concatenate the two vertices into one marked vertex of degree $l_0+l_i-1$. The orientation is inherited from the initial marked vertex, and the marked edge is chosen as the first edge to the left of the shrinked edge. It is also clear that we don't change the genus.
Since we forget the marking of the other face, we shall get a symmetry factor $l_i$, corresponding to the $l_i$ places where we glue to the $i^{\rm th}$ marked vertex.
We thus get a ribbon graph in $\mathbb \mathbb M^{(g)}_{l_0+l_i-1,L/\{l_i\}}$, and this is weighted with weight $l_i$.

\medskip

$\bullet$ on the other side of the marked edge, there is the same marked vertex $0$.
Again we shall shrink the marked edge, i.e. shrink the 2 single lines.
Several sub-situations may occur:

\smallskip
$*$ if the edge is untwisted, shrinking the 2 single lines splits the marked vertex of size $l_0+1$ into two vertices of size $l'$ and $l_0-l'-1$. They inherit their orientation and marked edge from the initial marked vertex. We have increased the number of marked vertices by 1.
The two new vertices are either connected together, or not.

$**$ If they are not connected, this means that the number of other marked vertices and  the genus simply add up.
We thus get two ribbon graphs in $\mathbb \mathbb M^{(g')}_{l',L'} \times \mathbb \mathbb M^{(g-g')}_{l_0-l'-1,L/L'}$, and this is weighted with weight $1/\beta$.

$**$ If they are connected, we see that we get a new ribbon graph, with one more vertex, 1 less edge, and we have not changed the connectivity of single lines. The genus has thus decreased by $1$.
We thus get a ribbon graph in $\mathbb \mathbb M^{(g-1)}_{l',l_0-l'-1,L}$, and this is weighted with weight $1/\beta$.

\smallskip
$*$ if the edge is twisted, shrinking the 2 single lines doesn't split the marked vertex.
Instead we get a new vertex of size $l_0-1$.
We assign to it the orientation of the half-vertex situated left of the marked edge, and we mark the edge left of the initial one.
We have decreased $q$ by $1$, and the genus is unchanged.
We thus get a ribbon graph in $\mathbb \mathbb M^{(g)}_{l_0-1,L}$, and this is weighted with weight $(1-1/\beta)\, l_0$ (indeed, there are $l_0$ places where we can glue the marked edge).


\bigskip

For the generating function, those bijections read:
\bea
x\,W_{k+1}^{(g)}(x,X_L) 
&=& \sum_{j=2}^{d}\, t_{j+1}\,x^{j}\,W_{k+1}^{(g)}(x,X_L) \cr
&& + {1\over \sqrt\beta}\,\sum_{i=1}^{k}\, \partial_{x_i}\,\,{W_{k}^{(g)}(x,X_{L/\{x_i\}})-W_{k}^{(g)}(x_i,X_{L/\{x_i\}})\over x-x_i} \cr
&& + {1\over \sqrt\beta}\, \sum_{g',L'\subset L} W_{1+\#L'}^{(g')}(x,X_{L'}) W_{1+k-\#L'}^{(g-g')}(x,X_{L/L'}) \cr
&& + {1\over \sqrt\beta}\,W_{k+2}^{(g-1)}(x,x,X_{L}) \cr
&& + (1-{1\over \beta})\,\partial_x\, W_{k+1}^{(g)}(x,X_L) \cr
&& + {1\over \sqrt\beta}\, P_{k+1}^{(g)}(x,X_L)
\eea
we define
\beq
V'(x) = { \sqrt\beta}\,\left(x-\sum_{j=2}^d t_{j+1} x^j\right)
\eeq
and the last term $P_{k+1}^{(g)}(x;X_L)$ accounts for all the boundary terms, and it is necessarily equal to:
\beq
P_{k+1}^{(g)}(x;X_L) = \left( V'(x)\,\,W_{k+1}^{(g)}(x;X_L)\right)_+.
\eeq
This can be rewritten:
\bea
V'(x)\,W_{k+1}^{(g)}(x,X_L) 
&=& \sum_{i=1}^{k}\, \partial_{x_i}\,\,{W_{k}^{(g)}(x,X_{L/\{x_i\}})-W_{k}^{(g)}(x_i,X_{L/\{x_i\}})\over x-x_i} \cr
&& + \sum_{g',L'} W_{1+\#L'}^{(g')}(x,X_{L'}) W_{1+k-\#L'}^{(g-g')}(x,X_{L/L'}) \cr
&& + W_{k+2}^{(g-1)}(x,x,X_{L}) \cr
&& + \hbar\,\partial_x\, W_{k+1}^{(g)}(x,X_L) \cr
&& + P_{k+1}^{(g)}(x,X_L)
\eea
where
$$
\hbar = {\sqrt\beta-1/\sqrt\beta}.
$$
In other words, the $W_n^{(g)}$'s defined in section \ref{secdefWngFg} provide a solution to Tutte's equations.
They are the generating functions counting our non-oriented ribbon graphs.
One just needs to find the polynomial $P_1^{(0)}(x)$, i.e. $U(x)$, and the choice of $\psi$ which is such that $W_1^{(0)}$ is a formal power series in $t_0$.

\section{Conclusion}
In this article, we have defined some "quantum" versions of quantities known in algebraic geometry and applied them to the resolution of the loop equations in the arbitrary $\beta$-random matrix model case, and in particular the enumeration of some non-orientable ribbon graphs.

Our formalism recovers standard algebraic geometry and the invariants of \cite{EynOM} in the classical limit $\hbar\to 0$.

Instead of an albebraic equation, we have to deal with a differential equation, which we interpreted as a "quantum spectral curve", and we were able to generalize the basic notions arising in classical algebraic geometry, like genus, sheets, branchpoints, meromorphic forms, of 1st kind, 2nd kind, 3rd kind, matrix of periods,...

It is surprising to see that the notion of branchpoints become "blurred", a branchpoint is no longer a point, but an asymptotic accumulation line. Also, there are two sheets, corresponding of the two possible large $x$ asymptotic behaviors of $\psi(x)\sim \exp{\pm V/2\hbar}$, but in fact any solution is a linear combination of these two, so that we could say that we are always in a "linear superposition" of two states like in quantum mechanics.

Another surprising thing, is that, in order for any cohomology theory to make sense, we need the cycle integrals of any forms to depend only on the homology class of the cycles, i.e. we need all forms to have vanishing residues at the $s_i$'s.
This "no-monodromy" condition is equivalent to a Bethe ansatz satisfied by the $s_i$'s, like in the Gaudin model \cite{BabBetGaudin}.
This provides a geometric interpretation of the Bethe ansatz, as the condition for cohomology to make sense.

\medskip

 However, we still lack of a complete understanding of the situation, since most of our results explicitely depend on an initial sector $S_0$ which we choose, whereas in algebraic geometry most of them only depend on the spectral curve and not on its parametrization.
For instance the genus itself depends on a choice of sector.
In some sense, the genus is no longer deterministic. 
 
\medskip

Moreover, we still lack the proper definition of the spectral invariants $F_{g}$, indeed we have defined the $F_g$'s only through solving a differential equation with respect to $\hbar$, which is not as explicit as \cite{EOFg} or \cite{ChekEynbeta}.
Out of the $F_g$'s, we could expect the possibility to make the link with integrable systems and define a "quantum Tau-function", like in \cite{EOFg}. 

\medskip
Also, we restricted ourselves to the case of hyperelliptic curves, i.e. second order differential equations, or also a 1-matrix model. 
In a forthcoming paper, we shall generalize all this construction to arbitrary linear differential equations of any order, and generalize to a 2-matrix model. This work is underway, almost finished and the article is being written at this time. As for the hyperelliptical case, the notions of genus, sheets, branchpoints, forms, $W_n^{(g)}$'s ... can be defined. Again there is a Bethe ansatz ensuring a no-monodromy condition so that all cycle integrals depend only on the homology class of cycles. So, there is no qualitative change, the difference is only technical, because the hyperelliptical case has big simplifications due to the involutive symmetry.
The difference between the hyperelliptical case and the general case is comparable to the difference between \cite{Eyn1loop} and \cite{CEO}, i.e. the definition of the kernel $K$ is really more complicated, and there is a rather "big" technical step.

\medskip
Then is would be interesting to see if the $F_g$'s have some sort of symplectic invariance, or more precisely some "canonical invariance", i.e. are unchanged under any change $(x,y)\to (\td x,\td y)$ such that $[\td y,\td x]=[y,x]=\hbar$.

\medskip

Finally, let us mention that we have developped a new geometrical approach to the study of D-modules, and it would be interesting to see how to relate it to more standard approaches, and also to the resurgence theory for studying the Schr\"odinger equation.


\section*{Acknowledgments}
We would like to thank O. Babelon, M. Berg\`ere, G. Borrot, P. Di Francesco, S. Guillermou, V. Pasquier, A. Prats-Ferrer, A. Voros
for useful and fruitful discussions on this subject.
The work of B.E. and O.M. is partly supported by the Enigma European network MRT-CT-2004-5652, ANR project GranMa "Grandes Matrices Al\'eatoires" ANR-08-BLAN-0311-01,  
by the European Science Foundation through the Misgam program,
by the Quebec government with the FQRNT. 
O. Marchal would like to thank the CRM (Centre de recheche math\'ematiques de Montr\'eal, QC, Canada) for its hospitality.


\setcounter{section}{0}

\appendix{Proof of the loop equation for $B(x,z)$}
\label{BergmannLoopEquation}

Let's first proove the first loop equation for $B(x,z)$:
Let's define:
\beq
\hat B(x,z) = {1\over 2}\,\partial_z\,(2{\psi'(z)\over \psi(z)}-\partial_z)\, \hat K(x,z)
\eeq
i.e. we have:
\beq
B(x,z) = \hat B(x,z) - \sum_{\alpha=1}^{d-1} v_\alpha(x)\, \oint_{\acycle_\alpha} \hat B(x'',z)dx''
\eeq

Since 
$(2{\psi'(x)\over \psi(x)}+\partial_x)\, v_\alpha(x) = h_\alpha(x)$
 is a polynomial of degree $\leq d-2$, it suffices to prove \eq{loopeqBx} for $\hat B(x,z)$.

Let us compute:
\bea
 (2{\psi'(x)\over \psi(x)}+\partial_x)\,\hat B(x,z)  
&=& {1\over 2}\,\partial_z\,(2{\psi'(z)\over \psi(z)}-\partial_z)\, {1\over x-z}\,  \cr
&=& {1\over 2}\,\partial_z\,(2{\psi'(z)\over \psi(z)(x-z)}-{1\over (x-z)^2})\,  \cr
&=& -{1\over (x-z)^3} + \partial_z\,{\psi'(z)\over \psi(z)(x-z)}   \cr
\eea
and therefore:
\beq
(2{\psi'(x)\over \psi(x)}+\partial_x)\,\left(\hat B(x,z)-{1\over 2(x-z)^2}\right) + \partial_z\,{{\psi'(x)\over \psi(x)}-{\psi'(z)\over \psi(z)}\over x-z} 
= 0
\eeq
This proves \eq{loopeqBx}, with:
\beq
 P_2^{(0)}(x,z) = - \sum_{\alpha=1}^g h_\alpha(x)\, \oint_{\acycle_\alpha} \hat B(x'',z)dx'' -\sum_{\alpha=g+1}^{d-1}\, h_\alpha(x)\, \oint_{\acycle_\alpha} dx''\,\partial_{x''}\,\psi^2(x'') \hat B(x'',z).
\eeq

\bigskip

Let's now proove the second loop equation for $B(x,z)$:
Similarly, let us compute $ (2{\psi'(z)\over \psi(z)}+\partial_z)\,\hat B(x,z)$:
\beq
(2{\psi'(z)\over \psi(z)}+\partial_z)\,\hat B(x,z) = {1\over 2} (2{\psi'(z)\over \psi(z)}+\partial_z)\,\partial_z\,(2{\psi'(z)\over \psi(z)}-\partial_z)\,\hat K(x,z)
\eeq

Notice that the operator $\hat U(z) = {1\over 2} (2{\psi'(z)\over \psi(z)}+\partial_z)\,\partial_z\,(2{\psi'(z)\over \psi(z)}-\partial_z)$, is equal to:
\beq
\hat U(z) = -{1\over 2}\,\partial_z^3 + 2 U(z) \partial_z + U'(z)
\eeq
which is also known in the litterature as the Gelfand-Dikii operator \cite{ZJDFG} (The Gelfand-Dikii differential polynomials $R_k(U)$ are computed recursively by $R_0=1$ and $\partial_z R_{k+1} = \hat U\,.R_k$), which plays a key role in the KdV hierarchy.

However, independently of any relationship with KdV, we get:
\bea
&& (2{\psi'(z)\over \psi(z)}+\partial_z)\,\hat B(x,z) \cr
&=& {1\over \psi^2(x)}\,\int_{\infty_0}^x\, \psi^2(x')\,dx'\,\,\hat U(x'). {1\over x'-z} \cr
&=& {1\over \psi^2(x)}\,\int_{\infty_0}^x\, \psi^2(x')\,dx'\,\,
\Big(-{3\over (x'-z)^4} + {2U(z)\over (x'-z)^2} + {U'(z)\over x'-z} \Big) \cr
\eea
We integrate the first term by parts three times, and we write $Y=\psi'/\psi$ (we have $Y'+Y^2=U$):
\bea
&& (2{\psi'(z)\over \psi(z)}+\partial_z)\,\hat B(x,z) \cr
&=& {1\over (x-z)^3} - {2\over \psi^2(x)}\,\int_{\infty_0}^x\, \psi^2(x')\,dx'\,\,{Y(x')\over (x'-z)^3}  \cr
&& + {1\over \psi^2(x)}\,\int_{\infty_0}^x\, \psi^2(x')\,dx'\,\,
\Big( {2U(z)\over (x'-z)^2} + {U'(z)\over x'-z} \Big) \cr
&=& {1\over (x-z)^3} + {Y(x)\over (x-z)^2} 
 - {1\over \psi^2(x)}\,\int_{\infty_0}^x\, \psi^2(x')\,dx'\,\,{Y'(x')+2Y^2(x')\over (x'-z)^2} \cr
&& + {1\over \psi^2(x)}\,\int_{\infty_0}^x\, \psi^2(x')\,dx'\,\,
\Big( {2U(z)\over (x'-z)^2} + {U'(z)\over x'-z} \Big) \cr
&=& {1\over (x-z)^3} + {Y(x)\over (x-z)^2} 
 + {1\over \psi^2(x)}\,\int_{\infty_0}^x\, \psi^2(x')\,dx'\,\,{Y'(x')\over (x'-z)^2} \cr
&& + {1\over \psi^2(x)}\,\int_{\infty_0}^x\, \psi^2(x')\,dx'\,\,
\Big( {2(U(z)-U(x'))\over (x'-z)^2} + {U'(z)\over x'-z} \Big) \cr
&=& {1\over (x-z)^3} + {Y(x)\over (x-z)^2} 
-{Y'(x)\over x-z} + {1\over \psi^2(x)}\,\int_{\infty_0}^x\, \psi^2(x')\,dx'\,\,{Y''(x')+2Y(x')Y'(x')\over (x'-z)^2} \cr
&& + {1\over \psi^2(x)}\,\int_{\infty_0}^x\, \psi^2(x')\,dx'\,\,
\Big( {2(U(z)-U(x'))\over (x'-z)^2} + {U'(z)\over x'-z} \Big) \cr
&=& {1\over (x-z)^3} - {\partial \over \partial x}\, {Y(x)\over x-z}  
 + {1\over \psi^2(x)}\,\int_{\infty_0}^x\, \psi^2(x')\,dx'\,\,
\Big( {2(U(z)-U(x'))\over (x'-z)^2} + {U'(z)+U(x')\over x'-z} \Big) \cr
\eea
This implies that:
\bea
&& (2{\psi'(z)\over \psi(z)}+\partial_z)\,\Big(\hat B(x,z)-{1\over 2(x-z)^2}\Big)
+ {\partial \over \partial x}\, {Y(x)-Y(z)\over x-z}  \cr
&=& {1\over \psi^2(x)}\,\int_{\infty_0}^x\, \psi^2(x')\,dx'\,\,
\Big( {2(U(z)-U(x'))\over (x'-z)^2} + {U'(z)+U(x')\over x'-z} \Big) \cr
&=& Q(z,x)
\eea
which is clearly a polynomial in $z$.
Taking integrals over $x$ along $\acycle_\alpha$ does not change its structure in $z$, and therefore:
\bea
&& (2{\psi'(z)\over \psi(z)}+\partial_z)\,\Big( B(x,z)-{1\over 2(x-z)^2}\Big)
+ {\partial \over \partial x}\, {Y(x)-Y(z)\over x-z}  \cr
&=& {1\over \psi^2(x)}\,\int_{\infty_0}^x\, \psi^2(x')\,dx'\,\,
\Big( {2(U(z)-U(x'))\over (x'-z)^2} + {U'(z)+U(x')\over x'-z} \Big) \cr
&=& \td{P}_2^{(0)}(z,x)
\eea
is of the required form.

By looking at the behavior of the various terms in the LHS of \eq{loopeqBz} when $z\to\infty$, we find that $\td{P}_2^{(0)}(z,x)$ is a polynomial of degree at most $d-2$ in $z$.

\appendix{Proof of theorem \ref{thpolessiWng}}
\label{approofthpolessiWng}

{\bf Theorem \ref{thpolessiWng}}
{\it Each  $W_n^{(g)}(x_1,\dots,x_n)$ with $2-2g-n<0$, is an analytical functions of all its arguments, with poles only when $x_i\to s_{j}$.
Moreover, it vanishes at least as $O\left(1/{x_i^2}\right)$ when $x_i\to\infty$ in all sectors.
It has no discontinuity across $\acycle$-cycles.}
\bigskip

{\bf proof:}

We proceed by recursion on $2g+n$.
The theorem is true for $W_2^{(0)}$.
Assume it is true up to $2g+n$, we shall prove it for $W_{n+1}^{(g)}(x_0,x_1,\dots,x_n)$.

\medskip

The integrand $U_n^{(g)}$ of theorem \ref{thWngdefintU} is singular only at $x=s_j$'s.
As long as $x_0$ is away from the $s_j$'s, we can continuously deform the $\acycle$-cycles and the contour ${\cal C}$ in order to have $x_0$ outside of the $\acycle$-cycles, and the integral can be evaluated and is analytical in $x_0$.
When $x_0$ approaches $s_i$,  we define $\hat{\cal C}_i$, a contour which surrounds all roots except $s_{i}$, i.e:
\beq
\oint_{{\cal C}} = \oint_{\hat{\cal C}_i} + 2i\pi\,\Res_{s_{i}}
\eeq
The integral over  $\hat{\cal C}_i$ can be evaluated and is convergent, thus it is analytical in $x_0$.

From the recursion hypothesis, all terms in the integrand are meromorphic in the vicinity of $s_{i}$, and thus the residue at $s_{i}$ can be computed by taking a finite Taylor expansion of $K(x_0,x)=\sum_k (x-s_i)^k\,K_{i,k}(x_0)$ in the vicinity of $x\to s_{i}$. The result is a finite sum of terms of the type $K_{i,k}(x_0)$. It is easy to see from the definition of $K$, that each $K_{i,k}(x_0)$ has only poles at $x_0=s_{i}$. Thus we have proved that $W_{n+1}^{(g)}$ has poles at the $s_i$'s in its first variable.

\smallskip

\medskip

In the other variables, the result comes from an obvious recursion.

$\square$

\appendix{Proof of theorem \ref{thWngPng}}
\label{approofthWngPng}

In this subsection we prove theorem \ref{thWngPng}, that all $W_n^{(g)}$'s satisfy
the loop equation.
\bea\label{loopeqPng}
 P_{n+1}^{(g)}(x,x_1...,x_n)
 &=&
2\hbar\frac{\psi'(x)}{\psi(x)}\overline{W}_{n+1}^{(g)}(x,x_1...,x_n) + \hbar \partial_{x}{\overline{W}_{n+1}^{(g)}(x,x_1,...,x_n}) \cr
&& + \sum_{I\subset J} \ovl{W}_{|I|+1}^{(h)}(x,x_I) \ovl{W}_{n-|I|+1}^{(g-h)}(x,J/I) +
\ovl{W}_{n+2}^{(g-1)}(x,x,J)  \cr
& &+ \sum_{j}
\partial_{x_j} \left( {{\ovl{W}_n^{(g)}(x,J/\{j\})-{\ovl{W}_n^{(g)}(x_j,J/\{j\})}} \over {(x-x_j)}}\right) \cr
\eea
is a polynomial in $x$ of degree at most $d-2$.

{\bf proof:}

From the definition we have:
\bea
W_{n+1}^{(g)}(x,J)&=&   {1\over 2i\pi} \oint_{{\cal C}}\, dz\,\,  K(x,z)\, U_n^{(g)}(z,J)\cr
&=&{1\over 2i\pi} \oint_{{\cal C}}\, dz\,\,  \hat{K}(x,z)\, U_n^{(g)}(z,J)\cr
&&-\sum_\alpha\, {v_\alpha(x)\over 2i\pi} \oint_{{\cal C}}\, dz\,\,  C_{\alpha}(z)\, U_n^{(g)}(z,J)
\eea
Then, notice that $\hat K(x,z)$ has a logarithmic cut along $]\infty_0,x]$, and the discontinuity across that cut is:
\beq
\delta \hat K(x,z) = {2i\pi\over \hbar}\, {\psi^2(z)\over \psi^2(x)}
\eeq
$U_n^{(g)}$ has no singularity outside of ${\cal C}$, and thus we can deform the contour into a contour enclosing only the logarithmic cut of $\hat K(x,z)$, and therefore:
\beq
{1\over 2i\pi} \oint_{{\cal C}}\, dz\,\,  \hat{K}(x,z)\, U_n^{(g)}(z,J) = -\,{1\over \hbar}\, \int_{\infty_0}^xdz\,\, {\psi^2(z)\over \psi^2(x)}\,U_n^{(g)}(z,J)
\eeq
We then apply the operator: $2\frac{\psi'(x)}{\psi(x)}+\partial_x$, that gives:
\beq
(2\hbar {\psi'(x)\over \psi(x)} + \hbar \partial_x)\,\,{1\over 2i\pi} \oint_{{\cal C}}\, dz\,\,  \hat{K}(x,z)\, U_n^{(g)}(z,J) = - U_n^{(g)}(x,J)
\eeq
and therefore:
\bea
P_{n+1}^{(g)}(x,J)
&=& (2\hbar {\psi'(x)\over \psi(x)} + \hbar \partial_x)\,W_{n+1}^{(g)}(x,J)+U_n^{(g)}(x,J) \cr
&=& -\,(2\hbar {\psi'(x)\over \psi(x)} + \hbar \partial_x)\,\sum_\alpha\, {v_\alpha(x)\over 2i\pi} \oint_{{\cal C}}\, dz\,\,  C_{\alpha}(z)\, U_n^{(g)}(z,J) \cr
&=& -\,\sum_\alpha\, h_\alpha(x) \oint_{{\cal C}}\, dz\,\,  C_{\alpha}(z)\, U_n^{(g)}(z,J)
\eea
which is indeed a polynomial of $x$ of degree at most $d-2$.

$\square$

\appendix{Proof of theorem \ref{thW3Krich}}
\label{approofthW3Krich}

{\bf Theorem \ref{thW3Krich}}
{\it 
The 3 point function $W_3^{(0)}$ is symmetric and we have:
\beq \label{appthW3Krich}
W_3^{(0)}(x_1,x_2,x_3) = {4\over 2i\pi} \oint_{{\cal C}}\, dx\,\, {B(x,x_1)B(x,x_2)B(x,x_3)\over Y'(x)}
\eeq
where $Y(x)=-2\hbar\frac{\psi'(x)}{\psi(x)}$}
\bigskip


{\bf proof:}

The definition of $W_3^{(0)}$ is:
\bea  
&& W_3^{(0)}(x_0,x_1,x_2)\cr 
&=&  {1\over i\pi}\oint_{{\cal C}}\, dx\,\, K(x_0,x)B(x,x_1)B(x,x_2) \cr
 &=&  {1\over 4i\pi} \oint_{{\cal C}}\, dx\,\,  K_0 \, G_1^{'} \, G'_2 \cr
 &=&  {1\over 4i\pi}\oint_{{\cal C}}\, dx\,\, K_0 \left( (\hbar K''_1 +  Y K'_1 +  Y'
 K_1)(\hbar K''_2 + YK'_2 +Y' K_2) \right) \cr
 &=&  {1\over 4i\pi} \oint_{{\cal C}}\, dx\,\, K_0 \, (\, \hbar^2 K''_1 K''_2 +  \hbar Y (K'_1
 K''_2+K''_1 K'_2) +  \hbar Y' (K''_1 K_2 +K''_2 K_1) \cr
 && +  Y^2 K'_1 K'_2+  Y Y' (K_1 K'_2+K'_1 K_2)+
{Y'}^2 K_1 K_2 \,) \cr
 \eea
where we have written for short $K_p = K(x_p,x)$, $G_p=G(x_p,x)$, and derivative are w.r.t. $x$. Note also that introducing $K_1$ and $K_2$ makes appear some additional and arbitrary logarithmic cuts from $x_1$ to $\infty_{0}$ and from $x_2$ to $\infty_{0}$. But these cuts can be chosen arbitrarily since from the definition of $W_3^{(0)}(x_0,x_1,x_2)$ it should not depend on that. Remember also that to use this definition of $W_3^{(0)}$ we need to assume that $x_1$ and $x_2$ are not circled by the contour $\mathcal{C}$. Therefore we can choose the logarithmic cut of $K_1$ and $K_2$ inside the contour $\mathcal{C}$ like we have done it for $x_0$. We now see that for example $K_0 K_1 K_2$ has no singularity outside $\mathcal{C}$ and thus will not contribute because of theorem \ref{NullityOfIntegrals}. Many other manipulations involving globally defined functions with no singularities outside $\mathcal{C}$ can be done.

For example, using the Ricatti equation $Y_i^2 = 2 \hbar Y_i' + 4U$, we may replace $Y_i^2$ by $2 \hbar Y_i'$ and $ Y_i Y_i'$ by $\hbar Y_i''$.
\bea
& & W_3^{(0)}(x_0,x_1,x_2)\cr
 &=&  {1\over 4i\pi}\oint_{{\cal C}}\, dx\,\, K_0 \, (  \hbar Y (K'_1
 K''_2+K''_1 K'_2) +  \hbar Y' (K''_1 K_2 +K''_2 K_1) \cr
 && + 2 \hbar Y' K'_1 K'_2+ \hbar Y'' (K_1 K'_2+K'_1 K_2)+
{Y'}^2 K_1 K_2 \,)\ \cr
 &=&  {1\over 4i\pi}\oint_{{\cal C}}\, dx\,\, K_0 \, (  \hbar Y (K'_1 K'_2)' +  \hbar Y' (K_1 K_2)'' + \hbar Y'' (K_1 K_2)'+  {Y'}^2 K_1 K_2 \,)\ \cr
 &=&   {1\over 4i\pi}\oint_{{\cal C}}\, dx\,\, {Y'}^2 K_0 K_1 K_2 +
 \hbar \big( Y'' K_0 (K_1 K_2)' - (Y K_0)' K'_1 K'_2 - (Y' K_0)' (K_1 K_2)' \big) \cr
 &=&  {1\over 2} {1\over 4i\pi}\oint_{{\cal C}}\, dx\,\, {Y'}^2 K_0 K_1 K_2 -
 \hbar \big(  (Y K_0)' K'_1 K'_2 + Y' K_0' (K_1 K_2)' \big) \cr
&=&  {1\over 4i\pi}\oint_{{\cal C}}\, dx\,\, {Y'}^2 K_0 K_1 K_2 - \hbar Y K'_0 K'_1 K'_2 - \hbar Y' (K_0 K'_1 K'_2+K'_0 K_1 K'_2+K'_0 K'_1 K_2) \cr
\eea
This expression is clearly symmetric in $x_0, x_1, x_2$ as claimed in theorem \ref{thsym}.

Let us give an alternative expression, in the form of the Verlinde or Krichever formula.
\beq\label{eq30Krichever}
W_3^{(0)}(x_0,x_1,x_2)=
{2\over i\pi}\oint_{{\cal C}}\, dx\,\, {B(x,x_1)B(x,x_2)B(x,x_3) \over Y^{'}(x)}
\eeq

\proof{
In order to prove formula \ref{eq30Krichever}, compute:
\beq
B(x,x_i) = -{1\over 2} G'(x,x_i) = -{1\over 2} G'_i = {1\over 2}(\hbar K''_i +  Y K'_i +  Y' K_i)
\eeq
thus:
\bea
&&  {1\over 2i\pi}\oint_{{\cal C}}\, dx\,\, {B(x,x_1)B(x,x_2)B(x,x_3) \over Y^{'}(x)} \cr
&=&  {1\over 16i\pi}\oint_{{\cal C}}\, dx\,\, {1\over Y'(x)}\, (\hbar K''_0 +  Y K'_0 +  Y' K_0)(\hbar K''_1 +  Y K'_1 +  Y' K_1)(\hbar K''_2 +  Y K'_2 +  Y' K_2)  \cr
&=&  {1\over 16i\pi}\oint_{{\cal C}}\, dx\,\,
 {\hbar^3\over Y'} K''_0 K''_1 K''_2 + \hbar^2 {Y\over Y'} (K'_0 K''_1 K''_2 + K''_0 K'_1 K''_2 + K''_0 K''_1 K'_2) \cr
&& + \hbar^2  (K_0 K''_1 K''_2 + K''_0 K_1 K''_2 + K''_0 K''_1 K_2)
+ \hbar {Y^2\over Y'} (K''_0 K'_1 K'_2+K'_0 K''_1 K'_2+K'_0 K'_1 K''_2) \cr
&& + \hbar Y (K_0 K'_1 K''_2 + K_0 K''_1 K'_2 + K'_0 K_1 K''_2 + K'_0 K''_1 K_2 + K''_0 K_1 K'_2 + K''_0 K'_1 K_2) \cr
&& + \hbar Y' (K''_0 K_1 K_2 + K_0 K''_1 K_2 + K_0 K_1 K''_2)
+ {Y^3\over Y'} K'_0 K'_1 K'_2  \cr
&& + Y^2  (K_0 K'_1 K'_2 + K'_0 K_1 K'_2 + K'_0 K'_1 K_2)
+ Y Y' (K'_0 K_1 K_2 + K_0 K'_1 K_2 + K_0 K_1 K'_2) \cr
&& + Y'^2 K_0 K_1 K_2  \cr
\eea

\bea
&&  {1\over 2i\pi}\oint_{{\cal C}}\, dx\,\, {B(x,x_1)B(x,x_2)B(x,x_3) \over Y^{'}(x)} \cr
&=&  {1\over 16i\pi}\oint_{{\cal C}}\, dx\,\,
 \hbar Y (K_0 K'_1 K''_2 + K_0 K''_1 K'_2 + K'_0 K_1 K''_2 + K'_0 K''_1 K_2 + K''_0 K_1 K'_2 + K''_0 K'_1 K_2) \cr
&& + \hbar Y' (K''_0 K_1 K_2 + K_0 K''_1 K_2 + K_0 K_1 K''_2)
+ {Y^3\over Y'} K'_0 K'_1 K'_2  \cr
&& + Y^2  (K_0 K'_1 K'_2 + K'_0 K_1 K'_2 + K'_0 K'_1 K_2)
+ Y Y' (K'_0 K_1 K_2 + K_0 K'_1 K_2 + K_0 K_1 K'_2) \cr
&& + Y'^2 K_0 K_1 K_2  \cr
\eea
Notice that $Y^2 = 2\hbar Y' + 4U$, thus we may replace $Y^3/Y'$ by $2\hbar Y$, and $Y^2$ by $2\hbar Y'$ and $Y Y'$ by $\hbar Y''$, for the same reasons as before. Thus:
\bea
&&  {1\over 2i\pi}\oint_{{\cal C}}\, dx\,\, {B(x,x_1)B(x,x_2)B(x,x_3) \over Y^{'}(x)} \cr
&=&  {1\over 16i\pi}\oint_{{\cal C}}\, dx\,\,
 \hbar Y (K_0 K'_1 K''_2 + K_0 K''_1 K'_2 + K'_0 K_1 K''_2 + K'_0 K''_1 K_2 + K''_0 K_1 K'_2 + K''_0 K'_1 K_2) \cr
&& + \hbar Y' (K''_0 K_1 K_2 + K_0 K''_1 K_2 + K_0 K_1 K''_2)
+ 2\hbar Y K'_0 K'_1 K'_2  \cr
&& + 2 \hbar Y'  (K_0 K'_1 K'_2 + K'_0 K_1 K'_2 + K'_0 K'_1 K_2)
+ \hbar Y'' (K'_0 K_1 K_2 + K_0 K'_1 K_2 + K_0 K_1 K'_2) \cr
&& + Y'^2 K_0 K_1 K_2  \cr
&=&  {1\over 16i\pi}\oint_{{\cal C}}\, dx\,\,
 \hbar Y (K_0 (K'_1 K'_2)' +  K_1 (K'_0 K'_2)' + K_2 (K'_0 K'_1)') \cr
&& + 2\hbar Y K'_0 K'_1 K'_2  + Y'^2 K_0 K_1 K_2  + \hbar (Y' (K'_0 K_1 K_2 + K_0 K'_1 K_2 + K_0 K_1 K'_2))' \cr
&=&  {1\over 16i\pi}\oint_{{\cal C}}\, dx\,\,
 \hbar Y (K_0 (K'_1 K'_2)' +  K_1 (K'_0 K'_2)' + K_2 (K'_0 K'_1)') \cr
&& + 2\hbar Y K'_0 K'_1 K'_2  + Y'^2 K_0 K_1 K_2  \cr
&=&  -{1\over 16i\pi}\oint_{{\cal C}}\, dx\,\, 3 \hbar Y K'_0 K'_1 K'_2 + \hbar Y' (K_0 K'_1 K'_2 + K'_0 K_1 K'_2 + K'_0 K'_1 K_2) \cr
&& - 2\hbar Y K'_0 K'_1 K'_2  - Y'^2 K_0 K_1 K_2  \cr
&=& {1\over 8i\pi}\oint_{{\cal C}}\, dx\,\, W_3^{(0)}(x_0,x_1,x_2)
\eea
}

\appendix{Proof of theorem \ref{thsym}}
\label{approofthsym}

{\bf Theorem \ref{thsym}}
{\it 
Each $W_n^{(g)}$ is a symmetric function of all its arguments.
}
\bigskip

{\bf proof:}

The special case of $W_3^{(0)}$ is proved in appendix \ref{appthW3Krich} above.

It is obvious from the definition that $W_{n+1}^{(g)}(x_0,x_1,\dots,x_n)$ is
symmetric in $x_1,x_2,\dots,x_n$, and therefore we need to show that (for $n\geq 1$):
\beq
W_{n+1}^{(g)}(x_0,x_1,J)-W_{n+1}^{(g)}(x_1,x_0,J)=0
\eeq
where $J=\{ x_2,\dots,x_n\}$.
We prove it by recursion on $-\chi=2g-2+n$. 

Assume that every $W_k^{(h)}$ with $2h+k-2\leq 2g+n$ is symmetric.
We have:
\bea
&& W_{n+1}^{(g)}(x_0,x_1,J) \cr
&=&\frac{1}{2\pi i} \oint_{\mathcal{C}} \,dx \,\, K(x_0,x)\,\, \Big(
W_{n+2}^{(g-1)}(x,x,x_1,J) + 2 \,\,\, B(x,x_1) W_{n}^{(g)}(x,J) \cr
&& + 2 \sum_{h=0}^g\sum'_{I\in J}\,\,\, W_{2+|I|}^{(h)}(x,x_1,I) W_{n-|I|}^{(g-h)}(x,J/I) \Big) \cr
\eea
where $\sum'$ means that we exclude the terms $(I=\emptyset, h=0)$ and $(I=J, h=g)$. Notice also that $\ovl{W}_{n+2}^{(g-1)}=W_{n+2}^{(g-1)}$ because $n\geq 1$.
Then, using the recursion hypothesis, we have:
\bea
&& W_{n+1}^{(g)}(x_0,x_1,J) \cr
&=& 2 \oint_{\mathcal{C}} \,dx \,\, K(x_0,x)\,\,  B(x,x_1) W_{n}^{(g)}(x,J) \cr
&& + \oint_{\mathcal{C}} \,dx \,\, \oint_{\mathcal{C}} \,dx' \,\, K(x_0,x) K(x_1,x')\,\, 
\Big(  W_{n+3}^{(g-2)}(x,x,x',x',J) \cr
&& + 2\sum_h\sum'_{I} W_{2+|I|}^{(h)}(x',x,I) W_{1+n-|I|}^{(g-1-h)}(x',x,J/I) \cr
&& + 2 \sum_h\sum'_{I} W_{3+|I|}^{(h)}(x',x,x,I) W_{n-|I|}^{(g-1-h)}(x',J/I) \cr
&& + 2 \sum_{h}\sum'_{I\in J}\,\,\, W_{n-|I|}^{(g-h)}(x,J/I)
\Big[ W_{3+|I|}^{(h-1)}(x,x',x',I)  \cr
&& + 2 \sum_{h'}\sum'_{I'\subset I} W_{2+|I'|}^{(h')}(x',x,I')   W_{1+|I|-|I'|}^{(h-h')}(x',I/I')  
\Big]\,
\Big) \cr
\eea
Now, if we compute $W_{n+1}^{(g)}(x_1,x_0,J)$, we get the same expression, with the order of integrations exchanged, i.e. we have to integrate $x'$ before integrating $x$.
Notice, by moving the integration contours,  that:
\beq
\oint_{\mathcal{C}} \,dx \,\, \oint_{\mathcal{C}} \,dx' - \oint_{\mathcal{C}} \,dx' \,\, \oint_{\mathcal{C}} \,dx =
- \oint_{\mathcal{C}} \,dx \frac{1}{2\pi i}\Res_{x'\to x} 
\eeq
Moreover, the only terms which have a pole at $x=x'$ are those containing $B(x,x')$.
Therefore:
\bea
&& W_{n+1}^{(g)}(x_0,x_1,J)-W_{n+1}^{(g)}(x_1,x_0,J) \cr
&=& 2 \oint_{\mathcal{C}} \,dx \,\, \left( K(x_0,x)\,\,  B(x,x_1)  - K(x_1,x)\,\,  B(x,x_0) \right) \, W_{n}^{(g)}(x,J) \cr
&& - 2 \oint_{\mathcal{C}} \,dx \,\, \frac{1}{2i\pi}\Res_{x'\to x}\,\, K(x_0,x) K(x_1,x')\,\, B(x,x')\,\,
\Big(     \cr
&& 2W_{1+n}^{(g-1)}(x',x,J)  + 2 \sum_{h}\sum'_{I\in J}\,\, W_{n-|I|}^{(g-h)}(x,J/I)     W_{1+|I|}^{(h)}(x',I)  
\Big) \cr
\eea
The residue $\Res_{x'\to x}$ can be computed:
\bea
&& W_{n+1}^{(g)}(x_0,x_1,J)-W_{n+1}^{(g)}(x_1,x_0,J) \cr
&=& 2  \oint_{\mathcal{C}} \,dx \,\, \left( K(x_0,x)\,\,  B(x,x_1)  - K(x_1,x)\,\,  B(x,x_0) \right) \, W_{n}^{(g)}(x,J) \cr
&& -  \oint_{\mathcal{C}} \,dx \,\, \,\, K(x_0,x) {\partial \over \partial x'}\Big( K(x_1,x')\,\, 
\Big(     \cr
&& 2W_{1+n}^{(g-1)}(x',x,J)  + 2 \sum_{h}\sum'_{I\in J}\,\, W_{n-|I|}^{(g-h)}(x,J/I)     W_{1+|I|}^{(h)}(x',I)  
\Big)\,\, \Big)_{x'=x} \cr
&=& 2 \oint_{\mathcal{C}} \,dx \,\, \left( K(x_0,x)\,\,  B(x,x_1)  - K(x_1,x)\,\,  B(x,x_0) \right) \, W_{n}^{(g)}(x,J) \cr
&& -   \oint_{\mathcal{C}} \,dx \,\, K(x_0,x) K'(x_1,x)\,\, 
\Big(     \cr
&& 2W_{1+n}^{(g-1)}(x,x,J)  + 2 \sum_{h}\sum'_{I\in J}\,\, W_{n-|I|}^{(g-h)}(x,J/I)     W_{1+|I|}^{(h)}(x,I)  
\Big)\,\,  \cr
&& -  \, \oint_{\mathcal{C}} \,dx \,\, K(x_0,x) K(x_1,x'){\partial \over \partial x'}\Big(      \cr
&& 2W_{1+n}^{(g-1)}(x',x,J)  + 2 \sum_{h}\sum'_{I\in J}\,\, W_{n-|I|}^{(g-h)}(x,J/I)     W_{1+|I|}^{(h)}(x',I)  
\,\, \Big)_{x'=x} \cr
&=& 2  \oint_{\mathcal{C}} \,dx \,\, \left( K(x_0,x)\,\,  B(x,x_1)  - K(x_1,x)\,\,  B(x,x_0) \right) \, W_{n}^{(g)}(x,J) \cr
&& -  \oint_{\mathcal{C}} \,dx \,\, K(x_0,x) K'(x_1,x)\,\, 
\Big(     \cr
&& 2W_{1+n}^{(g-1)}(x,x,J)  + 2 \sum_{h}\sum'_{I\in J}\,\, W_{n-|I|}^{(g-h)}(x,J/I)     W_{1+|I|}^{(h)}(x,I)  
\Big)\,\,  \cr
&& -  {1\over 2}  \oint_{\mathcal{C}} \,dx \,\, K(x_0,x) K(x_1,x) {\partial \over \partial x}\Big(      \cr
&& 2W_{1+n}^{(g-1)}(x,x,J)  + 2 \sum_{h}\sum'_{I\in J}\,\, W_{n-|I|}^{(g-h)}(x,J/I)     W_{1+|I|}^{(h)}(x,I)  
\,\, \Big) \cr
\eea
The last term can be integrated by parts, and we get:
\bea
&& W_{n+1}^{(g)}(x_0,x_1,J)-W_{n+1}^{(g)}(x_1,x_0,J) \cr
&=& 2 \oint_{\mathcal{C}} \,dx \,\, \left( K(x_0,x)\,\,  B(x,x_1)  - K(x_1,x)\,\,  B(x,x_0) \right) \, W_{n}^{(g)}(x,J) \cr
&& +{1\over 2}   \oint_{\mathcal{C}} \,dx \,\, \Big( K'(x_0,x) K(x_1,x)-K(x_0,x) K'(x_1,x)\Big)\,\, 
\Big(     \cr
&& 2W_{1+n}^{(g-1)}(x,x,J)  + 2 \sum_{h}\sum'_{I\in J}\,\, W_{n-|I|}^{(g-h)}(x,J/I)     W_{1+|I|}^{(h)}(x,I)  
\Big)\,\,  \cr
\eea
Then we use theorem \ref{thWngPng}:
\bea
&& W_{n+1}^{(g)}(x_0,x_1,J)-W_{n+1}^{(g)}(x_1,x_0,J) \cr
&=& 2  \oint_{\mathcal{C}} \,dx \,\, \left( K(x_0,x)\,\,  B(x,x_1)  - K(x_1,x)\,\,  B(x,x_0) \right) \, W_{n}^{(g)}(x,J) \cr
&& + \oint_{\mathcal{C}} \,dx \,\, \Big( K'(x_0,x) K(x_1,x)-K(x_0,x) K'(x_1,x)\Big)\,\, 
\Big(      P_{n}^{(g)}(x,J) \cr
&& + (Y(x) - \hbar \partial_x) W_{n}^{(g)}(x,J)  + \sum_j \partial_{x_j}
\Big( {W_{n-1}^{(g)}(x_j,J/\{x_j\})\over x-x_j} \Big)
\Big)\,\,  \cr
\eea
Since $P_{n}^{(g)}(x,J)$ and $W_{n-1}^{(g)}(x_j,J/\{x_j\})$ are entire functions of $x$, we can use the usual thorem \ref{NullityOfIntegrals} to say that they do not contribute. (Note again that we choose the logarithmic cut of $K_1$ inside the contour $\mathcal{C}$, and that we can do that because the contour $\mathcal{C}$ contains $x_1$.)
\bea
&& W_{n+1}^{(g)}(x_0,x_1,J)-W_{n+1}^{(g)}(x_1,x_0,J) \cr
&=& 2 \oint_{\mathcal{C}} \,dx \,\, \left( K(x_0,x)\,\,  B(x,x_1)  - K(x_1,x)\,\,  B(x,x_0) \right) \, W_{n}^{(g)}(x,J) \cr
&+& 2  \oint_{\mathcal{C}} \,dx \,\, \Big( K'(x_0,x) K(x_1,x)-K(x_0,x) K'(x_1,x)\Big)\cr
&&(Y(x) - \hbar \partial_x) W_{n}^{(g)}(x,J)    \cr
\eea
Notice that:
\beq
K'_0 K_1 - K_0 K'_1 = -{1\over \hbar}\,( G_0 K_1 - K_0 G_1)
\eeq
and $B=-{1\over 2}\, G'$, therefore:
\bea
&& W_{n+1}^{(g)}(x_0,x_1,J)-W_{n+1}^{(g)}(x_1,x_0,J) \cr
&=& - 2  \oint_{\mathcal{C}} \,dx \,\, \left( K_0 G'_1  - K_1 G'_0 \right) \, W_{n}^{(g)}(x,J) \cr
&& -{1\over \hbar}  2  \oint_{\mathcal{C}} \,dx \,\, \Big( G_0 K_1- K_0 G_1 \Big)\,\, 
(Y(x) - \hbar \partial_x) W_{n}^{(g)}(x,J)    \cr
\eea
we integrate the first line by parts:
\bea
&& W_{n+1}^{(g)}(x_0,x_1,J)-W_{n+1}^{(g)}(x_1,x_0,J) \cr
&=&  \oint_{\mathcal{C}} \,dx \,\, \left( K'_0 G_1  - K'_1 G_0 \right) \, W_{n}^{(g)}(x,J) \cr
&& +  \oint_{\mathcal{C}} \,dx \,\, \left( K_0 G_1  - K_1 G_0 \right) \, W_{n}^{(g)}(x,J)' \cr
&& -{1\over \hbar}  \oint_{\mathcal{C}} \,dx \,\, \,\, \Big( G_0 K_1- K_0 G_1 \Big)\,\, 
(Y(x) - \hbar \partial_x) W_{n}^{(g)}(x,J)    \cr
\eea
Notice that:
\beq
K'_0 G_1 - G_0 K'_1 = -{Y\over \hbar}\,( K_0 G_1 - G_0 K_1)
\eeq
So we find \beq  W_{n+1}^{(g)}(x_0,x_1,J)-W_{n+1}^{(g)}(x_1,x_0,J)=0\eeq

\vfill\eject

\end{document}